# Electronic and Thermoelectric Properties of Molecular Junctions Incorporating Organometallic Complexes: Implications for Thermoelectric Energy Conversion

*Joseane Santos Almeida,[1] Sergio González Casal,[2] Hassan Al Sabea,[1] Valentin Barth,[3] Gautam Mitra,[3] Vincent Delmas,[1] David Guérin,[2] Olivier Galangau,[1] Tiark Tiwary,[3] Thierry Roisnel,[1] Vincent Dorcet,[1] Lucie Norel,[1] Colin Van Dyck,[4] Elke Scheer,[3*] Dominique Vuillaume,[2*] Jérôme Cornil,[5] Stéphane Rigaut,[1*] Karine Costuas[1*]*

1 Univ Rennes, CNRS ISCR (Institut des Sciences Chimiques de Rennes), Rennes, F-35042, France

2 Institute for Electronics, Microelectronics and Nanotechnology (IEMN), CNRS, Villeneuve d'Ascq, F-59652, France.

3 Department of Physics, University of Konstanz, Konstanz, D-78464, Germany

4 Theoretical chemical physics group, Department of Physics, University of Mons, Mons, B-7000, Belgium

5 Laboratory for Chemistry of Novel Materials, University of Mons, Mons, B-7000, Belgium

*Email: elke.scheer@uni-konstanz.de; dominique.vuillaume@iemn.fr; stephane.rigaut@univ-rennes1.fr; karine.costuas@univ-rennes1.fr

ABSTRACT. The electronic and thermoelectric properties of molecular junctions formed from iron and ruthenium metal-acetylide were studied using complementary experimental techniques and quantum chemical simulations. We performed physical characterizations of single-molecule and self-assembled monolayer junctions of the same molecules that allowed meaningful comparisons between the Ru and Fe adducts. In the case of the Fe-containing junctions, two distinct oxidation states are present. These junctions exhibit one of the highest Seebeck coefficients ($S$ ~130 μV/K) reported to date for similar systems paired with broad electric conductance distribution and limited thermal conductance. As a result, the experimental thermoelectric figure of merit $ZT$ for Fe-containing junctions reaches up to 0.4 for junctions with relatively high conductance. This is one of the highest $ZT$ values reported for molecular systems at room temperature.

**INTRODUCTION**

The worldwide research investments in molecular electronics, opto-electronics and spintronics have produced major advancements of basic understanding and led to the discovery of novel physical phenomena.[1,2] In the course of developing nano-devices for opto-electronics, the control of energy flux at the nanoscale will be compulsory. Therefore, the developments of nano-units exploiting thermoelectric properties appear to be an appealing manner to regulate heat flux and to convert it to electrical power. The theoretical and experimental progress and perspectives of the thermoelectric properties of molecular junctions have been discussed by several authors.[3–6] It appears that heat transport characterization in molecular junctions is still in its infancy because (i) it remains quite challenging from the experimental point of view, and (ii) the large majority of experimental studies is performed on molecular junctions involving main group atoms.[7–9] There is an obvious need to extend the number and the diversity of experimental studies on molecular junctions in which the characterization of thermal and electron transport is combined with thermopower measurements performed in the same experimental conditions,[10] as exemplified by recent works on oligophenyleneethynylene derivatives.[11–14]

In recent years, organometallic molecular junctions achieved rising attention.[15–20] Several reasons motivate this interest, the main one being that the integration of metal centers surrounded by ligands offers the possibility to modify the electronic transmission properties. Indeed, several families of organometallic systems show a higher electric conductance than comparable full-organic analogues,[21,22] and more importantly, they provide additional functionalities (redox, magnetic, optical). Part of this advantage lies in the ease of accessing several oxidation states and spin states, which leads to specific transmission features not encountered in purely organic junctions.[17,23–24] Actually, quantum interference (QI) effects can also be triggered by coupling of the orbitals supported by the backbone to those localized on the ligands.[25] Moreover, they offer the possibility of fine tuning the electronic properties by chemical modifications of the metallic center, conjugated core or ligands.[26]

The pioneering work in that domain was the synthesis and measurement of the conductance properties of a series of Ru-containing crosswire junctions by some of us and the group of D.

Frisbie (*trans*-[Ru(dppe)$_2$-bis(σ-arylacetylide)]; dppe = diphenyl-diphosphino-ethane)).[27] Since then, molecular junctions including Ru bis(σ-acetylide) groups have been the subject of numerous studies.[17,20–22,28–32] Considering the large number of measurements performed on these systems, it appears to be interesting to provide an additional study not available to date that includes both heat conducting properties and Seebeck coefficient measurements from which the thermoelectric power factor and the figure of merit can be extracted. A comparison between published studies is currently hampered by critical differences in the experimental protocols, in the exact chemical structure or in the nature of the anchoring groups.

Here we report the comparison of the electrical and thermoelectric properties of junctions of two parent organometallic compounds with almost the same length and terminated with the same linkers (protected thiol groups) to the Au leads (Scheme 1). These molecular wires are composed of two arylacetylide groups linked in *trans* position to a metal center, e. g., a Ru or a Fe group. The Ru-containing system *trans*-[Ru(dppe)$_2$(C≡C-*p*-C$_6$H$_4$-S(EDMS))$_2$] (EDMS = ethyl(dimethyl)silane)) is part of the above-mentioned *trans*-Ru(dppe)$_2$ bis(σ-arylacetylide) family, in which the Ru center in a *pseudo*-octahedral environment is formally in the +II oxidation state in an eighteen-electron closed-shell configuration. In the second system, the metallic group, also in *pseudo*-octahedral symmetry, is a Fe(cyclam) (cyclam = 1,4,8,11-tetraazacyclotetradecane) center.[33] In that case, the precursor [Fe(cyclam)(C≡C-*p*-C$_6$H$_4$-S(EDMS))$_2$]$^+$ is charged +1 and paramagnetic. In this paramagnetic molecule, the Fe center oxidation state is formally +III (Fe$^{III}$). It has been shown that the spin density is not exclusively localized on the Fe atom, but spreads over the carbon atoms of the conjugated arylacetylide ligands.[33,34] Importantly, during the fabrication of the device and the contacting to electrodes, the molecule may either keep this redox state (**[Fe-cyclam]$^+$**) or to be reduced and then adopt a closed-shell electronic structure **[Fe-**

**cyclam]$^0$**. The electronic transmissions of molecular junctions fabricated using the precursors, labelled hereafter **Ru-dppe** and **Fe-cyclam** junctions, are thus expected to be noticeably different even if their length and the organic conjugated linkers are the same.

We provide below a detailed characterization of the thermal and electric conductance properties of those compounds as an *ensemble* of molecules i.e., in self-assembled monolayers (SAM), using Conductive Atomic Force Microscopy (C-AFM) and Scanning Thermal Microscopy (SThM). In addition, we studied the electric transport properties in *single-molecule junctions* using the Mechanically Controllable Break Junction (MCBJ) technique in vacuum and at low temperature to gain more insight into the transport behavior of individual junctions beyond statistical information and to decipher by comparison the effect of intermolecular interactions onto the electric transport properties. This comprehensive study offers one of the rare examples found in the literature to compare the transport properties of the same molecules measured by different techniques under different conditions.[35,36] To the best of our knowledge, this fundamental comparison is yet to be achieved for thermoelectric measurements. Importantly, quantum computing calculations are completing the study by providing relevant information on the electronic structures and electronic transmission features of model molecular junctions to better understand the experimental data.

The electronic transport properties were first analyzed in a single-molecule junction configuration. In particular, the influence of the bias voltage range on the shape of the current-voltage curves and the electrical conductance values upon elongation of the junction were investigated. The electrical conductance of SAM-based junctions of both systems shows differences and similarities with the single-molecule junction measurements that are discussed in the light of a thorough quantum chemical study of the transport properties of model junctions. The thermal conductance and thermoelectric properties of the SAMs were calculated allowing for a complete analysis of the electronic and thermoelectric properties of parent organometallic junctions.

**RESULTS AND DISCUSSION**

**Chemical synthesis and characterization of Ru(dppe)$_2$(C≡C-*p*-C$_6$H$_4$-S(EDMS))$_2$ and [Fe(cyclam)(C≡C-*p*-C$_6$H$_4$-S(EDMS))$_2$](OTf)**

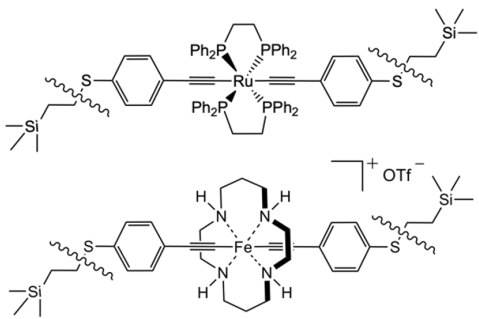

**Scheme 1.** Schematic representation of the *trans*-[Ru(dppe)$_2$(C≡C-*p*-C$_6$H$_4$-S(EDMS))$_2$] (top) and *trans*-[Fe(cyclam)(C≡C-*p*-C$_6$H$_4$-S(EDMS))$_2$](OTf) (bottom) molecule.

The systems presented in Scheme 1 were synthesized by modifying the known synthetic procedures of their parent compounds incorporating the EDMS protecting end-groups. The new Ru compound *trans*-[Ru(dppe)$_2$(C≡C-*p*-C$_6$H$_4$-S(EDMS))$_2$] was prepared in high yield (92 %) according to the classical procedure[37] consisting in the reaction of *cis*-[RuCl$_2$(dppe)$_2$] and 2-(trimethylsilyl)ethyl-4'-(ethynyl)phenyl sulfide (**1**) in the presence of sodium hexafluorophosphate and triethylamine, as shown in Scheme 2 and was fully characterized (see SI). Noteworthy is the FT-IR spectrum with the C≡C stretching at $\upsilon_a = 2051$ cm$^{-1}$ and the singlet at 52.76 ppm in the $^{31}$P NMR spectrum corresponding to the four phosphorus atoms of the dppe moiety in the *trans* arrangement.[37] Concerning the new Fe complex with the crown ligand, based on published procedures for related complexes,[33,38] the terminal alkyne **1** and the triflate precursor [Fe(cyclam)(OTf)$_2$](OTf) were mixed in THF containing four equivalents of n-BuLi to lead to the formation of *trans*-[Fe(cyclam)(C≡C-*p*-C$_6$H$_4$-S(EDMS))$_2$](OTf) in moderate to good yields (75 %) (Scheme 2). The complex was characterized by HR-MS spectrometry (SI). Raman spectra reveal features characteristic for *trans*-isomers with a single band that falls in the C≡C stretching region at $\upsilon_s$ = 2083 cm$^{-1}$.[33,28] Importantly, we could obtain crystals of good quality of both compounds to perform X-ray diffraction characterizations.

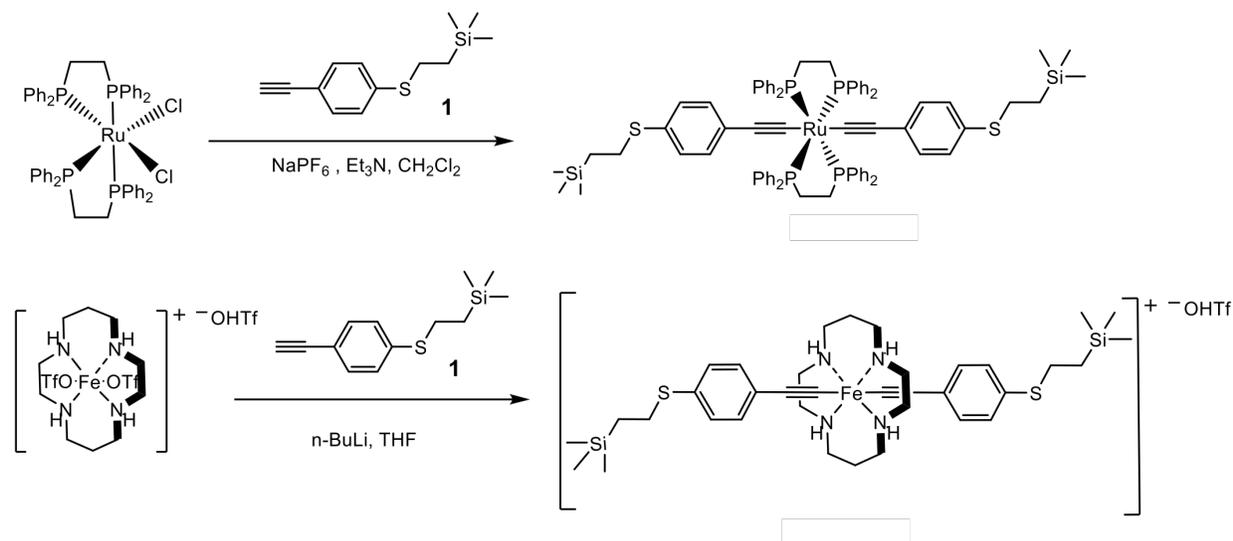

**Scheme 2**. Preparation of *trans*-[Ru(dppe)$_2$(C≡C-*p*-C$_6$H$_4$-S(EDMS))$_2$] and *trans*-[Fe(cyclam)(C≡C-*p*-C$_6$H$_4$-S(EDMS))$_2$](OTf).

The *trans*-[Ru(dppe)$_2$(C≡C-*p*-C$_6$H$_4$-S(EDMS))$_2$] compound crystallizes in a triclinic crystal system within the P$\bar{1}$ space group; the crystallographic data are presented in the supplementary information (Table S1). The structural solid-state arrangement shown in Figure 1 evidences the *trans*-isomer formation with the hexa-coordinated Ru center. The four phosphorus atoms form a plane orthogonal to the bis-acetylide moieties, which are almost aligned (torsion angle formed by the two first carbon atoms of each acetylide = 0.77°). All bond lengths are consistent with analogous Ru complexes and deserve no further comment.[39] The *trans*-[Fe(cyclam)(C≡C-*p*-C$_6$H$_4$-

S(EDMS))$_2$](OTf) complex crystallizes in a monoclinic crystal system within the P2$_1$ space group. The coordination geometry of the transition metal center is pseudo-octahedral with a slight distortion from a perfect octahedron. All bond lengths and angles are consistent with published data on analogous complexes.[33,38] We observe the *trans*-isomer configuration with the cyclam coordinated to the metal center through the four N atoms that lie perpendicular to the C-rich axis. The C-M-C angle involving the two C atoms of the acetylide moieties reaches linearity (179.70(2)°). The two bis-acetylide moieties are not aligned but bent with respect to each other by 31.06°.

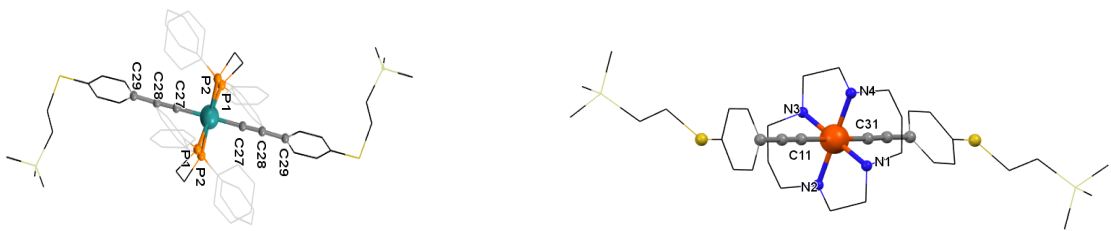

**Figure 1.** X-ray structure of *trans*-[Ru(dppe)$_2$(C≡C-*p*-C$_6$H$_4$-S(EDMS))$_2$] (left) and *trans*-[Fe(cyclam)(C≡C-*p*-C$_6$H$_4$-S(EDMS))$_2$](OTf) (right) in two viewing orientations. H atoms are omitted for clarity. S, N, C, P, Fe and Ru are shown as gold, blue, grey, orange, red and green spheres, respectively, while the terminal parts of the structure are shown in wireframe style.

Cyclic voltammetry was conducted on the two complexes in dichloromethane containing 0.1 M [TBA](PF$_6$). The *trans*-[Fe(cyclam)(C≡C-*p*-C$_6$H$_4$-S(EDMS))$_2$](OTf) complex, in which the Fe center is formally in +III oxidation state (17-electron metal center), undergoes two different electrochemical processes: (i) a Fe$^{III}$/Fe$^{IV}$ irreversible oxidation process at E = 1.05 V vs. SCE (0.61 V vs. Fc/Fc$^+$, 0.1 V.s$^{-1}$) not observed by T. Ren et al. in related complexes (Figure S1) and that reveals the enhanced richness of this Fe complex[33] and (ii) a Fe$^{III}$/Fe$^{II}$ chemically reversible reduction wave at E$^0_{1/2}$ = -0.71 V vs. SCE (E$^0_{1/2}$ = -1.15 V vs. Fc/Fc$^+$) but featuring an irreversible electrochemical (slow) process ($\Delta E_p$ = 100 mV). The *trans*-[Ru(dppe)$_2$(C≡C-Ph-S(EDMS))$_2$] complex, in which the Ru center is in an eighteen-electron environment (Ru$^{II}$), undergoes a first electrochemical reversible oxidation process at $E^0_{1/2}$ = 0.34 V vs. SCE ($\Delta E$ = 65 mV, $E^0_{1/2}$ = -0.10 V vs. Fc/Fc$^+$, 0.1 Vs$^{-1}$) (Figure S2). The oxidation weakly involves the Ru orbitals, whereas it actually strongly involves the carbon-rich ligands, as proven by theoretical studies for analogous Ru bis-acetylide complexes.[34] A second irreversible oxidation process commonly observed in analogous complexes was observed at $E^0_{pa}$ = 1.04 V vs. SCE (0.6 V vs. Fc/Fc$^+$, 0.1 Vs$^{-1}$).[39]

**Single-Molecule Conductance Measurements**

Measurements of single-molecule junctions at low temperature were performed using the lithographically fabricated MCBJ method using gold as electrode material. The molecules were dissolved in tetrahydrofuran (THF) or in a mixture of 20 % THF and 80 % ethanol. To remove the protection groups, a droplet of tetrabutylammoniumfluorid (TBAF) was added to the solution just before the molecular deposition which is done either using the drop-casting or the immersion method. Samples prepared with the immersion method were stored in the molecular solution as specified in the Supporting Information (SI). For the drop-cast method, a single drop of the solution was poured onto the junction. In both cases the connection to the electrical circuit was done after the solvent was completely evaporated and the sample blown dry in gaseous nitrogen. The electric transport measurements were performed in cryogenic vacuum at 4.2 K. Details of the sample preparation, the measurement setup as well as the deposition parameters for each sample are summarized in Table S2 in the SI.

A first set of data was obtained by repeatedly stretching and relaxing the contacts, by bending the substrate, to determine statistically the electric conductance *G* of a single-molecule junction. The number of stretching-relaxing cycles that can be recorded is a few hundreds per sample (limited by the mechanical movement speed). $G$ is determined as the ratio between current and voltage across the junction at a fixed applied bias of 100 mV and with a serial resistance for limiting the maximum current. The $G$ values of all stretching traces obtained for all samples are collected into a (one-dimensional) conductance histogram (Figure 2-a for **Fe-cyclam** and Figure 2-c for **Ru-dppe**). For further analysis, the conductance and displacement data are also compiled in two-dimensional histograms (Figure 2-b and Figure 2-d).

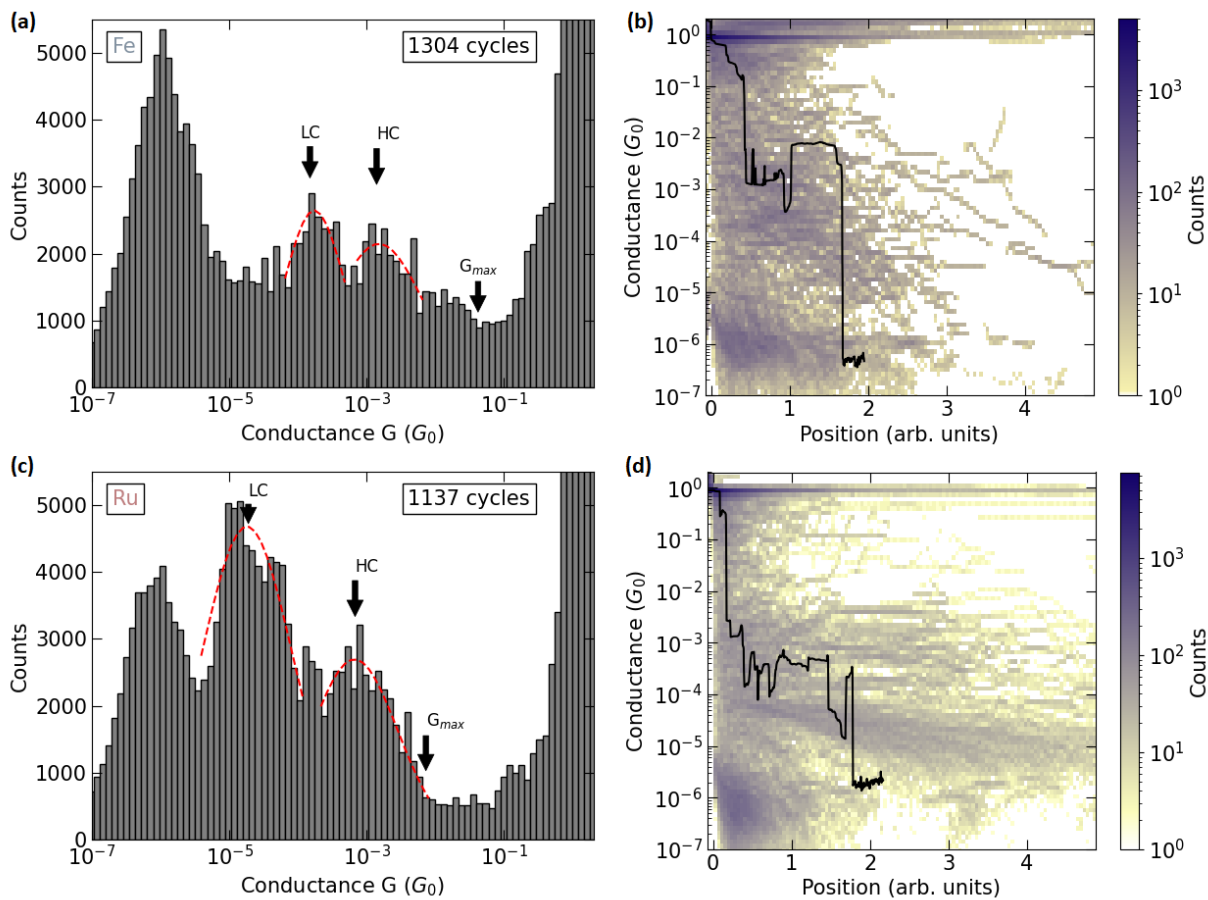

Figure 2. Conductance histograms of stretching curves gathering all stretching traces obtained for all samples for both molecules. (a) One-dimensional histogram of Fe-cyclam generated through 1304 stretching curves. LC (HC) denotes low (high) conductance peaks, $G_{max}$ the high-conductance edge of the range with increased count rate. The red dashed lines indicate the Gaussian fitting of the LC and HC peaks. (b) Two-dimensional histogram of the same dataset. One unit of the position axis is of the order of one nanometer. The black curve shows an example for a stretching trace. (c) and (d) represent equivalent data for Ru-dppe (1137 stretching traces). All measurements were performed with an applied voltage of 100 mV and at 4.2 K.

Both histograms show peaks around $10^{-6}$ $G_0$ (conductance quantum $G_0 = 2e^2/h \sim 77.5$ $\mu S$) corresponding to vacuum tunneling though disconnected electrodes. Figure 2-a reveals two additional regions with enhanced number of counts around $G = 1.7 \times 10^{-4}$ $G_0$ and at $G = 1.5 \times 10^{-3}$ $G_0$ for **Fe-cyclam**, determined by fitting the histogram peaks with a Gaussian distribution. These high-conductance (HC) and low-conductance (LC) maxima are not very pronounced, and further shallow maxima are found at even higher $G$ values. The count rate drops at $G_{max} \sim 4.2 \times 10^{-2}$ $G_0$. This "upper shoulder" $G_{max}$ of broad conductance distributions may be interpreted as the conductance of an optimally and symmetrically bound molecular junction according to Eq. S4 (see below). The peak around 1 $G_0$ is indicative for Au single-atom contacts. The pronounced tail towards $10^{-1}$ $G_0$ is typical for lithographic MCBJs at low temperature.[39] These contacts may correspond either to disordered Au contacts or multi-molecule contacts.[2] In the two-dimensional representation of the data (Figure 2-b), longer plateaus can be identified in the broad conductance range from around $G = 8 \times 10^{-5}$ $G_0$ to about $G = 6 \times 10^{-3}$ $G_0$ while for lower and higher conductance the plateaus are shorter. The black curve in Figure 2-b is one example of a stretching curve that shows a relatively short plateau in the HC range and then an abrupt break-down to the tunneling regime. We assign the whole range as possible conductance values of single-molecule junctions with the two most probable values, HC and LC, seen in Figure 2-a. A possible origin of these two preferred conductance regions could be the different oxidation states of the Fe-core (see below).

Figure 2-c and Figure 2-d show equivalent data for **Ru-dppe**. The conductance peaks and plateaus at $10^{-6}$ and 1 $G_0$ correspond to tunneling through broken junctions and atomic contacts, respectively. In the intermediate conductance range enhanced count rates are observed in the one-dimensional histogram (Figure 2-c) around $G_{LC} = 1.8 \times 10^{-5}\ G_0$ and $G_{HC} = 6.9 \times 10^{-4}\ G_0$ and $G_{max} \sim 7.3 \times 10^{-3}\ G_0$. The two-dimensional histogram shows that the contacts in the LC regime tend to be longer, indicating fully stretched junctions. Furthermore, the example stretching trace shown in Figure 2-d shows that in the HC regime several pronounced conductance jumps occur which may hallmark electronic changes or rearrangements of the junction.[2]

To verify that the recorded data indeed corresponds to molecular junctions, we performed inelastic electron tunneling spectroscopy (IETS) for selected stable junctions. The details are given in the SI (Figure S9 and associated text). We also checked the effect of the application of a magnetic field on the electronic transport properties. Indeed, the **Fe-cyclam** junctions might be magnetic in the case of **[Fe-cyclam]⁺** molecules which comprise one unpaired electron. However, no significant effect of an applied magnetic field on the transport properties of **Fe-cyclam** junctions, nor of the **Ru-dppe** junctions could be detected (see SI).

To gain more insight into the transport mechanism of these junctions, current-voltage (*I-V*) curves were recorded over two different bias ranges. The measured *I-V* curves are plotted in Figure 3. Junctions revealing *I-V* characteristics with sudden jumps are mostly observed in the LC range for ultimately stretched molecular junctions. This non-reversible behavior of *I-V*s is a hallmark of unstable junctions (e.g., due to reconfiguration of the bonding to the electrodes). Only the *I-V* curves of junctions stable over the selected bias range were kept for further analysis. Furthermore, for **Ru-dppe**, we focused the analysis on the conductance range between $G_{HC} \sim 10^{-3}$ and $G_{max} \sim 10^{-2}\ G_0$ to study the possible origins of the multi-jumpy behavior of the stretching curves shown in

Figure 2-d. This implies that the statistical weight of the *I-V*s is not the same as in the conductance histograms. Figure 3-a and Figure 3-c show the *I-V* curves for **Fe-cyclam** which reveal the typical and mainly symmetric shape for off-resonant tunneling through molecular junctions. The spread of the data reflects the wide conductance range already observed in the histograms in Figure 2. The red curves indicate numerical averages over all data shown in the respective panel to reflect the typical behavior. The low-bias curves shown in Figure 3-a mostly have a similar shape, e.g., the curvature occurs at a similar voltage value despite the large conductance range spanned. This indicates that the electronic structures of these junctions are similar. The data for higher bias shown in Figure 3-c points to at least two families of curves, one of which is similar to the low-bias data from Figure 3-a. The second one shows less curvature and also slightly asymmetric behavior as shown in the inset of Figure 3-c.

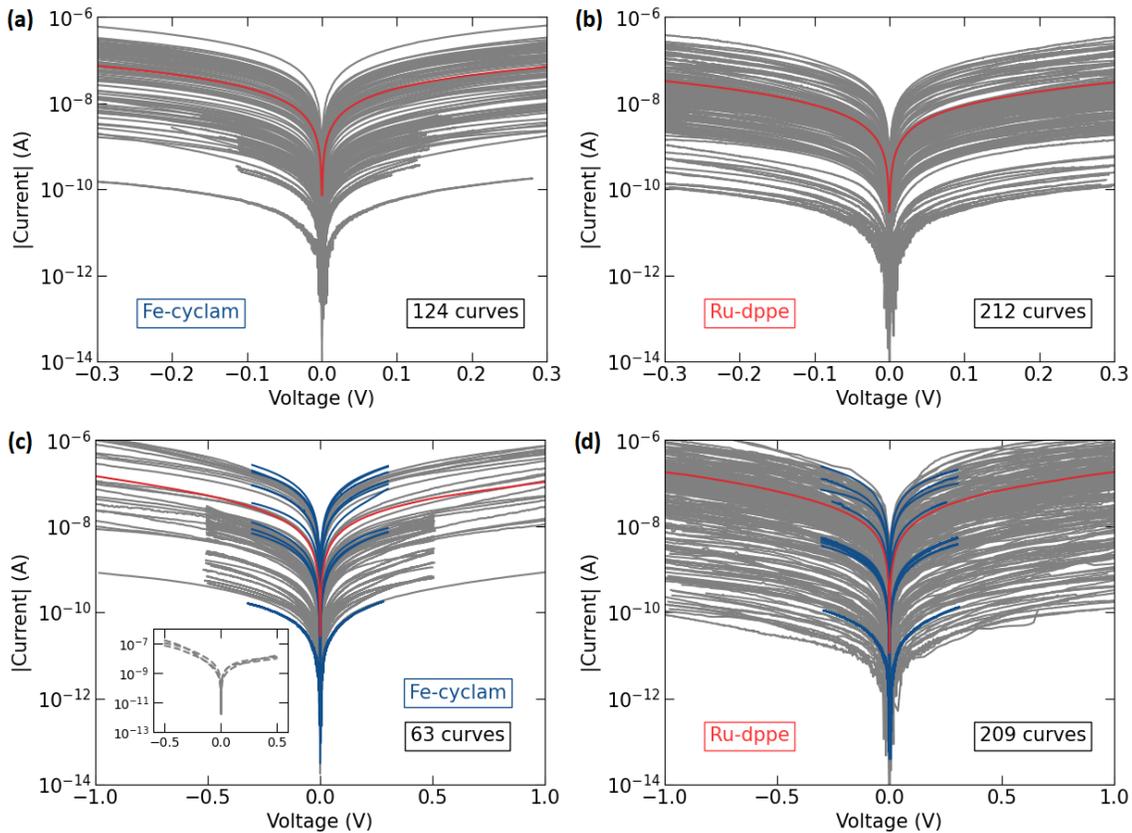

**Figure 3.** Current-voltage (*I-V*) measurements for (a) **Fe-cyclam** and (b) **Ru-dppe.** *I-V* characteristic over a wider voltage range for (c) **Fe-cyclam** and (d) **Ru-dppe.** The number of *I-V* curves in each figure is indicated in the boxes in the lower right corners. The red lines represent the average of the individual data sets corresponding to a zero-bias conductance of (a) $G = 2.84 \times 10^{-3}\ G_0$, (b) $G = 1.28 \times 10^{-3}\ G_0$, (c) $G = 1.10 \times 10^{-3}\ G_0$, (d) $G = 4.4 \times 10^{-3}\ G_0$. The blue curves in (c) and (d) are selected curves from (a) and (b), respectively, to compare the small-bias range with the high-bias range data. The inset in (c) represents examples of asymmetric curves extracted from (c). Unstable *I-V*s are not shown in this figure.

For **Ru-dppe** (Figure 3-b and Figure 3-d), we observe a similar shape of the *I-V*s also over a wide spread current range. As mentioned above, the majority of data has been sampled on purpose in the intermediate to higher conductance range to investigate possible origins of the jumpy stretching curves shown in Figure 2-d. Nevertheless, also for relatively low conductance, stable *I-V*s are observed over the whole bias range, in agreement with the observations from the histograms shown in Figure 2-c and Figure 2-d.

The individual stable *I-V* curves are analyzed using the single-level model (SLM) (Eq. S4) which assumes that a single molecular orbital couples to the electrodes such that its transmission function adopts a Lorentzian shape the maximum of which is located at the energy $\epsilon_0$ with respect to the Fermi energy (called the level alignment and the electronic couplings with the electrodes are labelled $\Gamma_1$ and $\Gamma_2$ (labeled $\Gamma$ for symmetric junctions). We systematically performed fits with the SLM for all experimental *I-V* curves. Examples of individual experimental *I-V* fits for both molecules are shown in Figure S4. The analysis shows that the fit results depend on the voltage range used for fitting the data. This is not unexpected since the SLM is a small-bias approximation. On the other hand, a meaningful determination of the parameters requires the bias range large enough to show a non-linear *I-V*. A detailed discussion also addressing the limitations of the SLM is given in the SI. Fits for small (|V| < 0.5 V) are given in Figure 4 and large (|V| > 0.5 V) bias ranges are provided in Figure S5 (Section Fitting results in the SI).

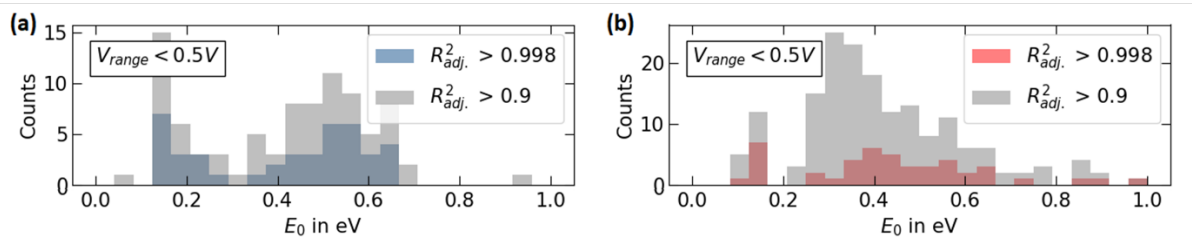

**Figure 4.** $\epsilon_0$ statistics of all *I-V* curves fitted with the SLM separated by the goodness of fit ($R^2_{adj}$) for bias range ($V_{range}$) inferior to 0.5 V. (a) **Fe-cyclam**; 56 *I-V*s have $R^2_{adj} > 0.998$. (b) **Ru-dppe**; 231 *I-V*s have $R^2_{adj} > 0.998$.

The result of the fits shown in Figure S4-a and Figure S4-b suggests that for both molecules the *I-V*s can be well described by the SLM, as confirmed by a high goodness-of-fit[40] $R^2_{adj}$ value. The fits of the *I-V* measurements of **Ru-dppe** with $R^2_{adj} > 0.998$ result in a distribution of $\epsilon_0$ with a narrow peak around 0.15 eV few junctions and a broad peak centered at roughly 0.4 eV with a tail towards higher values. Fits over the larger voltage range $V_{range} > 0.5$ V confirm the latter while the smaller $\epsilon_0$ is not recovered. For **Fe-cyclam**, in the small-bias range, the $\epsilon_0$ histogram reveals two broad maxima, one located at $\epsilon_0 = 0.15$ eV and the other around $\epsilon_0 = 0.55$ eV. The two different $\epsilon_0$ values emerging from this analysis are in line with expectation of two different oxidation states of the molecule. The small curvature of the *I-V* curves for **Fe-cyclam** (Figure S4-a) leads to an important loss in accuracy preventing further conclusion for higher bias ranges (Figure S5). The coupling constant extraction is given in the SI (Figure S6 and associated text).

**SAM junctions: Electronic and Heat Conductance and Seebeck Coefficient Measurements**

***Physical characterization of the SAMs.*** The self-assembled monolayers (SAMs) of the **Ru-dppe** and **Fe-cyclam** molecules were formed by dipping ultra-flat template-stripped Au electrodes ($^{TS}$Au) in solutions of [Ru(dppe)$_2$(C≡C-Ph-S(EDMS))$_2$] and [Fe(cyclam)(C≡C-Ph-S(EDMS))$_2$](OTf) at 10$^{-3}$ M in THF/ethanol (50/50) for one day (for more details see the SI). The SAMs were physically characterized by tapping-mode AFM (TM-AFM), ellipsometry, X-ray photoemission spectroscopy (XPS) and ultraviolet photoemission spectroscopy (UPS). The topographic TM-AFM images (Figure S10) show that the SAMs are homogeneous and flat and free of gross defects; the measured rms roughness for the **Ru-dppe** SAM is ≈ 0.7 nm and ≈ 0.5 nm for the **Fe-cyclam** SAM. Both values are close to the roughness observed for the $^{TS}$Au substrates (ca. 0.4 nm),[41] which indicates the good formation of a monolayer. The SAM thicknesses measured by ellipsometry are 1.5 ± 0.2 nm and 1.8 ± 0.2 nm for the **Ru-dppe** and **Fe-cyclam**, respectively. Since these values are smaller than the length of the molecule (1.88 nm and 1.84 nm, respectively), we estimated a tilt angle of ≈ 38° and ≈ 10° from the Au surface normal, respectively (SI, Table S4). The molecules in the **Ru-dppe** SAM are less packed than in the **Fe-cyclam** SAM, as expected since the **Ru-dppe** is bulkier because of the phenyl groups of the dppe ligand. The tilt angle for the **Ru-dppe** molecules is similar to that already measured for similar Ru-derivatives.[42] The XPS spectra in the S2p region (Figure S11) clearly show four peaks corresponding to the doublet (2p$_{1/2}$, 2p$_{3/2}$) of the S-Au bond and to the S atom unbounded to Au (i.e., S-C) at the top surface of the SAMs. The peak amplitude ratios [S-Au]/[S-C] are ≈ 0.65 for the two SAMs, while a ratio of 1 is expected if all molecules in the SAM are chemisorbed on Au. This is not surprising considering the steric hindrance in these molecules with the dppe or cyclam ligands around the metal atom. It is likely that a fraction of the molecules in the SAMs is wedged between neighboring chemisorbed molecules without forming an S-Au bond with the substrate. For the **Ru-dppe** SAM, the Ru3d$_{3/2}$ peak is observed (Figure S12-a) with a small amplitude in agreement with other works on parent Ru$^{II}$(dppe)$_2$-containing SAMs.[27,42] The **Fe-cyclam** SAM displays the peaks associated to Fe2p in the Fe$^{II}$ and Fe$^{III}$ states (Figure S12-b), with an estimated ratio [Fe$^{II}$]/[Fe$^{III}$] ≈ 2. We note that no Si2p peak was observed for the two SAMs meaning that the sulfur protecting groups (EDMS) were removed during the SAM formation. On the basis of usual assumptions that are discussed below, the highest occupied molecular orbital (HOMO) energy ($\varepsilon_h^{UPS}$) can be estimated at 0.34 eV and 0.75 eV below the Au electrode Fermi energy from the UPS spectra of **Ru-dppe** and **Fe-cyclam** SAMs, respectively (Figure S13-a). From the secondary electron cutoff (Figure S13-b), we determine a decrease in the work function (WF) by 0.40 and 0.55 eV for the **Ru-dppe** and **Fe-cyclam** SAMs with respect to the naked Au, respectively.

**Electrical conductance.** Figure 5-a shows the typical *I-V* dataset of the **Ru-dppe** SAM measured by conductive AFM (C-AFM, measurement details and protocol in the SI). A similar behavior was observed for two other **Ru-dppe** SAMs (Figure S21). The zero-bias conductance of the SAM, $G_{SAM}(0)$, was determined from the slope of the *I-V* traces around 0 V and more precisely from another dataset measured between -75 and 75 mV (Figure 5-b). The statistical distribution of $G_{SAM}(0)$ is shown in Figure 5-c. The values follow a log-normal distribution with a mean $\bar{G}_{SAM}(0)$ = 7.9 × 10$^{-10}$ S. All the *I-V* traces of the dataset in Figure 5-a were fitted with the analytical SLM to determine the energy position $\epsilon_0^{SAM}$ (with respect to the Fermi energy of the electrodes) of the molecular orbital (MO), here the HOMO, involved in the electron transport across the molecular junctions (see details in the SI, Eq. S4). We obtained an energy level $\epsilon_0^{SAM}$ = 0.32 ± 0.12 eV (Figure 5-d). We note a very good agreement of the energy position with the one resulting from the MCBJ analysis and with the UPS determination of the HOMO onset (Figure S13-a) at $\varepsilon_h^{UPS}$ ≈ 0.34 eV for the **Ru-dppe** SAM (*vide supra*).

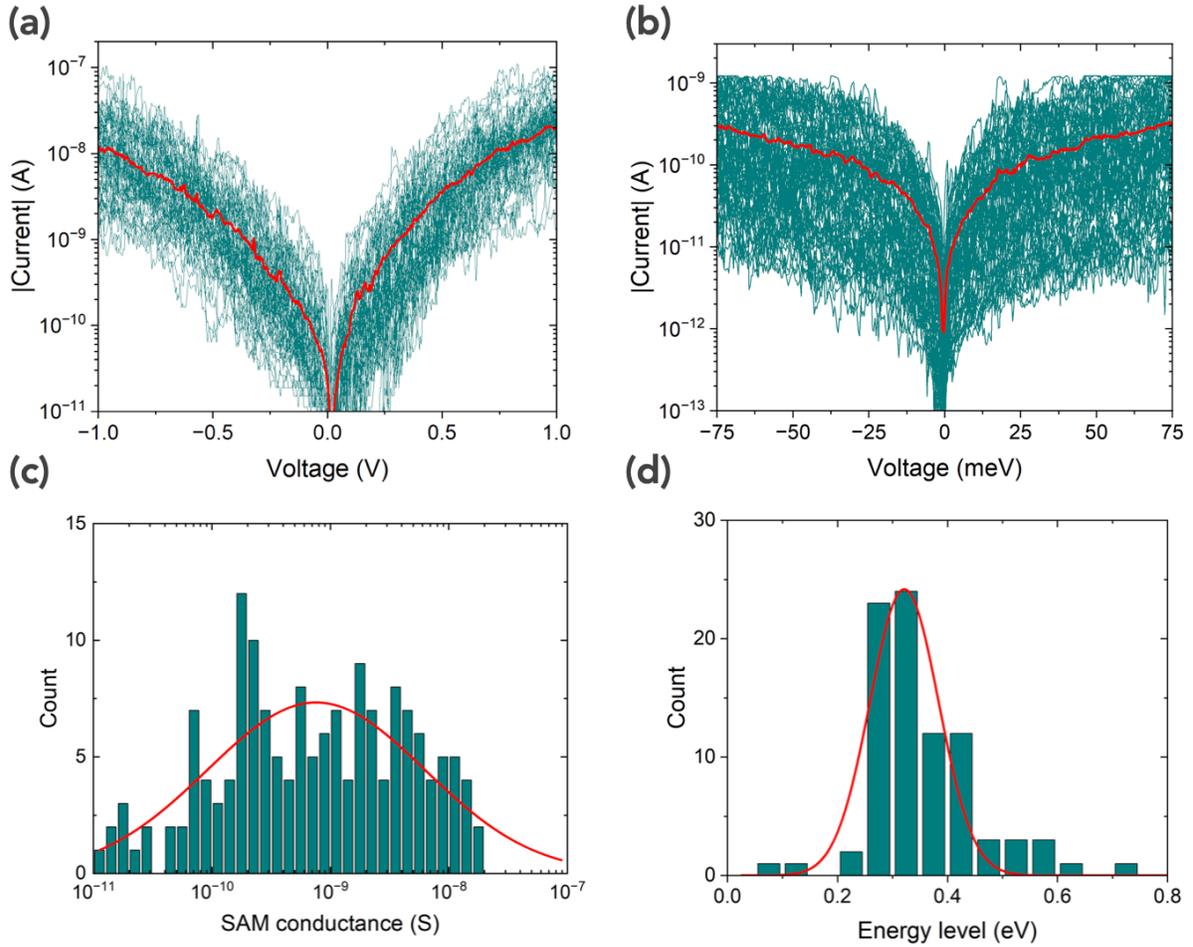

**Figure 5.** (a) Current-voltage (*I-V*) dataset for the **Ru-dppe** SAM (82 *I-V* traces, absolute current values, voltage step δ$V$ = 4 mV). (b) *I-V* dataset (84 *I-V* traces) acquired around zero bias with a better resolution (voltage step, δ$V$ = 0.4 mV). In panels a and b, the red lines are the mean current. (c) Zero-bias SAM conductance, $G_{SAM}(0)$, from datasets shown in panels a and b. (d) Histogram of the energy level $\epsilon_0^{SAM}$ determined by fitting the SLM (Eq. S4) on all *I-V* traces shown in the panel a. In panels c and d, the red lines are the fits of a log-normal distribution (log-mean = -9.1 (i.e. $\bar{G}_{SAM}(0) = 7.9 \times 10^{-10}$ S); log-standard deviation = 0.9) and of a Gaussian distribution (mean 0.32 eV, standard deviation 0.12 eV), respectively. The other SLM parameters (molecule-electrode coupling energies) are given in Figure S16.

The same measurements were performed for the **Fe-cyclam** SAM (Figure 6, and for two other SAMs see Figure S22). We clearly observe lower currents and zero-bias conductance than for the Ru-dppe SAMs. The mean zero-bias conductance is $\bar{G}_{SAM}(0) = 6.3 \times 10^{-11}$ S. The energy level histogram is broad as observed with the MCBJ technique (Figure S5-a); it can be fitted with a double-Gaussian distribution with two average electronic levels $\epsilon_0^{SAM}(1) = 0.33 \pm 0.06$ eV and $\epsilon_0^{SAM}(2) = 0.60 \pm 0.15$ eV, albeit the count for the $\epsilon_0^{SAM}(1)$ peak is low. The first one could be ascribed to the HOMO level of the molecule with the $Fe^{II}$ state and the second to the $Fe^{III}$ state, respectively (*vide infra*). Since the contact area of the C-AFM is small (≈ 18 nm², see the SI, C-AFM measurement protocol), it is likely that depending on the position of the tip on the surface, some locations show a majority of **[Fe-cyclam]⁰** molecules ($Fe^{II}$), while at other locations the **[Fe-cyclam]⁺** dominates ($Fe^{III}$). Moreover the spatial distribution of the **[Fe-cyclam]⁰** and **[Fe-cyclam]⁺** species is unknown, and it is likely that some nanoscale domains are formed. From UPS, since the UV spot is very large ca. mm², only an average value (≈ 0.75 eV, Figure S13-a) is obtained, which lies in the distribution of the main $\epsilon_0^{SAM}(2)$ peak deduced from the fits of the *I-V* curves (Figure 6-d).

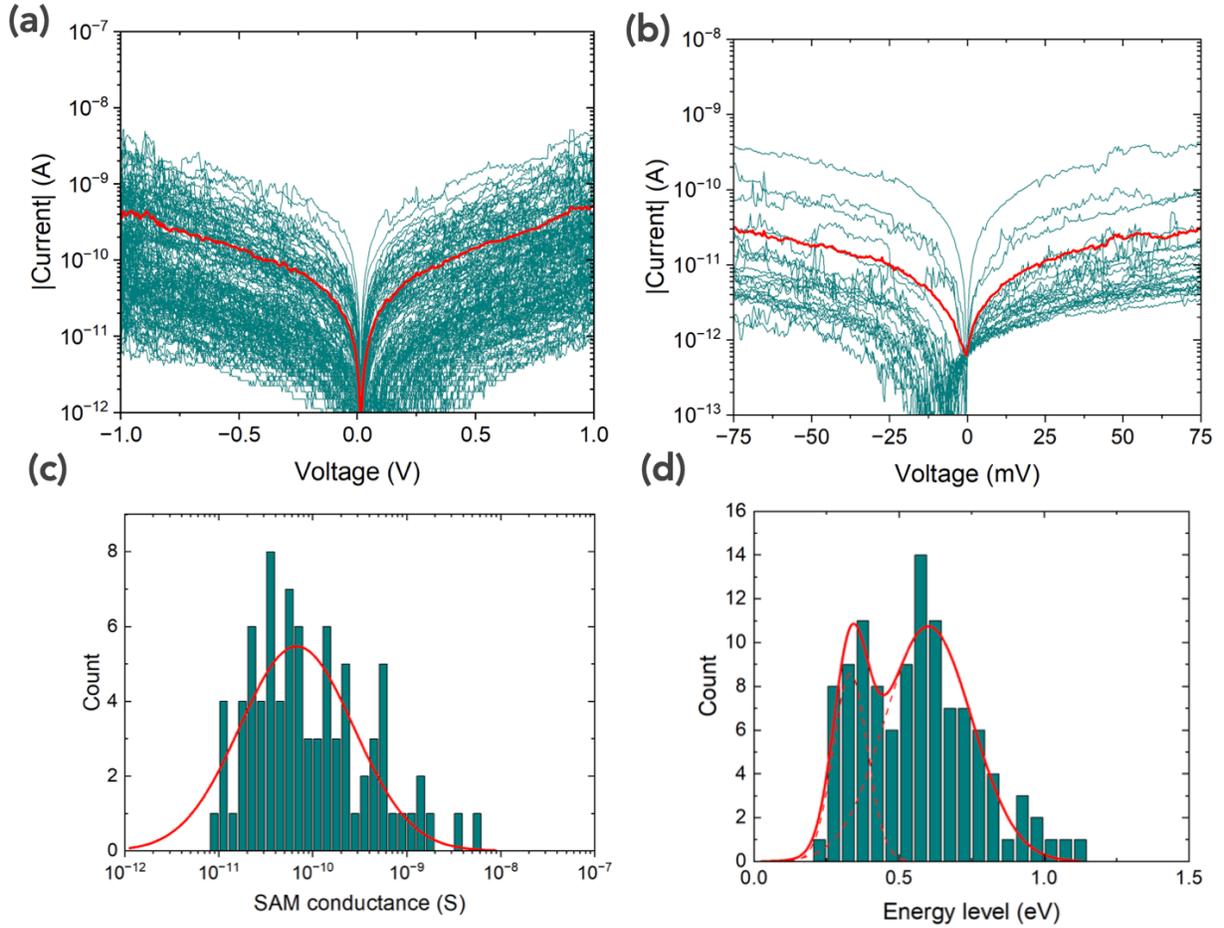

**Figure 6.** (a) Current-voltage (*I-V*) dataset for the **Fe-cyclam** SAM (127 *I-V* traces, current in absolute value, voltage step δ$V$ = 4 mV). (b) *I-V* dataset (27 *I-V* traces) acquired around zero bias with a better resolution (voltage step, δ$V$ = 0.4 mV). In panels a and b, the red lines are the mean current. (c) Zero-bias SAM conductance, $G_{SAM}(0)$, from datasets shown in panels a and b. (d) Histogram of the energy level $\epsilon_0^{SAM}$ determined by fitting the SLM (Eq. S4) on all *I-V* traces shown in the panel a. In panels c and d, the red lines are the fits of a log-normal distribution (log-mean = -10.2 (i.e. $\bar{G}_{SAM}(0) = 6.3 \times 10^{-11}$ S); log-standard deviation = 0.6) and of a double Gaussian distribution ($\epsilon_0^{SAM}(1)$= 0.33 ± 0.06 eV and $\epsilon_0^{SAM}(2)$= 0.60 ± 0.15 eV), respectively. The other SLM parameters (molecule-electrode coupling energies) are given in Figure S16.

The *I-V* dataset of the **Ru-dppe** SAM (Figure 5-a) is in good quantitative agreement with previous C-AFM measurements on SAMs of other Ru or Fe-containing molecules (zero-bias conductance between $10^{-10}$ and $10^{-6}$ S).[27,42] These previous studies have shown that the contact resistances are 10 to 100 times smaller than the lowest measured resistance for a Ru-dppe MJs and consequently they were neglected in the present study. However, those molecules are not strictly similar, since they feature different anchor groups and a $CH_2$ or $O(CH_2)_6$ moiety between the phenyl unit and the anchor.[27,42] The C-AFM setup details are not similar either, the tip radius being smaller in the present study ($\approx$ 20 nm vs. 50 nm). To tentatively compare the data with the single-molecule results of this study and previous reports,[20] we estimated a single-molecule conductance from the SAM measurements by dividing the value of $G_{SAM}(0)$ by the estimated number of molecules under the C-AFM tip, $N \approx 5$ and $\approx 15$ for the **Ru-dppe** and **Fe-cyclam** SAMs, respectively (SI, "C-AFM contact area" section). By doing so, we also assume that the SAM is formed of independent molecules conducting in parallel, *i.e.*, we do not consider intermolecular interactions and other factors, which can modify the total conductance of an ensemble of molecules (see Discussion section).[43,44] The **Ru-dppe** single-molecule conductance obtained from the mean $\bar{G}_{SAM}(0)$ is $\bar{G}_{mol}(0) \approx 2 \times 10^{-6}\ G_0$, with a maximum of the distribution at $\approx 5 \times 10^{-5}\ G_0$ (Figure 5-c).

**Thermal conductance.** The thermal conductance of the SAMs was measured by the null-point scanning thermal microscopy (NP-SThM) method,[45] which is a differential method measuring the temperature difference ($T_{NC}$ - $T_C$) between the tip and the sample just before ($T_{NC}$) and at contact ($T_C$) of the tip, see Figure 7-a. This contact induces a sudden jump from $T_{NC}$ to $T_C$ measuring the additional heat flux passing through the molecular junctions (note that the measured tip temperature starts decreasing slowly before the contact because the heat transfer through the air gap is increased when approaching the tip to the sample). This method removes the parasitic contributions (air thermal conduction, radiation...).[45] By varying the heat flux passing through the $^{TS}$Au-SAM/tip junctions (the tip is heated by increasing the voltage applied on the Wheatstone bridge $V_{WB}$, see the SI), we can plot the $T_C$ vs. ($T_{NC}$ - $T_C$) curves for the **Ru-dppe** and **Fe-cyclam** SAMs (Figure 7-b and Figure 7-c, respectively). The slope gives the thermal conductance of the SAM/Au samples, $G_{th}(SAM/Au)$, with the equation:[45]

$$T_C - T_{amb} = \left(\alpha \frac{4 r_{th}}{G_{th}(SAM/Au)} + \beta\right)(T_{NC} - T_C) \quad \text{Equation 1}$$

where $T_{amb}$ is the room temperature (RT) (22.5°C in our air-conditioned laboratory) and $r_{th}$ is the thermal contact radius of the SThM tip ≈ 20 nm (see SI). The calibration parameters, $\alpha$ and $\beta$, depend on the actual tip and equipment. They were systematically measured before all measurements shown in Figure 7 (more details in SI).

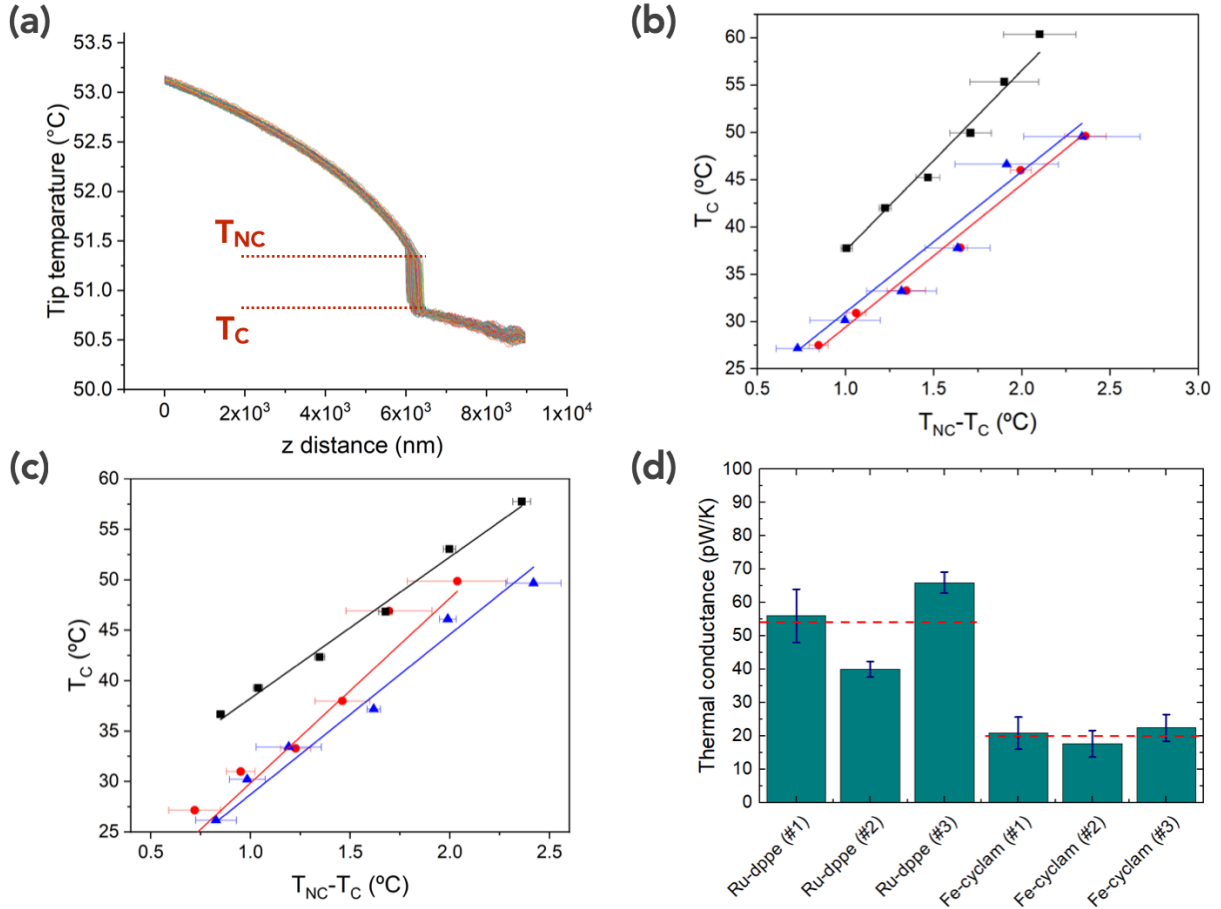

**Figure 7.** (a) Typical tip temperature vs. tip-surface distance measured at $V_{WB} = 0.8$ V during the approach z-scan (0 corresponds to the tip retracted), 25 traces acquired sequentially at the same location on the SAM (Fe-cyclam, SAM#1). (b) Tip temperature at contact, $T_C$, versus the temperature jump, $T_{NC}$-$T_C$ measured at various heating of the tip ($T_C$ increases with the voltage applied on the Wheatstone bridge, $V_{WB}= 0.6$ to $1.1$ V). The measurements were done at 3 locations marked by the 3 different colors, one for each location, randomly selected on the **Ru-dppe** SAM (#3) and the given $G_{th}$(SAM) is an averaged of these measurements. (c) Similar $T_C$ vs. $T_{NC}$-$T_C$ data for the **Fe-cyclam** SAM (#3). (d) Thermal conductance $G_{th}$(SAM) for three SAMs with **Ru-dppe** and **Fe-cyclam** molecules (additional data for SAMs #1 and #2 in the SI, Figure S22).

Since the Au electrode has a high thermal conductance and the SAM is very thin, the thermal conductance of the SAM, $G_{th}$(SAM), is obtained by correcting the measured values for the contribution of the underneath Au electrode (see SI). Figure 7-d shows $G_{th}$(SAM) for three different samples of the **Ru-dppe** and **Fe-cyclam** SAMs (all fabricated and measured in the same conditions, additional data in the SI, Figure S23). The results are reproducible and, more importantly, there is no significant difference between the two molecules. The average $G_{th}$(SAM) is around 20 nW/K, the same order of magnitude as previously measured on SAMs of alkylthiols by SThM and other techniques.[4,8,46,47] Considering that around 400 (10³) **Ru-dppe** (**Fe-cyclam**) molecules are contacted by the SThM tip (which is wider than the C-AFM tip), we can estimate a single-molecule thermal conductance of 50 pW/K for **Ru-dppe** and 20 pW/K for **Fe-cyclam**, fairly of the same order of magnitude as the previously measured values for other molecules in molecular electronics: ≈ 14 pW/K for octane chains in SAMs,[8] ≈ 26 pW/K and 37 pW/K for alkyl chains in single-molecule experiments,[48,49] ≈ 23 pW/K for a three-ring oligophenyleneethynylene (OPE3),[50] and ≈ 15 pW/K for a thiolated benzothieno-benzothiophene.[14]

**Seebeck coefficient.** The thermovoltages were measured from the temperature-dependent voltage offset of the *I-V* curves.[50] These *I-V*s were recorded around zero bias (voltage range ± 25 mV) using the C-AFM setup. We determined the voltage at null current ($V_{NC}$) and the shift of $V_{NC}$ is followed versus the temperature difference established between the Au electrode (substrate) and the tip (see details in the SI). The difference $V_{NC}(\Delta T)-V_{NC}(\Delta T=0)$ gives the thermovoltage.[51] Figure 8-a shows the histograms of $V_{NC}$ measured at several temperature differences applied on the **Ru-dppe** SAM, and Figure 8-b shows the evolution of the thermovoltage as function of the temperature difference. The slope $\Delta V_{NC}/\Delta T$ is -9 ± 4 µV/K. Taking into account the Au film substrate, tip and connecting wires,[50,51] (slope $(\Delta V_{NC}/\Delta T)_{Au/tip}$ = 5 ± 1 µV/K, see Figure S18), the Seebeck coefficient is $S_{Ru\text{-}dppe}$ = -[$\Delta V_{NC}/\Delta T$ - $(\Delta V_{NC}/\Delta T)_{Au/tip}$] = 14 ± 5 µV/K. The same measurements for a second SAM give $S_{Ru\text{-}dppe}$ = 28 ± 5 µV/K (Figure S15). This value of the Seebeck coefficient for the **Ru-dppe** molecule (≈ 14-28 µV/K) is in agreement with results in the literature for a similar molecule terminated by thiomethyl anchoring groups (27 µV/K)[18] or even a slightly different molecule in which the Ru is coordinated to four triethyl phosphite group (P(OEt)$_3$) (7 µV/K).[20] For the **Fe-cyclam** SAM, we measured higher values (Figure 8-c and Figure 8-d) $S_{Fe\text{-}cyclam}$ = 145 ± 23 µV/K, and 128 ± 71 µV/K for a second sample (Figure S20).

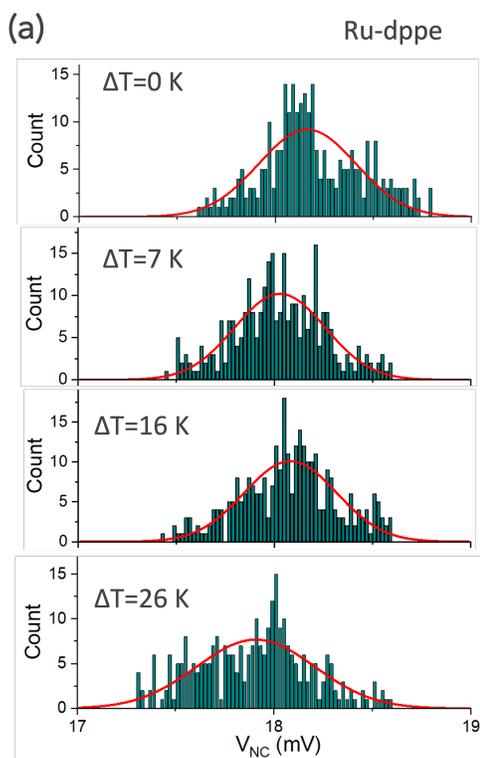
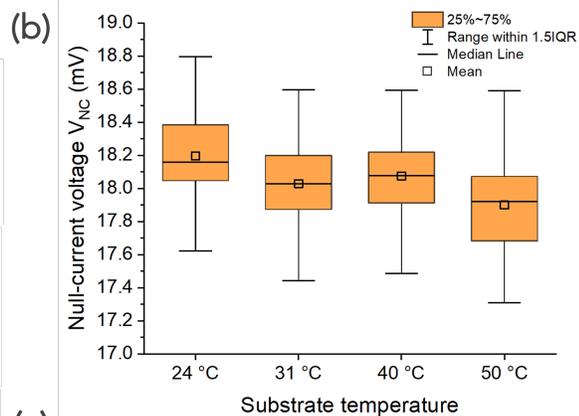
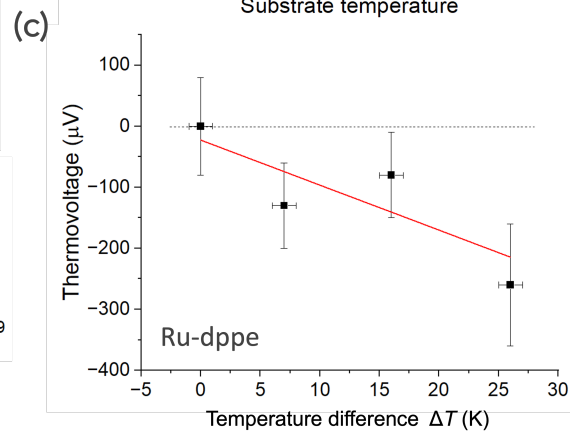
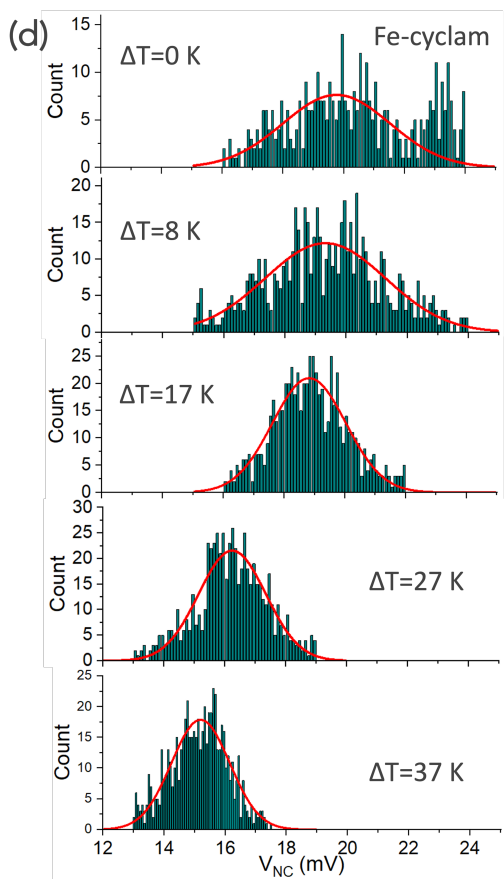
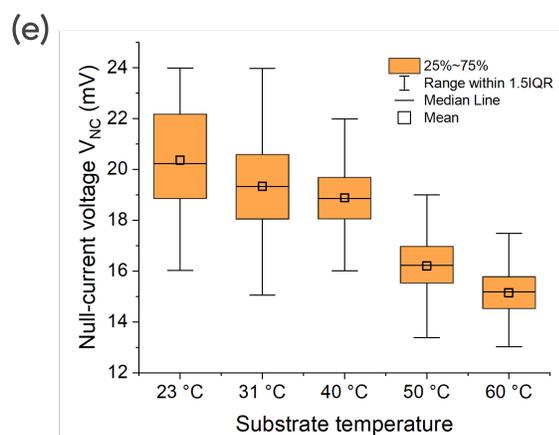
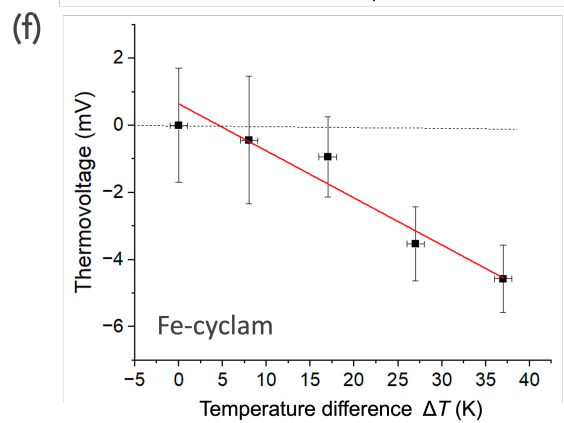

Figure 8. (a) Histograms of the voltage at null current $V_{NC}$ at several temperature differences applied on the **Ru-dppe** SAM (ca. 300 measurements at each temperature). The red lines are the fits with a Gaussian distribution. At $\Delta T = 0°C$, the measured offset voltage is due to the equipment (amplifier, cables, contacts...). (b) Box plots of $V_{NC}$ measurements showing the mean and median values, the 25-75 % percentiles and the limits at 1.5 IQR (interquartile range). (c) Evolution of the mean thermovoltage (the mean $V_{NC}$ normalized to the value at room temperature; error bar from the standard deviation of the histograms in panel (a)) versus the temperature difference and linear fit (red line) with a slope $\Delta V_{NC}/\Delta T = -9 \pm 4$ µV/K. (d), (e), (f): results for the **Fe-cyclam** SAM (400 to 640 measurements, depending on temperature). The slope in (f) is $\Delta V_{NC}/\Delta T = -140 \pm 22$ µV/K.

**Computational study**

The methodology used was to optimize first the geometry of each isolated molecular structure (thiol terminated), to insert them in junctions, and to calculate their electron transport properties in Au|molecule|Au junction configurations (fully optimized or not, see SI) at the Density Functional Theory (DFT) level using the QuantumATK software.[52] Additional information related to the computational details are provided in the Computational Section of the SI. The **Ru-dppe**-type junctions have been previously studied computationally with polyyne linkers using a simplified model to compute the transmission properties (using Au clusters to mimic the leads and a broadening of levels),[29] with SMe as anchoring group (no conductance value provided).[22] Considering that the nature of the anchoring group and contact geometry has a major influence,[50] new calculations that integrate a complete description of the junction and its conducting properties with semi-infinite electrodes are needed for the **Ru-dppe** system.

First, the geometries of the isolated [Ru(dppe)$_2$(C≡C-Ph-SH)$_2$] and [Fe(cyclam)(C≡C-Ph-SH)$_2$]$^0$, [Fe(cyclam)(C≡C-Ph-SH)$_2$]$^+$ complexes were calculated. The molecular junctions are built by grafting the molecule to two Au electrodes via S-Au bonds on a hollow site of the Au (111) surface (with the H atom of the thiol removed). Indeed, the protecting groups (EDMS) coordinated to the S atoms of the precursors are removed during the device fabrication (absence of Si2p peak in the XPS measurements). In our model junctions, the molecules are thus standing perpendicular to the surface of the electrodes. The effect of the tilting was evaluated (see SI). The transmission spectra have been simulated in the coherent regime at the DFT level coupled to the non-equilibrium Green's function formalism at zero bias ($V = 0$ V) (see SI for additional information). Considering that a quantitative description of the level alignment is not achievable at our level of theory,[53,54] we provide hereafter the conductance and the Seebeck coefficient as a function of an energy shift of the transmission spectrum with respect to the Fermi energy ranging between -0.10 and +0.10 eV (see SI).[55]

The molecular electronic levels of the isolated molecules are globally maintained in the device (Figure 9). In all cases, a π-delocalized molecular level lies close to the Fermi level. They exhibit a metal d character that supports the conjugation over the backbone and leads to a closer alignment with the Fermi level compared to organic analogues.[19,34] Importantly, in all cases, another molecular level lies nearby presenting also a delocalization character over the molecule with a different symmetry and spatial extension. In the case of the **Fe-cyclam** junctions, a molecular level localized on the Fe(cyclam) central part is also observed close in energy of those two levels. The [Fe(cyclam)(C≡C-$p$-C$_6$H$_4$-SH)$_2$]$^+$ complex is a paramagnetic molecule presenting an open-shell electronic structure. It results from the electronic and geometric reorganization upon oxidation (removal of one electron α) of [Fe(cyclam)(C≡C-$p$-C$_6$H$_4$-SH)$_2$]$^0$. It is thus treated within an unrestricted formalism that gives rise to differences in spatial extension and energy of spin-up and spin-down orbitals, as shown in Figure 9. Combinations of the three highest MOs of the [Fe(cyclam)(C≡C-$p$-C$_6$H$_4$-SH)$_2$]$^0$ are found in the spin-up orbital diagram while the shape of the orbitals its globally unchanged for the spin-down levels, with the former HOMO being depopulated.

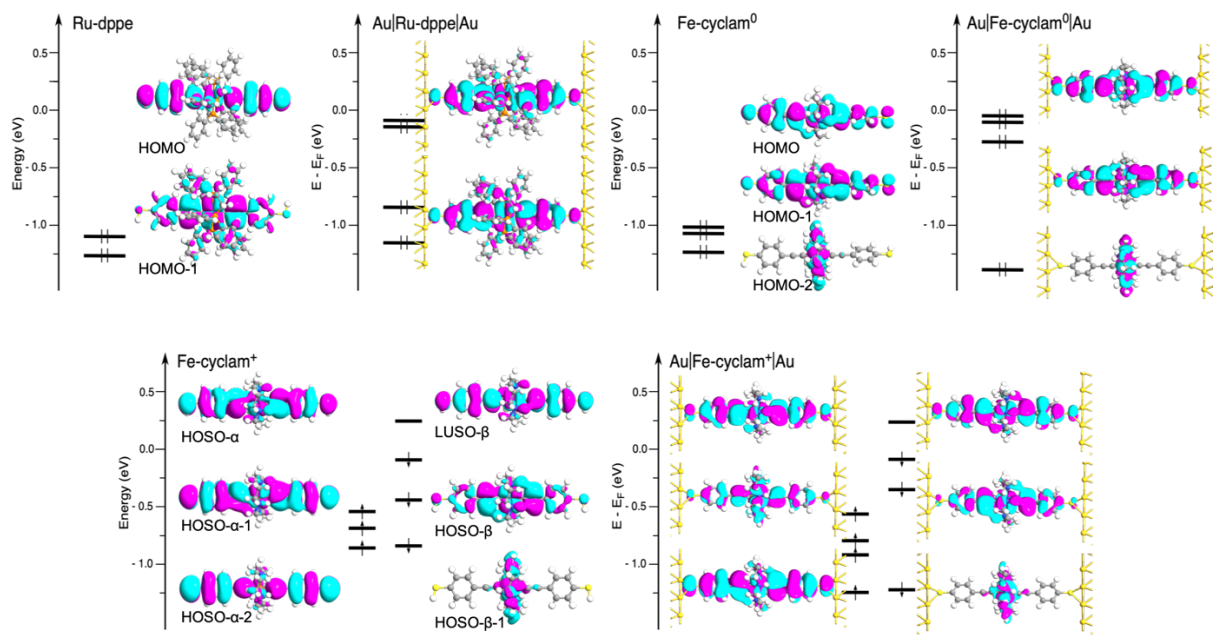

**Figure 9.** Molecular energy diagram of [Ru(dppe)$_2$(C≡C-$p$-C$_6$H$_4$-SH)$_2$], [Fe(cyclam)(C≡C-$p$-C$_6$H$_4$-SH)$_2$]$^0$, [Fe(cyclam)(C≡C-$p$-C$_6$H$_4$-SH)$_2$](Br) and molecular projected self-consistent Hamiltonians (MPSH) energy spectra of **Ru-dppe**, **[Fe-cyclam]$^0$** and **[Fe-cyclam]$^{+1}$** junctions. The isovalues of the contour plots are ±0.02 (e/Å$^3$)$^{½}$. Highest occupied molecular orbital = HOMO; Lowest unoccupied molecular orbital = LUMO; Highest occupied spin-orbital = HOSO-(α or β); Lowest unoccupied spin-orbital = LUSO-(α or β).

The calculated adiabatic ionization potential of the parent molecules in the gas phase [Fe(cyclam)(C≡C-Ph-SH)$_2$]$^0$, [Fe(cyclam)(C≡C-Ph-SH)$_2$]$^+$, [Ru(dppe)$_2$(C≡C-Ph-SH)$_2$] are 3.43, 5.25 and 3.71 eV following the trend of the measured electrochemical oxidation potentials in solution: Fe(cyclam)(C≡C-Ph-S(EDMS))$_2$(OTf) ($E^0_{1/2}$(Fe$^{II}$Fe$^{III}$) = -1.15 V vs. Fc/Fc$^+$; $E^0_{1/2}$(Fe$^{III}$/Fe$^{IV}$) = 0.61 V vs. Fc/Fc$^+$) and [Ru(dppe)$_2$(C≡C-Ph-S(EDMS))$_2$] (E$^0_{1/2}$(Ru$^{II}$/Ru$^{III}$) = -0.10 V vs. Fc/Fc$^+$). The change in the work function of the Au electrode upon deposition of SAMs based on these grafted molecules was calculated in order to enlarge the computational/experimental comparison (see "Work function shift" in the SI). These calculations reveal that the presence of adsorbed molecules on the Au(111) surface shifts its work function down by 2.01 eV, 1.05 eV, and 1.34 eV for **[Fe-cyclam]$^0$**, **[Fe-cyclam]$^{+1}$**, and **Ru-dppe**, respectively. Interestingly, the shift for **[Fe-cyclam]$^0$** is larger than for **Ru-dppe** but the opposite is the case for **[Fe-cyclam]$^{+1}$**. Experimentally, the measured shifts are less pronounced with 0.40 eV for the **Ru-dppe** SAM and 0.55 eV for **Fe-cyclam** SAM which is roughly composed of 2/3 **[Fe-cyclam]$^0$** and 1/3 of **[Fe-cyclam]$^{+1}$**. Considering the electrochemical properties of the isolated molecules, the trends between experiment and theory are consistent with both oxidation states of **Fe-cyclam** on the surface; the sole presence of Fe$^{II}$ complexes would lead to a larger difference in the work function shift according to the calculations.

The calculated transmission functions are shown in Figure 10. In all cases, the shape of the transmission peaks deviates from an idealized Lorentzian shape, because of QI effects due to the presence of molecular levels energetically close to the most conducting ones.[27] Interestingly, the transmission functions of **Ru-dppe** and **[Fe-cyclam]$^0$** are showing destructive QI features just below the maximum of transmission which is more pronounced for **[Fe-cyclam]$^0$** than for **Ru-dppe**. For [**Fe-cyclam**]$^{+1}$, the spin-up main transmission peak maximum is at lower energy than the **[Fe-cyclam]$^0$** one (0.55 eV vs 0.11 eV) and is almost Lorentzian in shape. In the spin-down channel, an antiresonance profile is found in the transmission function below but very close to the Fermi level followed by a high transmission peak in the LUMO region (see Figure 10). These transmission properties can be understood in terms of a molecular projected self-consistent Hamiltonians (MPSH) analysis of the central scattering region of the molecular junctions by studying the transmission eigenstate composition. The main transmission peak for the **Ru-dppe** junction, which is labeled **c** in Figure 10, is composed of 84.4 % of the HOMO and 7.6 % of HOMO-1, both MOs being delocalized over the molecule. The second transmission peak, which is labeled as **a**, originates mainly from HOMO-1 which contributes by 96.4 % (completed by 1.3 % from HOMO-2). The valley in the transmission function (**b**) that exhibits a common signature of destructive QI is the out-of-phase combination between 11.6 % of HOMO and 79.6 % of HOMO-1. For the **[Fe-cyclam]⁺** junction, for the α spin transmission channel, the main transmission peak (energy **d**, Figure 10) is due to the HOSO-α (94.7 %) as one would expect (transmission eigenstates at the energies **d**, Figure 11). The highest transmission peak in the β channel, labelled **e** below the Fermi energy and **g** in Figure 10), are composed of 99.6 % of the HOSO and 98.2 % of LUSO-β, respectively. Interestingly, the reduction of transmission observed in between (**f**) reveals a composition of 57.8 % of HOSO-β, 5.5 % of HOSO-2-β and 11.8 % of LUSO-β, which leads to a small S character on one of the two anchor groups that is responsible for the pronounced decrease in transmission. This analysis reveals that transmission eigenstates are not molecular orbitals (i.e., MPSH) that are lying at the same energy, but they result from combinations between molecular levels close in energy that can produce in some cases QI features.

In contrast to most organic conjugated molecules, the non-Lorentzian shape of the transmission peaks limits the precision with which $\epsilon_0$ and $\Gamma$ can be extracted by fitting the SLM of Eq. S4 to the experimental data. Indeed, small deviations from the ideal Lorentzian shape can lead to important deviation, in particular when antiresonances or Fano peaks are present.[56,57]

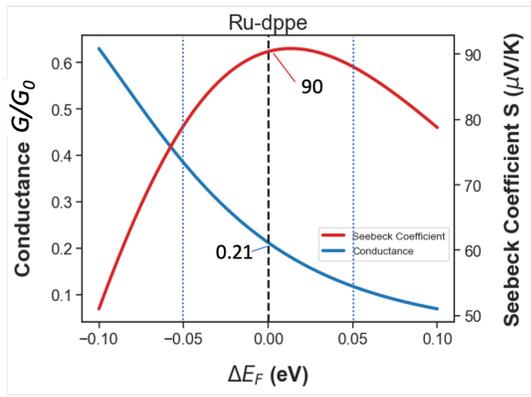
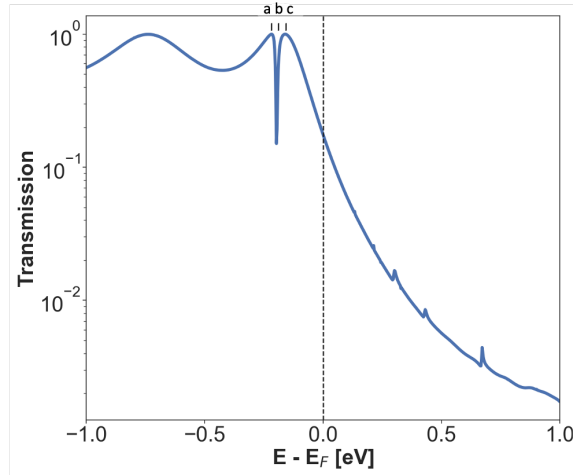
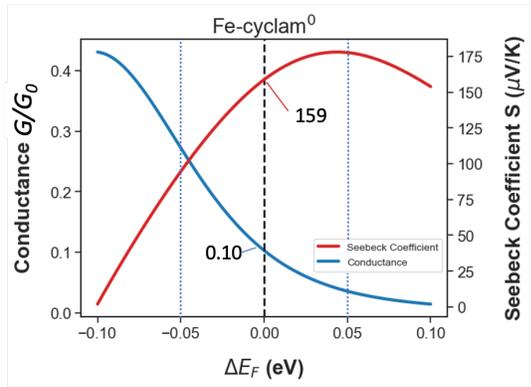
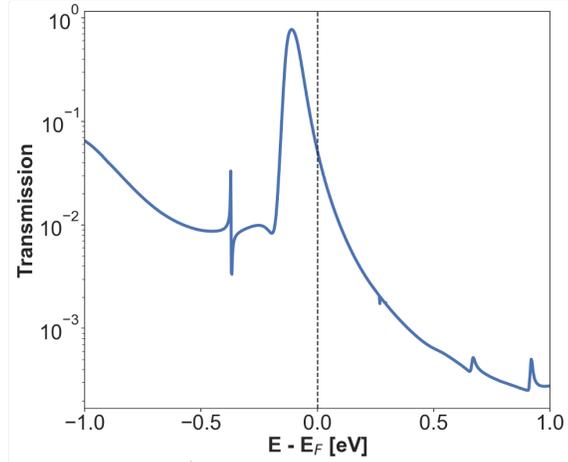
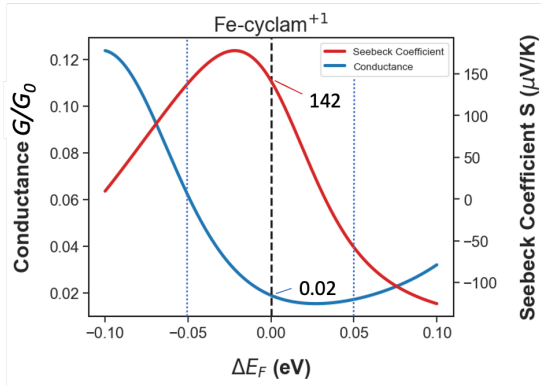
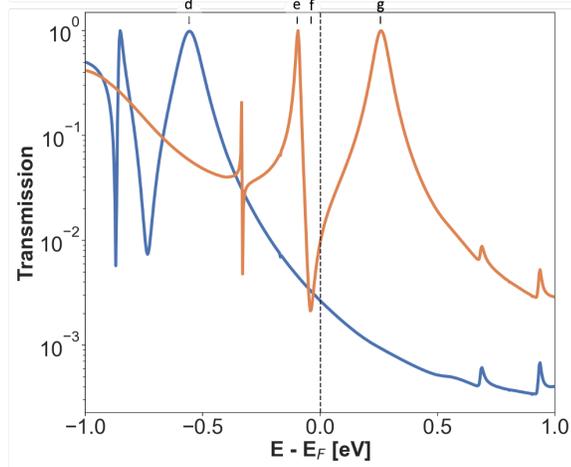

**Figure 10.** Conductance and Seebeck coefficient at 300 K as a function of the energy shift of the transmission spectrum by negative and positive $\Delta E_F$ with respect to the Fermi level (left), as extracted from the calculated electronic transmission spectra at zero bias voltage (right) of the flat

configuration-type **Ru-dppe**, **[Fe-cyclam]⁰**, **[Fe-cyclam]⁺** (spin-up and down transmissions in blue and orange, respectively) junctions.

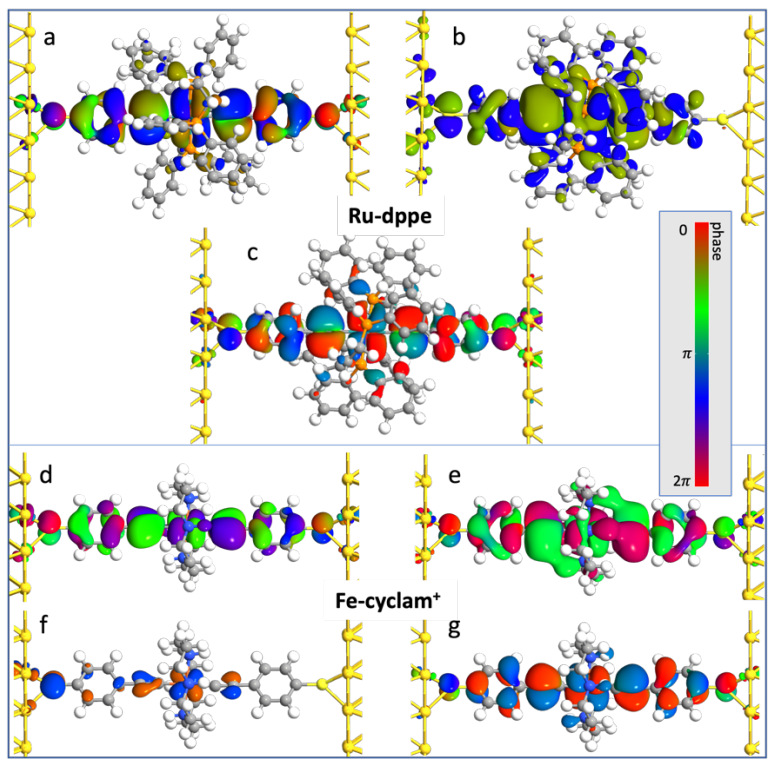

**Figure 11.** Phase-colored transmission eigenstate plots issued from the electronic transmission calculations of the **Ru-dppe** junction at the energy **a**, **b** and **c** indicated in Figure 10, and of **[Fe-cyclam]⁺** junction at **d** (spin-up) and **e**, **f** and **g** (spin-down). The iso-contour value is 0.01 (e/Å³)^½.

The problem of positioning adequately the molecular levels with respect to the Fermi level in DFT is often taken into account by applying an energy shift of the transmission spectrum ($\Delta E_F$) with respect to the Fermi level, which allows for a partial consideration of computational uncertainty (see "HSE functional" section of the SI).[55] This was done here to calculate the electronic conductance and the Seebeck coefficient (Figure 10). Variations by up to 50 % in absolute value of the Seebeck coefficient are found for shifts of 0.05 eV in all cases, with a change in sign for **[Fe-cyclam]⁺**. Absolute values are thus not meaningful, but trends can be given.

Interestingly, in the case of the open-shell configuration of the **[Fe-cyclam]⁺**, the main conducting channel can be driven by spin-up (at energy **d**) or spin-down (at energy **e**) for very small changes in energies of the incoming electrons (Figure 10). Upon bias, the energy and spin of the incoming electron will determine its probability of tunneling, thus creating a partial spin filtering upon injection.[58] The calculated conductance of **Ru-dppe** junction ($2.1 \times 10^{-1}\ G_0$) is slightly higher than that of **[Fe-cyclam]⁰** ($1.0 \times 10^{-1}\ G_0$) and **[Fe-cyclam]⁺** ($2.0 \times 10^{-2}\ G_0$) by roughly a factor two and one order of magnitude, respectively (Figure 10).

Since the application of a bias can polarize electronically the system and modify its coupling with the electrodes and hence the conducting properties, we have recomputed the transmission spectra upon application of a 100, 300, and 500 mV voltage for **[Fe-cyclam]⁰** and **Ru-dppe** junctions with flat electrode surfaces (Figure 12).[59] Reliable calculations under bias for **[Fe-cyclam]⁺** junction unfortunately could not be carried out due to the impossibility of maintaining the one-electron spin-polarization. The calculated changes in the electronic transmission properties of **[Fe-cyclam]⁰** and **Ru-dppe** junctions are small at 100 mV. The changes are more pronounced for higher bias. The shapes of the transmission functions are globally preserved but an overall decrease of transmission (from 0.77 to 0.40 for **[Fe-cyclam]⁰**; from 1.00 to 0.84 for **Ru-dppe**) and a shift of the conducting peaks away from the Fermi level are observed. The maximum of transmission close to the Fermi energy is shifting from –0.11 to –0.21 eV for **[Fe-cyclam]⁰**, and from –0.16 to –0.27 eV for **Ru-dppe**. When imposing a bias of 500 mV, the conductance of **Ru-dppe** and **[Fe-cyclam]⁰** junctions decreases by a factor of 2.6 and 6.2, respectively ($G_{500\ mV} = 7.8 \times 10^{-2}\ G_0$ for **Ru-dppe**; $G_{500\ mV} = 1.6 \times 10^{-2}\ G_0$ for **[Fe-cyclam]⁰**), highlighting the more pronounced effect of bias voltage for **[Fe-cyclam]⁰** junctions.

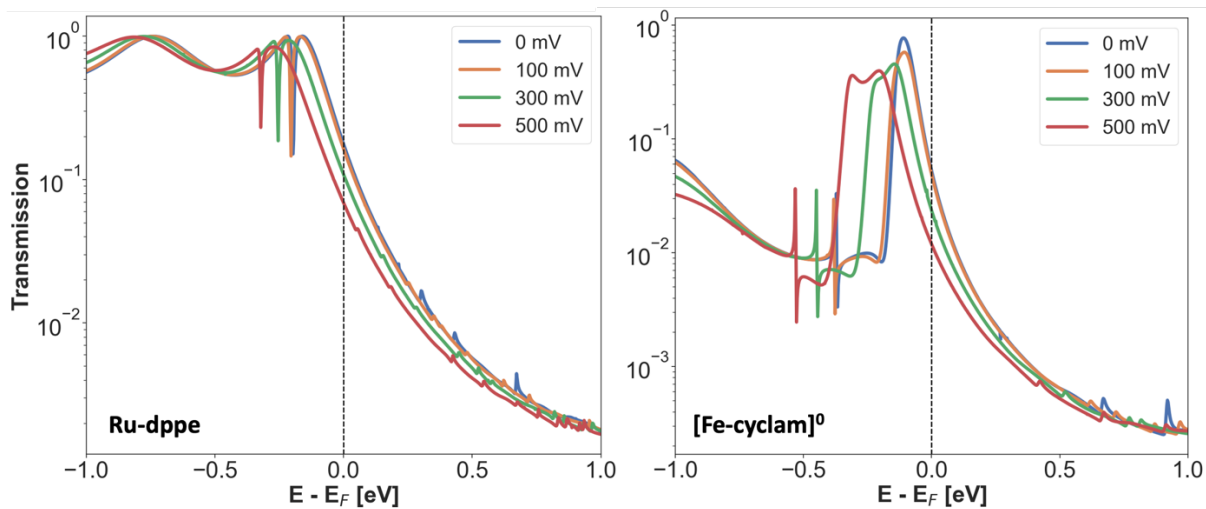

**Figure 12.** Calculated transmission spectra at zero, 100, 300 and 500 mV bias of flat configuration-type **Ru-dppe** and **[Fe-cyclam]⁰** junctions.

The absolute value of the calculated conductance is higher by at least two orders of magnitude compared to the experimental measurements. In fact, our calculations deal with an ideal model in which the electrode surface is simulated by a perfect Au (111) flat surface and the molecule/Au electronic coupling is maximized. In order to evaluate the robustness of this model, several alternative configurations were tested and analyzed. These studies are given in the SI in the section entitled "Evaluation of the computational model". The use of tip-shaped electrodes decreases the conductance by a factor ranging from 1.6 to 3.0 (Figure S26) while the shape of the transmission is globally maintained; however, the transmission peaks are narrower and shifted down with respect to the Fermi level because of the decrease in the coupling constant. We also increased the distance between one electrode and the coordinated sulfur anchor group to simulate partly-grafted molecules (Figure S34). In that case, the slope of the transmission close to the Fermi level is slightly affected but the conductance is significantly decreased by more than one order of magnitude upon an elongation of 1.5 Å. For **Ru-dppe**, we have optimized a junction configuration in which the molecule is tilted to simulate the experimental observations (details in SI, Figure

S35). This configurational change results in an increase of conductance from $2.12 \times 10^{-1}$ to $6.80 \times 10^{-1}$ $G_0$ and a decrease of the Seebeck coefficient from 90 to 51 μV/K.

The calculated Seebeck coefficients at 300 K are reaching values in the range of ∼150 μV/K for both **[Fe-cyclam]$^0$** and **[Fe-cyclam]$^+$** junctions and ∼90 μV/K for **Ru-dppe** junctions for a moderate shift of the Fermi level (by 0.025 eV). The order of magnitude and ordering of the Seebeck coefficient of **Fe-cyclam** versus **Ru-dppe** are in line with the experiments. The measured Seebeck coefficient is ≈ 14-28 μV/K for **Ru-dppe** junctions and ≈ 128-145 μV/K for the **Fe-cyclam** molecules. As shown in **Figure 10**, the difference in Seebeck coefficient between **Ru-dppe** and **[Fe-cyclam]$^0$** junctions is directly linked to the difference of the slope of the transmission coefficient $T(E)$ at the Fermi energy which arises from the slight difference in the electronic configurations. For **[Fe-cyclam]$^+$** junctions, the large Seebeck coefficient originates from transmission functions of both spin channels which show more pronounced changes around the Fermi energy (**Figure 10**). Altogether, the power factor $S^2G$ and figure of merit $ZT$ can be deduced from our computational study by considering a phonon thermal conductance of 20 pW/K for **Fe-cyclam** junctions and 50 pW/K for **Ru-dppe** junctions (vide supra SThM section). They are presented in Figure S40. The computed power factor of the **Ru-dppe** and **[Fe-cyclam]$^0$** junctions are both over 100 pW m$^{-1}$ K$^{-2}$, reaching even values higher than 200 pW m$^{-1}$ K$^{-2}$ for the **[Fe-cyclam]$^0$** junction. Even if **[Fe-cyclam]$^+$** presents a high calculated Seebeck coefficient, it is a less efficient energy converter due to its lower electronic conductance. The $ZT$s calculated from the quantum chemical studies are estimated to be 0.8 for the **Fe-cyclam** junctions and below 0.2 for **Ru-dppe** (0.1 for the **Ru-dppe** bent configuration).

**DISCUSSION**

The single-molecule conductance of **Ru-dppe** and **Fe-cyclam** junctions extracted from measurements of C-AFM (SAM) and MCBJ (single molecule) are summarized in Table 1. The **Ru-dppe** MCBJ conductance histogram reveals a LC peak around $1.8 \times 10^{-5}$ $G_0$ and a HC one $6.9 \times 10^{-4}$ $G_0$, with the conductance of the longer plateaus ranging from $10^{-4}$ $G_0$ to $10^{-3}$ $G_0$. The conductance of the thiomethyl-anchored analogue single-molecule junction was reported at $2.2 \times 10^{-4}$ $G_0$ from STM measurements (bias voltage 100 mV), but the conductance histograms and the plateau lengths were not provided.[20] This latter value lies in the lower part of the conductance range of the longer traces of the MCBJ data (Figure 2-c and Figure 2-d). The C-AFM average value $\bar{G}_{mol}$ is $2 \times 10^{-6}$ $G_0$ with a maximum at $5 \times 10^{-5}$ $G_0$, i.e., one to two orders of magnitude smaller. This underestimation of the conductance values of a single-molecule junction based on ensemble measurements compared to single-molecule junction measurements is typical.[60] This difference is usually attributed to electrostatic effects in the ensemble measurements,[36] to dielectric screening of the medium[61] or to intermolecular interaction in the quantum tunneling regime.[62,63] Moreover, the number of molecules which are directly probed by ensemble measurement techniques is difficult to estimate and irregularities are likely to form at the top contact, both affecting the single-molecule conductance evaluation.[64,65] The stochastic fluctuations in the number of contacted molecules influences also the statistical distributions of the energy level of the dominant molecular orbital and of the molecule-electrode coupling energies.[66] This is indeed validated by the XPS spectra analysis of the **Ru-dppe** SAMs which reveals that 35% of the molecules are not chemisorbed on the surface. Another source of discrepancy is the tilt angle of 38° between the **Ru-dppe** molecules with respect to the Au surface normal in the SAMs.

**Table** 1. Comparison of the experimental conductance values (in unit of $G_0$). $G_{mol}$ is the estimated value per molecule deduced from zero-bias conductance the SAMs (C-AFM, see text). The MCBJ

$G$ are obtained from the histograms of conductance peak (Figure 2). The MCBJ average $G$ are obtained from the MCBJ $I$-$V$ curves stable over two different bias range (0.3 and 1.0 V).

|  | C-AFM | | MCBJ | | | MCBJ (bias dependent $I$-$V$s) Stable junctions | |
|---|---|---|---|---|---|---|---|
|  | Average $G_{mol}$ | Maximum $G_{mol}$ [a] | LC | HC | $G_{max}$ [b] | LC | HC |
| **Ru-dppe** | $2 \times 10^{-6}$ | $5 \times 10^{-5}$ | $1.8 \times 10^{-5}$ | $6.9 \times 10^{-3}$ | $\sim 7.3 \times 10^{-3}$ | $1.3 \times 10^{-3}$ | $4.4 \times 10^{-3}$ |
| **Fe-cyclam** | $5.4 \times 10^{-8}$ | $1.7 \times 10^{-6}$ | $1.7 \times 10^{-4}$ | $1.5 \times 10^{-3}$ | $\sim 4.2 \times 10^{-2}$ | $2.8 \times 10^{-3}$ | $1.1 \times 10^{-3}$ |
| Bias (V) | 0.075 | 0.075 | 0.100 | 0.100 | 0.100 | 0.300 | 1.000 |

(a) Maximum of the distribution shown in Figure 5-c Figure 6-c. (b) Maximum estimated from the "cut-off" of the highest molecule conductance peak (Figure 2-a and Figure 2-c), i.e., the intercept of the molecule peak tail with the tail of the $G_0$ peak (conductance at the minimum of the U-shaped distribution between the two peaks).

For the **Fe-cyclam** junctions, we extract $\bar{G}_{mol}(0) \approx 5.4 \times 10^{-8}$ $G_0$ from C-AFM measurements with a maximum value $\approx 1.7 \times 10^{-6}$ $G_0$. The same percentage of molecules as for **Ru-dppe** (35 %) is not chemically attached via a S-Au bond to the surface. Considering that in MCBJ, the trapped molecules are either **[Fe-cyclam]⁰** or **[Fe-cyclam]⁺** contrarily to C-AFM which probes ensemble of **[Fe-cyclam]⁰** or **[Fe-cyclam]⁺** and considering all factors previously cited, the comparison between the two data sets is not conceivable at the single-molecule level. The MCBJ conductance measurements of **Fe-cyclam** junctions show traces with longer plateaus uniformly from around $G = 8 \times 10^{-5}$ $G_0$ to about $G = 6 \times 10^{-3}$ $G_0$ (Figure 2-a and Figure 2-b); only roughly one tenth of the junctions is stable over the selected bias (Figure 3). Interestingly, when considering only bias stable junctions in MCBJ, the **Ru-dppe** and **Fe-cyclam** show similar average conductance at $\sim 10^{-3}$ $G_0$ (Table 1). The computational studies performed on idealized molecular junctions show that **Ru-dppe** and **[Fe-cyclam]⁰** junctions yield a conductance of $\sim 10^{-1}$ $G_0$ while it is one order of magnitude less for **[Fe-cyclam]⁺** junctions ($2.0 \times 10^{-2}$ $G_0$). The oxidation state of each **Fe-cyclam** junction probed in MCBJ is not known, nor the proportion of each oxidation state within a batch of tested molecules, thus preventing a more detailed comparison. Interestingly, some MCBJ elongation traces of **Ru-dppe** junctions are presenting jumps that signal changes in the configuration when elongating the junction.[2] We assume that these higher conducting **Ru-dppe** junctions correspond to similarly bent configurations as those determined for **Ru-dppe** SAMs by ellipsometry-based thicknesses measurements. Indeed, when imposing a tilted configuration to the **Ru-dppe** junction in which the S atoms are chemically bonded to both leads, the computational study shows enhanced conductance associated with smaller Seebeck coefficient values. Upon elongation, this configuration can rearrange to a linear one without losing the anchor chemical bonds.

The SLM fitting applied to the experimental *I-V* curves is providing interesting results that allow for a deeper analysis. Table 2 reports the electronic characteristics of the molecular junctions deduced from the different approaches applied in this study. The agreement between the $\epsilon_0$ values of **Ru-dppe** junctions (energy of the conducting level) obtained by different approaches is particularly good. The $\epsilon_0^{SAM}$ (C-AFM), $\epsilon_0^{MCBJ}$ and $\epsilon_0^{DFT}$ for the same bias range are 0.32 ± 0.10 eV, 0.40 eV and 0.27 eV, respectively. The energy of the HOMO level extracted from UPS ($\varepsilon_h^{UPS}$ = 0.34 eV) is consolidating these values. The computational studies highlight that the bias voltage range has a noticeable influence on the positioning of the $\epsilon_0$ and on the amplitude of the electrode coupling, and thus also on the conductance values. This aspect is vastly neglected in the literature. In the case of the **Fe-cyclam**, the influence of bias is probably different between **[Fe-cyclam]⁰** and **[Fe-cyclam]⁺** junctions but our studies do not allow segregating it. Additionally, **[Fe-cyclam]⁰** oxidation is quite low and could be triggered by bias, as detailed by Lörtscher and co-workers for other organometallic systems.[67] Indeed, the calculated adiabatic ionization potential of **[Fe-cyclam]⁰** is 3.43 eV to be compared with 3.71 eV for **Ru-dppe**, and 5.25 eV for **[Fe-cyclam]⁺**.

**Table 2.** Calculated energy of the maximum of the transmission peak closest to the Fermi energy ($\epsilon_0^{DFT}$) that is compared to $\epsilon_0^{SAM}$ extracted from C-AFM and $\epsilon_0^{MCBJ}$ extracted from MCBJ experimental data using the SLM, and to the UPS HOMO value ($\varepsilon_h^{UPS}$). Values are in eV.

|  | $\epsilon_0^{DFT}$ | $\epsilon_0^{SAM}$ | $\epsilon_0^{MCBJ}$ | $\varepsilon_h^{UPS}$ |
|---|---|---|---|---|
| **Ru-dppe** | 0.16-0.27[a] | 0.32 ± 0.12 | 0.40[d] | ≈ 0.34 |
| **Fe-cyclam** | N.A. | N.A. | N.A. | ≈ 0.75 |
| **[Fe-cyclam]⁰** | 0.11-0.21[a] | 0.33 ± 0.06 | 0.15[c] | N.A. |
| **[Fe-cyclam]⁺** | 0.55[b] | 0.60 ± 0.15 | 0.55[c] | N.A. |

[a] Range between 0 to 500 mV bias. [b] Energy of the spin-α transmission peak maxima since the **[Fe-cyclam]²⁺** ground state is calculated to be a triplet state.[68] [c] Two maxima extracted for $V_{\text{range}}$ < 500 mV. [d] A narrow peak centered at 0.15 eV is in the distribution but corresponds to a limited number of junctions (Figure S5-b).

The $\epsilon_0^{SAM}$ and $\epsilon_0^{MCBJ}$ histograms are clearly showing to sets of values which can be fitted with two Gaussians. On the basis of the computational study (Table 2), we attribute $\epsilon_0^{SAM}$ = 0.33 ± 0.06 eV and $\epsilon_0^{MCBJ}$ = 0.15 eV to **[Fe-cyclam]⁰** junctions ($\epsilon_0^{DFT}$ = 0.21 eV). The higher values $\epsilon_0^{SAM}$ = 0.60 ± 0.15 eV, $\epsilon_0^{MCBJ}$ = 0.55 eV and $\epsilon_0^{DFT}$ = 0.55 eV (zero-bias) thus correspond to **[Fe-cyclam]⁺** junctions. This analysis leads to consider that, when probing an ensemble of molecules, the transmission is mostly driven by one of the two depending on the contacted molecules (**[Fe-cyclam]⁺** and **[Fe-cyclam]⁰**) to the tip, or that the surface is inhomogeneous and is composed of islands of **[Fe-cyclam]⁺** and **[Fe-cyclam]⁰**. This also tends to indicate that the oxidation state of the grafted molecules is preserved during the measurements.

The thermal conducting properties of **Fe-cyclam** and **Ru-dppe** SAMs measured by SThM allow for the estimation of the molecular thermal conductance: 50 pW/K for **Ru-dppe** and 20 pW/K for **Fe-cyclam**. The Seebeck coefficients measured for two different samples for each molecule are $S_{\text{Ru-dppe}}$ 14 ± 5 µV/K and 28 ± 5 µV/K, and $S_{\text{Fe-cyclam}}$ = 145 ± 23 µV/K and 128 ± 71 µV/K. In the case of **Fe-cyclam** junctions, the proportion [Fe$^{II}$]/[Fe$^{III}$] molecules should also be considered for analyzing the Seebeck coefficient and the power factor. The phonon conductance is much less affected since the vibrational properties of both oxidation states are highly similar. The computed $S$ are slightly overestimating these values but the order between $S_{\text{Fe-cyclam}} > S_{\text{Ru-dppe}}$ is reproduced. Interestingly, even though presenting QI in the transmission spectrum, the thermoelectric power of the **[Fe-cyclam]⁺** junction does not exceed that of **[Fe-cyclam]⁰** because of the partial spin filtering favoring the spin-up path at the Fermi energy. The measured $S_{\text{Fe-cyclam}}$ outperforms the previously reported values for comparable molecular junctions, i.e., showing electronic conductance operating in the tunneling regime with comparable length without applying external magnetic or electric field and operating at room temperature.[3,4,5,20,24] As a reminder, the **Fe-cyclam** and **Ru-dppe** SAM thicknesses are slightly below 2 nm. The higher Seebeck coefficients which were recently reported are 73 µV/K for a 4 nm-length tri-Ru system[18] and more recently 307 ± 16 µV/K for an electrografted (bis-tpy)Ru-containing film of 2 nm thickness (tpy = terpyridine).[69] In these two studies, the increase in the thickness of the film yields an increase in the Seebeck coefficient to the detriment of the electronic conductance. Substantial electron conductance is required to apply these systems for energy conversion. Moreover, the heat conductance needs to be known to estimate $ZT$.[6,70] This is not provided in the two cited studies.

In our study, we perform these evaluations by combining several techniques. This allowed us to provide an experimental estimation of the $ZT$ by considering the value of the HC peak from the MCBJ measured distribution (see Table 1), the thermal conductance per molecule (50 pW/K and 20 pW/K – **Ru-dppe** and **Fe-cyclam**) and the highest average Seebeck coefficient 28 µV/K (**Ru-dppe**) and 145 µV/K (**Fe-cyclam**). Doing so, we obtain $ZT = 6.2 \times 10^{-4}$ and 0.01 with associated power factor of $4.2 \times 10^{-5}$ pW m$^{-1}$ K$^{-2}$ and $2.4 \times 10^{-3}$ pW m$^{-1}$ K$^{-2}$ for the **Ru-dppe** and **Fe-cyclam** molecular junctions, respectively. The estimation of $ZT$ based on the HC value is justified because junctions in this conductance range have shown to be stable. If we consider instead the $G_{max}$ value measured in MCBJ (see Table 1), we can estimate a maximum $ZT$ of $6.6 \times 10^{-3}$ for **Ru-dppe** and 0.4 for **Fe-cyclam**, this latter value being close to the computed value. This is the highest value reported to date. A value of 0.7 was reported by van der Zant and co-workers for Au|Gd(tpy-SH)$_2$(SCN)$_3$|Au molecular junctions but it was estimated for an operating temperature at 2 K (without heat transport measurements) for a Seebeck coefficient obtained upon voltage gating of the molecular junction.[12]

**CONCLUSION**

In the present work, we provide a comprehensive experimental and theoretical study of the electronic properties in single-molecule and SAM-based junctions of two organometallic compounds differentiated by the nature of the central ditopic metallic group and both substituted by the same arylacetylide thiol linker. **Ru-dppe** and **Fe-cyclam** junctions yield an important spreading of electronic conductance values in C-AFM and MCBJ. The deep analysis of the single-molecule stretching traces and *I-V* measurements in MCBJ reveals that the junction configurations are numerous and that for **Ru-dppe**, the configuration can change upon elongation without changes in oxidation state. In contrast, the Fe complex can be reduced during the fabrication of the devices. The differentiation between **[Fe-cyclam]⁰** and **[Fe-cyclam]⁺** junctions is accessed by fitting of the *I-V* curves. Indeed, the MCBJ, C-AFM and computational studies coincide in the evaluation of $\epsilon_0$ for the two compounds. The **Fe-cyclam** SAMs are presenting high Seebeck coefficients (> 130 µV/K) and low thermal conductance (20 pW/K) which make the **Fe-cyclam** junctions better candidates than **Ru-dppe** junctions for energy conversion. The computational study reveals that **[Fe-cyclam]⁺** presents quite different electronic properties than **[Fe-cyclam]⁰** since it is a paramagnetic system; its transmission properties show QI effects close to the Fermi level but it yields to moderately improved thermoelectric properties compared to the closed-shell system because of the predominance of one conducting spin-channel over the other. We computed a *ZT* of 0.8 for the **Fe-cyclam** molecular junctions while estimating an experimental value 0.4. The **Fe-cyclam** SAMs show higher thermoelectric abilities than the **Ru-dppe** SAMs. We demonstrate by both experimental and computational means that these improved Seebeck coefficient (∼130 µV/K) and *ZT* up to 0.4 can be reached by molecular changes performed on purpose to modify the electronic transmission properties within the constraints of remaining synthesizable, operative and importantly stable at room temperature. Contrary to ref. 12 on junctions of a Gd-containing metal complex where the peak value of S = 414 $\mu$V/K was reported (bias of 3.4 mV and –0.86 V of gate voltage), we discuss here the average values. To the best of our knowledge, the *ZT* for Fe-cyclam reported here is the highest one for molecular junctions from metal complexes operated at room temperature and without additional gating or field tuning.[24] Indeed, thermal stability is compulsory for further energy conversion application. The future developments, if application of external stimulus is excluded, rely on chemical modifications in order to modify the electronic transmission such that QI features are present at the Fermi energy in the main spin channel at this energy with a high transmission peak within the bias window. Coordination chemistry offers an almost unlimited catalog to achieve this.

ASSOCIATED CONTENT

**Supporting Information**. The Supporting Information is available free of charge at https://pubs.acs.org/doi/10.1021/acsnano.xxx. It contains further details of the experimental procedures and of the computational details. Additional characterizations and results are provided (Supporting Figures S1–S41; Tables S1–S4; Equations S1-S4) (PDF). The X-ray crystallographic data for the structures reported in this article have been deposited at the Cambridge Crystallographic Data Centre (CCDC) under deposition numbers CCDC 2344014 and 2344016 for $trans$-[Ru(dppe)$_2$(C≡C-$p$-C$_6$H$_4$-S(EDMS))$_2$] and Fe(cyclam)(C≡C-$p$-C$_6$H$_4$-S(EDMS))$_2$]$^+$, respectively (CIF).

AUTHOR INFORMATION

**Corresponding Authors**

* Elke Scheer,[3*] Dominique Vuillaume,[2*] Stéphane Rigaut,[1*] Karine Costuas[1*]

**Funding Sources**

The authors from the French institutions acknowledge the ANR – FRANCE (French National Research Agency) for its financial support of the project HotElo (ANR-21-CE30-0065). J.S.A. thanks the Région Bretagne,Rennes Métropole, and the Collège doctoral de Bretagne» for a mobility grant to the University of Mons, Mons, Belgium. J.C. is an FNRS research director. The research in Mons is supported by the Belgian National Fund for Scientific Research (FRS-FNRS), within the Consortium des Équipements de Calcul Intensif – CÉCI (grant number U.G.018.18), and by the Walloon Region (LUCIA Tier-1 supercomputer; grant number 1910247). We thank the Deutsche Forschungsgemeinschaft (DFG) for funding through SFB1432.

ACKNOWLEDGMENT

We thank the staff of the scanning probe microscopy platform at IEMN for technical support and X. Wallart (IEMN) for the XPS/UPS measurements. We thank T. Huhn and L. Holz for their contributions and discussion, the staff of the nano.lab at the University of Konstanz for technical support.

Supporting information

# Electronic and Thermoelectric Properties of Molecular Junctions Incorporating Organometallic Complexes: Implications for Thermoelectric Energy Conversion

Joseane Santos Almeida, [a] Sergio González Casal, [b] Hassan Al Sabea, [a] Valentin Barth, [c] Gautam Mitra, [c] Vincent Delmas, [a] David Guérin, [b] Olivier Galangau, [a] Tiark Tiwary, [c] Thierry Roisnel, [a] Vincent Dorcet, [a] Lucie Norel, [a] Colin Van Dyck, [d] Elke Scheer,*[c] Dominique Vuillaume,*[b] Jérôme Cornil, [e] Stéphane Rigaut,*[a] Karine Costuas*[a]

a) Univ Rennes, CNRS ISCR (Institut des Sciences Chimiques de Rennes), Rennes, France.
b) Institute for Electronics, Microelectronics and Nanotechnology, CNRS, Villeneuve d'Ascq, France.
c) Department of Physics, University of Konstanz, Konstanz, Germany.
d) Theoretical chemical physics group, Department of Physics, University of Mons, Mons, Belgium
e) Laboratory for Chemistry of Novel Materials, University of Mons, Mons, Belgium

Corresponding author's email addresses: elke.scheer@uni-konstanz.de, dominique.vuillaume@iemn.fr, stephane.rigaut@univ-rennes1.fr, karine.costuas@univ-rennes1.fr

# TABLE OF CONTENTS

## 8 – Computational details

## References

# 1 - Chemical Synthesis and Characterization of the Molecules

**General methods.** All manipulations were performed in Schlenk-type flasks under dry nitrogen. Solvents were dried by conventional methods and distilled immediately prior to use. Routine $^1$H and $^{31}$P{$^1$H} spectra were recorded on a Bruker AVANCE III 400 MHz. Mass spectra were recorded either on a Bruker MicroTOF spectrometer (ESI-TOF) using $CH_2Cl_2$, $CH_3CN$ or $CH_3OH$ as the solvent. Raman spectra were collected using a confocal Raman spectrometer LabRAM HR800 (HORIBA Scientific, Jobin-Yvon) using a 632.8 nm He-Ne laser directed on the sample through an objective (OLYMPUS, ULWD 100x, N.A. 0.8). All measurements were carried out under ambient conditions at room temperature. Elemental analyses were performed by the CRMPO (Centre régional de mesures physiques de l'Ouest). All commercial reagents were used as supplied. *cis*-[Ru(Cl)$_2$(dppe)$_2$][1] (dppe = diphenyl-diphosphino-ethane), [Fe(cyclam)(OTf)$_2$](OTf)[2,3] (cyclam = 1,4,8,11-tetraazacyclotetradecane), 2-(trimethylsilyl)ethyl-4'-(ethynyl)phenyl sulfide (**1**)[4] precursors were prepared according to the literature. EDMS is used as acronym of ethyl(dimethyl)silane.

***trans*-[Ru(dppe)$_2$(C≡C-*p*-C$_6$H$_4$-S(EDMS))$_2$]**. Under argon, a Schlenk tube was charged by *cis*-[Ru(dppe)$_2$(Cl)$_2$] (0.155 g, 0.16 mmol) and NaPF$_6$ (0.083 g, 0.5 mmol). In parallel, in another Schlenk tube under argon, compound **1** (0.125 g, 0.5 mmol) was dissolved in $CH_2Cl_2$ (20 mL) and Et$_3$N (0.14 mL) and kept stirring for 5 minutes. Then, the solution was added over the solids and allowed to stir during 4 hours. The solution mixture was then filtered into a clean Schlenk by cannula and evaporated under reduced vacuum pressure. The obtained solid was further dissolved in degassed $CH_2Cl_2$ (5 mL). The slow addition of pentane yielded a yellow precipitate of the desired complex (0.200 g, 92 %). **$^1$H NMR** (400 MHz, CD$_2$Cl$_2$, 297 K) δ = 7.51 (m, 16H, H$_{PPh}$), 7.19 (m, 8H, H$_{PPh}$), 7.14 (d, $^3J$ = 8 Hz, 4H, H$_{a/b}$), 6.97 (t, $^3J$ = 8 Hz, 16H, H$_{PPh}$), 6.60 (bd, $^3J$ = 8 Hz, 4H, H$_{a/b}$), 2.98 (m, 4H, H$_c$), 2.63 (m, 8H, PCH$_2$CH$_2$P), 0.94 (m, 4H, H$_d$), 0.07 (s, 18H, 6SiCH$_3$). **$^{13}$C NMR** (126 MHz, CD$_2$Cl$_2$, 297K) δ = -1.7 (6SiCH$_3$), 17.5 (CH$_2$), 30.7(CH$_2$), 31.8 (m, |$^1J_{PC}$ + $^3J_{PC}$ = |24 Hz, PCH$_2$CH$_2$P), 127.4 (PPh), 127.9 (SPh), 129.1 (PPh), 129.3 (SPh), 131.0 (SPh), 132.7 (SPh), 134.6 (PPh), 137.6 (C$_q$, PPh). **$^{31}$P NMR** (162 MHz, CD$_2$Cl$_2$, 297K) δ = 52.76 (PCH$_2$CH$_2$P). **HR-MS** ESI (CH$_2$Cl$_2$): *m/z* 1364.3390 [M]$^+$, Calcd. (C$_{78}$H$_{82}$S$_2$P$_4$RuSi$_2$) 1364.3385. **IR** ν = 2051 cm$^{-1}$ (C≡C).

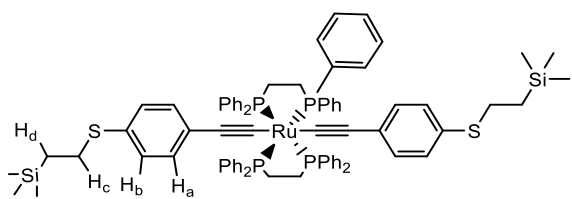

**[Fe(cyclam)(C≡C-*p*-C$_6$H$_4$-S(EDMS))$_2$](OTf).** Under argon, a dried Schlenk tube was charged by [Fe(cyclam)(OTf)$_2$]OTf (0.182g, 0.25 mmol) and compound **1** (0.121 g, 0.50 mmol), and freshly distilled THF (15 mL). After the mixture was cooled down in a dry ice/ethanol bath for 10 minutes, n-butyllithium (2.5 M in THF, 1 mmol) was added through dropwise. The reaction mixture was stirred in the cold bath for 5 minutes and allowed to warm to room temperature with stirring for 2 h. The mixture was quenched by 1 drop of water, and the reaction mixture was loaded directly onto silica gel (4 cm in fritted glass funnel) and eluted by using a (70:30) (v/v) mixture of dichloromethane and acetonitrile (until the eluent is colorless). The solvents were removed using rotary evaporation. The dried powder was then dissolved in a minimum amount of THF (4 mL) in a test tube and good quality blue crystals were obtained using pentane gas diffusion and filtered to afford **Fe** (43 % yield, 80 mg) **HR-MS** ESI (CH$_2$Cl$_2$): *m/z* 722.2985 [M]$^+$, Calcd. (C$_{36}$H$_{58}$S$_2$FeN$_4$Si$_2$) 722.29854. **Raman** v = 2083 cm$^{-1}$ (C≡C).

**X-Ray measurements**. X-Ray measurements were performed at 150(2) K with crystals mounted with a cryoloop on the goniometer head of a D8 Venture Bruker AXS diffractometer (multilayer monochromator, Mo Kα radiation, λ = 0.710 73 Å) at 150 K. The structures were solved by dual-space algorithm using the *SHELXT* program[5], and then refined with full-matrix least-squares methods based on $F^2$ (*SHELXL*)[6]. All non-H atoms of the non-disordered parts of the molecules were refined with anisotropic atomic displacement parameters. H atoms were finally included in their calculated positions and treated as riding on their parent atom with constrained thermal parameters, except for NH groups of Fe(cyclam)(C≡C-Ph-S(EDMS))$_2$(OTf) that were introduced in the structural model through Fourier difference map analysis. CCDC 2344014 and 2344016 contains the supplementary crystallographic data associated to this paper. These data can be obtained free of charge from The Cambridge Crystallographic Data Centre via www.ccdc.cam.ac.uk/data_request/cif. The main crystallographic data are given in Table S1.

Table S1. Main crystallographic data for *trans*-[Ru(dppe)$_2$(C≡C-Ph-S(EDMS))$_2$] and Fe(cyclam)(C≡C-Ph-S(EDMS))$_2$(OTf)

|  | *trans*-[Ru(dppe)$_2$(C≡C-Ph-S(EDMS))$_2$] | Fe(cyclam)(C≡C-Ph-S(EDMS))$_2$(OTf) |
|---|---|---|
| Formula | C$_{78}$ H$_{82}$ P$_4$ Ru S$_2$ Si$_2$ | C$_{36}$ H$_{58}$ Fe N$_4$ S$_2$ Si$_2$, CF$_3$ O$_3$ S |
| FW | 1364.68 | 872.08 |
| Cryst. Syst. | triclinic | monoclinic |
| Space group | P$\bar{1}$ | P2$_1$ |
| a (Å) | 9.882(2) | 12.0439(11) |
| b (Å) | 13.496(4) | 11.7117(9) |
| c (Å) | 14.580(4) | 16.0291(14) |
| α (°) | 100.143(6) | 90 |
| β (°) | 90.238(7) | 99.481(3) |
| γ (°) | 111.063(6) | 90 |
| V (Å$^3$) | 1781.5 (7) | 2230.1(3) |
| Z | 2 | 2 |
| D$_{ca}$ (g·cm$^{-3}$) | 1.272 | 1.299 |
| T (K) | 150 | 150 |
| Final R | 0.07 | 0.05 |
| R$_w$ | 0.22 | 0.12 |

**Cyclic voltammetry.** Electrochemical studies were carried out under argon using an Eco Chemie Autolab PGSTAT 30 potentiostat (CH$_2$Cl$_2$, 0.2M Bu$_4$NPF$_6$), the working electrode was a Pt disk, and ferrocene the internal reference.

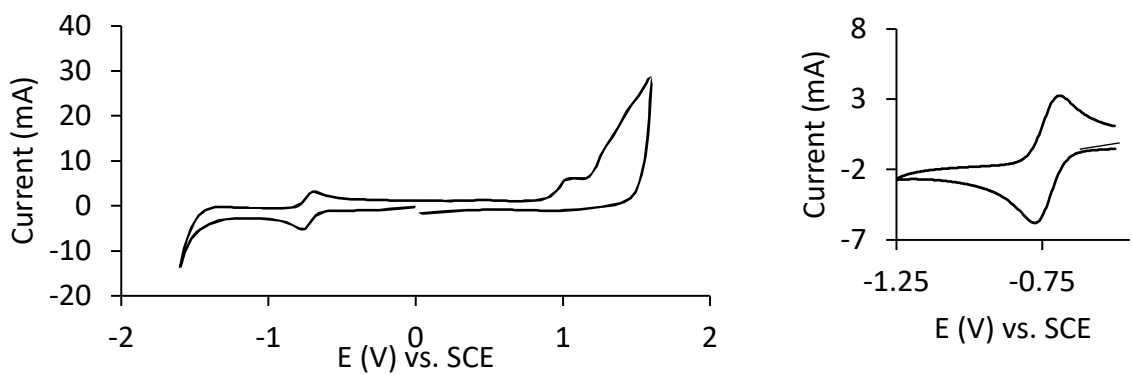

Figure S1. Cyclic voltammogram of a solution of Fe(cyclam)(C≡C-Ph-S(EDMS))2(OTf) at 0.1V/s vs. SCE (0.1 M [TBA](PF$_6$) in CH$_2$Cl$_2$)

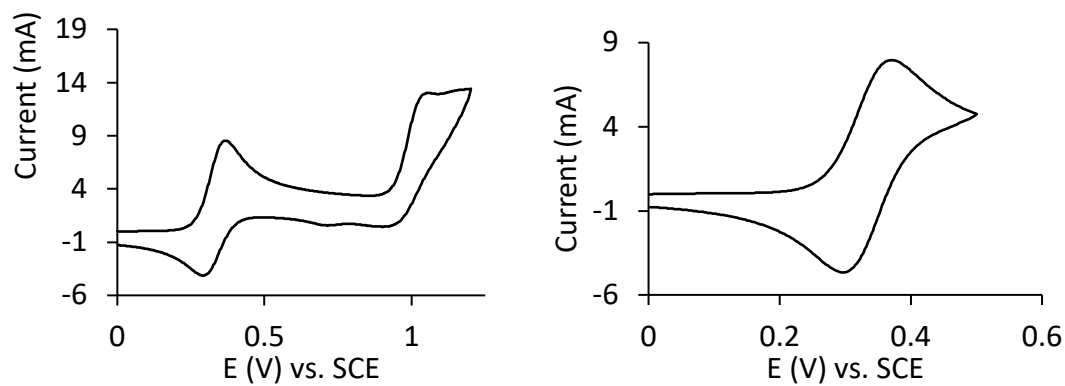

Figure S2. Cyclic voltammograms of a solution of Ru(dppe)$_2$(C≡C-Ph-S(EDMS))$_2$] at 0.1V/s vs. SCE (0.1 M [TBA](PF$_6$) in CH$_2$Cl$_2$)

## 2 – Single-Molecule Conductance Measurements

**Device fabrication.** Break junction devices were fabricated on a polished bronze substrate with a thickness of 500 µm. A polyimide layer approximately 2 µm thick was spin-coated onto the substrate, serving as both an insulating and sacrificial layer. The wafer was subsequently soft-baked at 130°C for 5 minutes and hard-baked at 430°C for 90 minutes under vacuum conditions. Two resist layers, consisting of MMA-MAA and PMMA, were spin-coated onto the wafer and baked at 170°C for 30 minutes. The wafer was subsequently diced into individual samples with dimensions of 3 × 18 mm². Structuring was performed using electron beam lithography on a Zeiss Cross Beam system operated at an acceleration voltage of 10 kV. Development was carried out in a 1:3 MIBK:IPA solution for 30 seconds, followed by rinsing with pure IPA. A gold (Au) layer with a thickness of 80 nm was deposited using electron beam evaporation. To create a freestanding bridge, the polyimide layer was etched using anisotropic reactive ion etching machine (parameters: 50 sccm $O_2$, $T$ = 100°C, $P_{RF}$ = 20 W, $P_{ICP}$ = 400 W, $P$ = 100 mTorr). This process removed approximately 500 nm of the polyimide layer, controlled by an interferometer.

**Molecular deposition.** After sample fabrication, molecules were deposited onto the prepared samples. The deposition parameters are summarized in Table S2. Initially, the molecules were dissolved in the appropriate solvent. For drop-cast deposition, a drop of the molecular solution was applied to the junction. Once the solvent evaporated, the sample was connected. For immersion deposition, the sample was immersed in the molecular solution for a specified duration and subsequently dried. In both deposition methods, the sample was connected to the setup using copper cables and silver paste. To ensure enhanced mechanical stability, the cables were additionally adhered to the sample.

Table S2. Overview of the different samples and the parameters used in the fabrication of the samples.

| Molecule | Concentration | Deposition | Time | Solvent |
|---|---|---|---|---|
| **Fe-cyclam** | 1 mMol | drop-cast | - | 100 % THF |
| **Fe-cyclam** | 1 mMol | drop-cast | - | 100 % THF |
| **Fe-cyclam** | 1 mMol | mmersion | 20 min | 20 % THF, 80 % ethanol |
| **Fe-cyclam** | 1 mMol | immersion | 3 min | 20 % THF, 80 % ethanol |
| **Fe-cyclam** | 1 mMol | immersion | 60 min | 20 % THF, 80 % ethanol |
| **Fe-cyclam** | 5 mMol | immersion | 60 min | 20 % THF, 80 % ethanol |
| **Ru-dppe** | 1 mMol | drop-cast | - | 100 % THF |
| **Ru-dppe** | 2 mMol | drop-cast | - | 100 % THF |
| **Ru-dppe** | 1 mMol | drop-cast | - | 100 % THF |
| **Ru-dppe** | 0.1 mMol | drop-cast | - | 100 % THF |
| **Ru-dppe** | 4 mMol | drop.cast | - | 100 % THF |
| **Ru-dppe** | 0.1 mMol | drop-cast | - | 100 % THF |

**Transport measurements.** All measurements were conducted using a custom-built cryogenic vacuum dipstick incorporating a mechanically controlled break junction (MCBJ). The samples were mounted in a three-point bending mechanism operated by a DC stepper motor, with the displacement of the pushing rod finely controlled by a differential screw featuring a 50 µm pitch difference. The electrical setup consisted of homemade coaxial cables, shielded with a copper cap to reduce the influence of stray electromagnetic fields. Electronic transport measurements were performed using a Yokogawa 7651 DC voltage source, while current and voltage signals were captured with Femto DLCPA and DLPVA low-noise amplifiers and recorded using Agilent 34410A multimeters (Figure S3).

Prior to measurements, the samples were cooled, and amplifier offsets were corrected in the open configuration. Differential conductance (d*I*/d*V*) was measured using HF2LI Zurich Instruments lock-in amplifiers with high precision.

The entire measurement process, including device operation, was automated and remotely controlled using a custom Python-based software program.

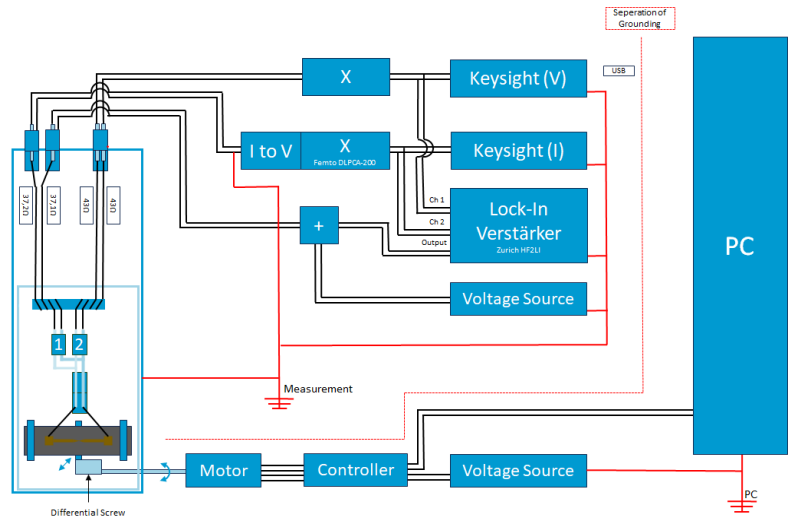

Figure S3. Electronics of the mechanically controlled break junction setup.

**Fitting results**

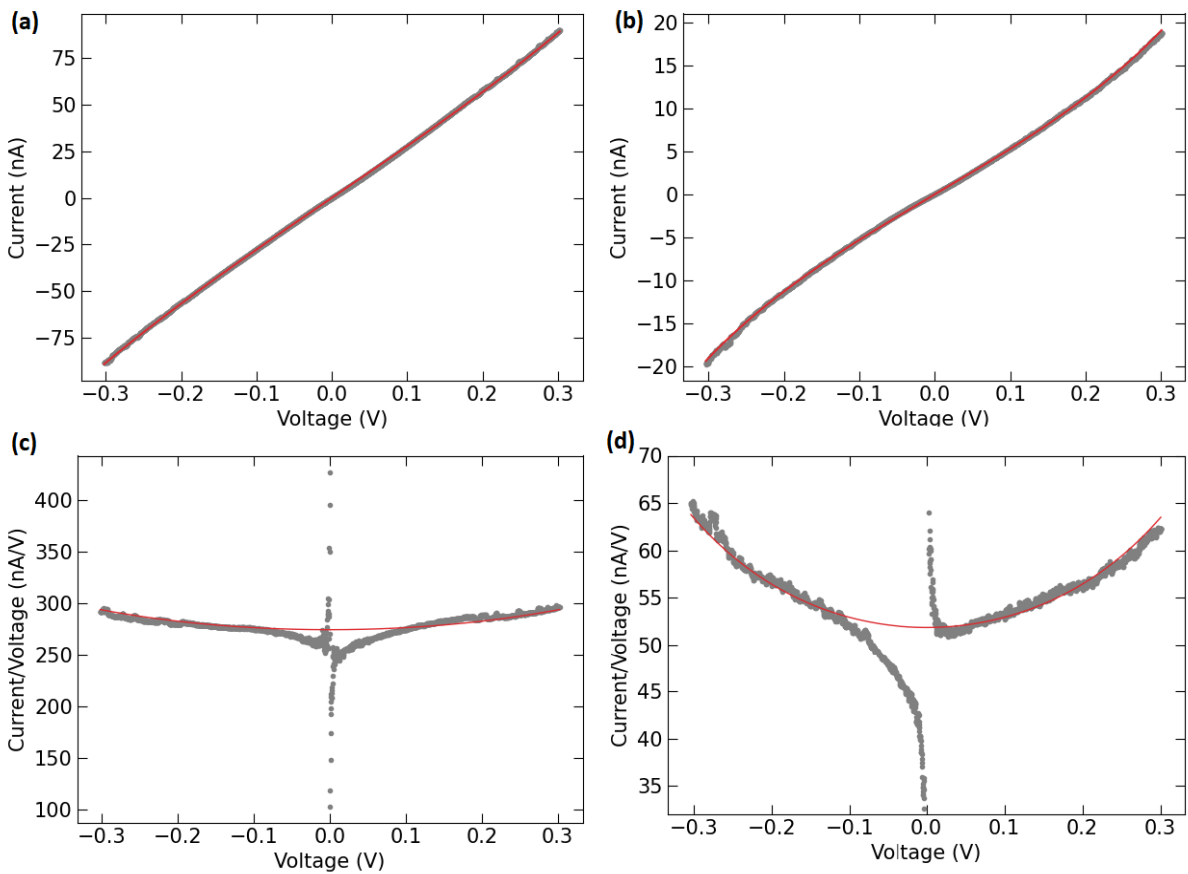

Figure S4. Example of *I-V* curves for **Fe-cyclam** (a) and **Ru-dppe** (b), fitted with the SLM (red line). (c) and (d) are the *I/V-V* characteristic of the datasets used in (a) and (b) respectively; the red line represents the same fit. Fit parameters in (a) and (c): $\epsilon_0$ = 0.58 eV, $\Gamma_L$ = 0.018 eV, $\Gamma_R$ = 0.017 eV; Fit parameters in (b) and (d): $\epsilon_0$ =0.35 eV, $\Gamma_L$ = 0.0053 eV, $\Gamma_R$ = 0.0039 eV.

The individual stable I-V curves are analyzed using the single-level model (SLM) (Eq. S4) which assumes that a single molecular orbital couples to the electrodes such that its transmission function adopts a Lorentzian shape. Examples for these fits to individual experimental I-Vs for both molecules are shown in Figure 4. The quality of the fit is determined by the goodness-of-fit $R^2_{adj}$.[7]

The result of the fits shown in Figure S4-a and Figure S4-b suggest that for both molecules the I-Vs can be well described by the SLM, as confirmed by a high $R^2_{adj}$ value of 0.9999 for **Fe-cyclam** and of 0.9997 for **Ru-dppe**. Figure S4-c and Figure S4-d show the same data but plotted as *I/V* vs *V* to reveal the remaining systematic deviations between the fit and the data. These experimental curves show anomalies around zero bias arising from the division through a small number and small offsets of the current and voltage amplifiers. These deviations do not affect the fitting quality markedly. However, despite the high $R^2_{adj}$ values, systematic deviations are observed that cannot be explained by experimental aspects, but indicate the limitations of the SLM. These arise from the simplified nature of the SLM and may indicate that in the real system either more than a single orbital contributes to the current or the coupling is more complex. It also indicates that fits with lower $R^2_{adj}$ have to be interpreted with caution.

We systematically performed fits with the SLM for all experimental I-V curves and found that 36 % of them could be fitted with a $R^2_{adj}$ of 0.9 or higher for **Fe-cyclam** and about only 25 % for **Ru-dppe**. The analysis shows that the fit results depend on the voltage range used for fitting the data. To exemplify When using the I-Vs recorded over large bias were analyzed in smaller bias range, the fit results systematically varied with a trend to retrieve larger $\epsilon_0$ for larger bias range. Because of the limited database, we did not analyze the bias dependence systematically but the fit results are plotted separately for small (|*V*| < 0.5 V) and large (|*V*| > 0.5 V) bias ranges. We also found that only the fits with $R^2_{adj}$ > 0.998 show negligible systematic deviations from the experimental data and therefore color-coded them differently. Figure 4-a, Figure S5-left and Figure S6 show the results for the level alignment $\epsilon_0$ and the coupling constants for **Fe-cyclam** while Figure 4-b, Figure S5-right and Figure S7 summarize the equivalent data for **Ru-dppe**.

For **Fe-cyclam**, in the small bias range, the $\epsilon_0$ histogram reveals two broad maxima, one located at $\epsilon_0$ = 0.15 eV and the other around $\epsilon_0$ = 0.55 eV. Curves with $\epsilon_0$ higher than the bias range have to be taken with caution, because in this range the *I-V*s reveal only small curvature, limiting the accuracy of the resulting fit values. Two different $\epsilon_0$ values emerge from this analysis in line with expectation of two different oxidation states of the Fe-core of the molecule. The analysis of high bias range shows only very few *I-V*s that can be fitted with $R^2_{adj} > 0.998$. When relaxing this criterion to $R^2_{adj} > 0.9$, the maximum around $\epsilon_0 = 0.55$ eV is confirmed and the one at smaller $\epsilon_0$ is slightly shifted to higher values. In addition, we observe a third maximum around 0.9 which contains a few counts only.

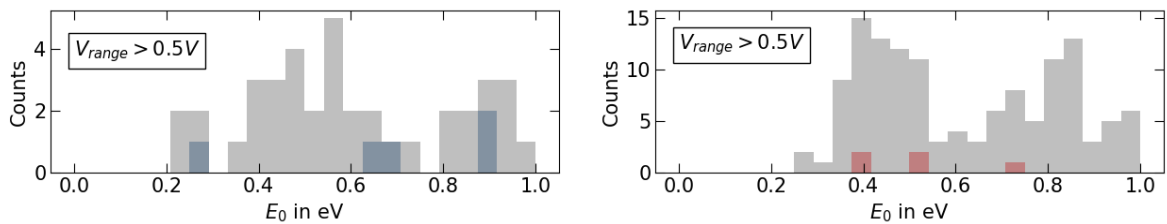

Figure S5. $\epsilon_0$ statistics of all *I-V* curves fitted with the SLM separated by the goodness of fit ($R^2_{adj}$) for bias range ($V_{range}$) superior to 0.5 V. (left) **Fe-cyclam**; 6 *I-V*s have $R^2_{adj} > 0.998$. (right) **Ru-dppe**; 182 *I-V*s have $R^2_{adj} > 0.998$. Statistics for $V_{range}$ inferior to 0.5 V are given in Figure 4 (main text).

I-V measurements of **Ru-dppe** result in a distribution with a narrow peak around $\epsilon_0 = 0.15$ eV and a broad peak at around $\epsilon_0 = 0.4$ eV with a tail towards higher values. Fits over the larger voltage range (i.e. $V_{range} > 0.5$ V) confirm the latter, when again relaxing the goodness-of-fit criterion to $R^2_{adj} > 0.9$, but not the one at small $\epsilon_0$. The **Ru-dppe** is expected to have a well-defined electronic state and, hence, we expect only one peak.

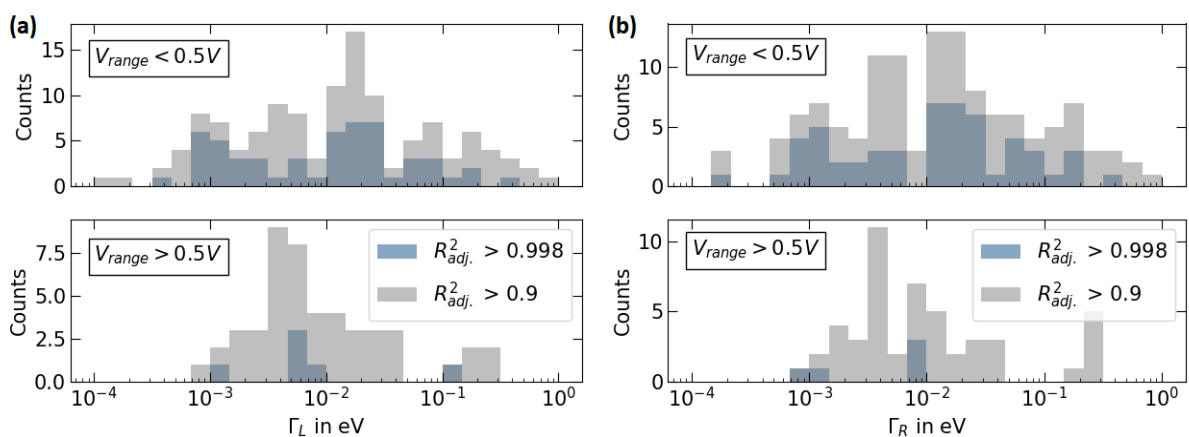

Figure S6. $\Gamma_{1,2}$ determined by the single-level model for **Fe-cyclam**. The data sets used are the same as in Figure S5. The separation of the dataset is done by the bias range ($V_{range}$) and color-coded according to $R^2_{adj}$.

Figure S6-a and Figure S6-b show a rather symmetric distribution of the coupling constants ($\Gamma_{1,2}$) for Fe-cyclam as expected for a symmetric molecule. The broad distribution for both bias ranges indicates that a range of junction configurations is possible in agreement with the broad distribution of observed

conductance values (Figure 2, main text). Comparing the ratio $\Gamma_1/\Gamma_2$ for individual junctions retrieves values between 0.9 and 1.1 for all fittable curves, in line with the symmetry of the I-V curves.

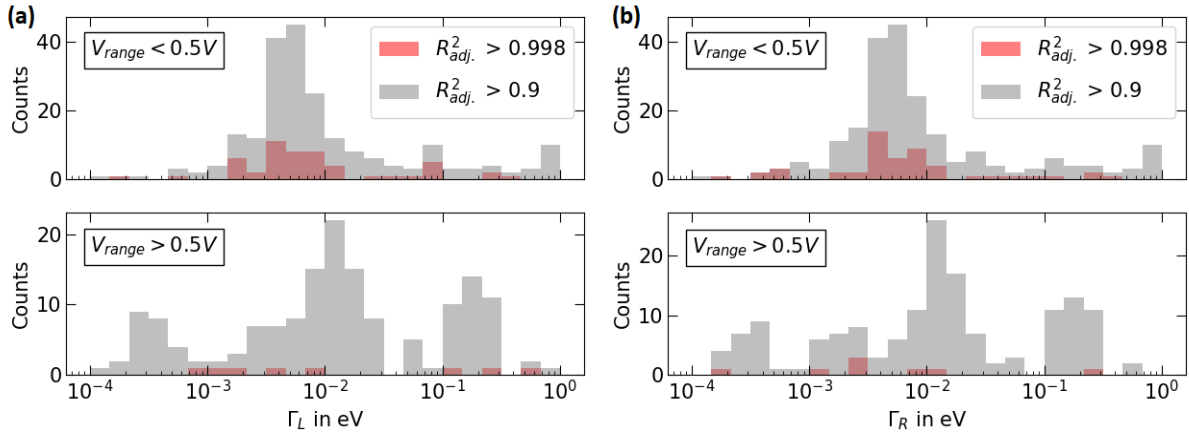

Figure S7. $\Gamma_{1,2}$ determined by the single-level model for the measured I-V curves of **Ru-dppe**. The datasets used are the same as in Figure S5. The separation of the dataset is done by $R^2_{adj}$ and $V_{range}$.

Figure S7 shows the coupling constant distributions for **Ru-dppe** which are also symmetric, as expected. The peaks slightly below $10^{-2}$ eV arise from the non-uniform sampling of the data mentioned earlier.

**Transport measurements in magnetic field.** Additional measurements of single-molecule junctions (SMJs) under magnetic field were performed. To this end, a SMJ was created in the manner described for I-V measurements. I-V curves were taken under different magnetic field applied out-of-plane. All these measurements were performed at T = 4.2 K.

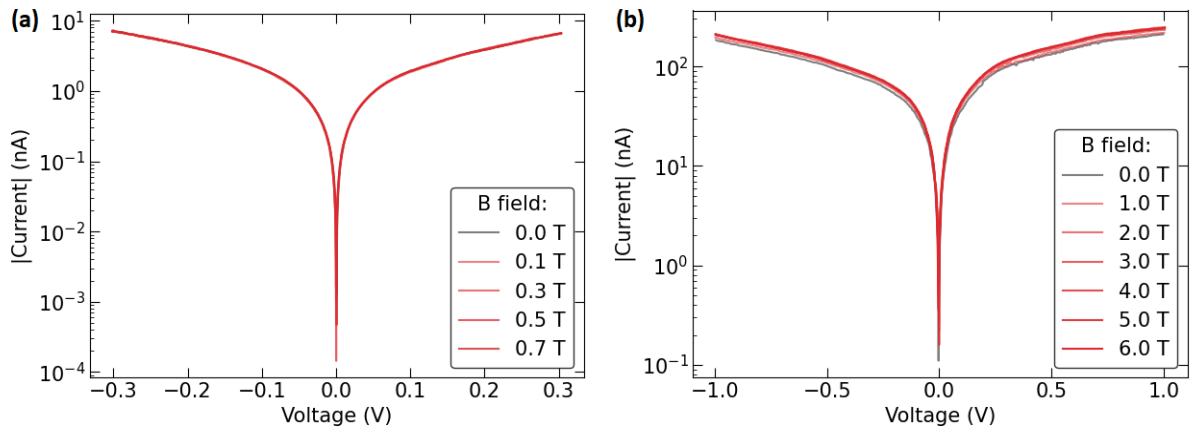

Figure S8. Transport measurements in magnetic field on single-molecule contacts of (a) **Fe-cyclam** and (b) **Ru-dppe**. The magnetic field was applied out-of-plane.

The influence of magnetic fields on the electronic transport properties of **Fe-cyclam** and **Ru-dppe** SMJs was investigated by recording current-voltage (I-V) characteristics under varying magnetic field strengths, as shown in Figure S8. For **Fe-cyclam**, no significant variation was observed in the I-V curves across different magnetic field intensities. Even at higher magnetic field strengths, the conductance remained constant. **Ru-dppe** junctions exhibit a slight variation in the absolute current values across the *I-V* curves under different magnetic field conditions. However, the overall shape of the curves remained unchanged.

**Inelastic Electron Tunneling Spectroscopy (IETS).**

Table S3. Infrared (IR) and Raman spectroscopy phonon modes identified in the literature.

| Phonon mode | Peak position in mV | Reference |
|---|---|---|
| Au-Au | 10-20 | 8 |
| Au-S | 37-43 | 9 |
| Ring vibrations | 80-85, 102-105, 119-126, 180-183 | 10 |
| C-C | 134 | 11 |
| C≡C | 250-270 | 11 |
| C-S | 131-150 | 12 |
| Fe-N | 27-32, 51-52 | 13 |
| Fe-C≡C (for Fe$^{III}$) | 258 | 3 |
| C-N | 60-75, 165-180 | 14 |
| N-H | 100-110 | 14 |
| Ru-C≡C | 254 | this work |

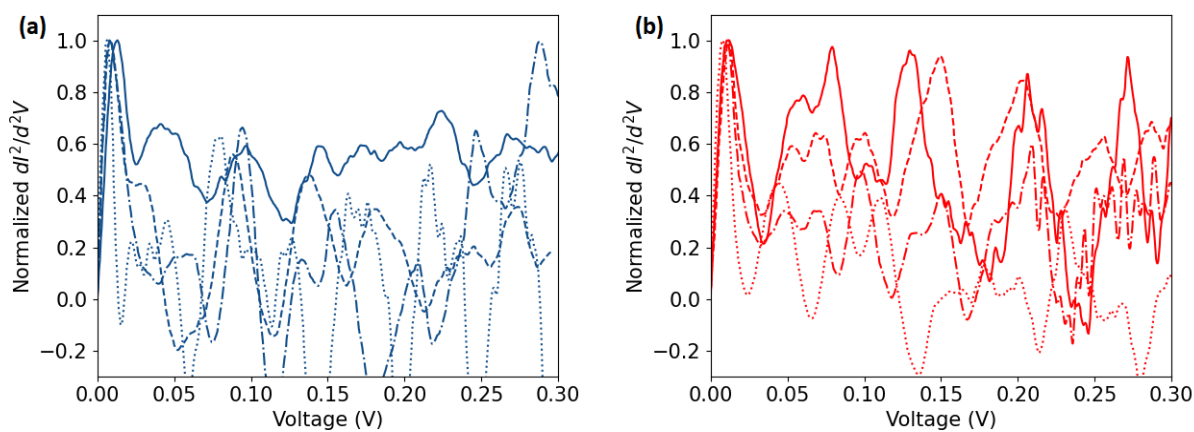

Figure S9. Example of inelastic electron tunneling spectra (IETS) for (a) **Fe-cyclam** and (b) **Ru-dppe**, Record mainly in the conductance range around $10^{-3}$ $G_0$. The measurements were performed using the lock-in technique. The amplitude is normalized to the peak height of the Au phonon mode (~ 20 mV) for better comparison.

Finally, to verify that the recorded data indeed corresponds to molecular junctions, we performed inelastic electron tunneling spectroscopy (IETS) for selected stable junctions (Figure S9). Due to the limited signal to noise ratio, we concentrated on the conductance range around $10^{-3}$ $G_0$. The IETS spectra of the junctions reflect the vibrational excitations of the intermolecular bonds as well as of the bonds to the metal electrodes and the phonon modes of the electrodes close to the junction. These are directly measured by lock-in technique recording the 2$^{nd}$ harmonic of the excitation frequency. The peak positions in the plot indicate the excitation energies according to $eV_{peak} = hf_{vib}$. The amplitudes of the peaks depend on the electron-vibron coupling strength which is sensitive to details of the contact configuration.[15] By comparing the peak positions to Raman and IR spectroscopy (Table S3), the modes can be identified. In Figure S9, for better comparison and visibility, only the spectra for positive voltages are shown and their amplitudes are normalized to the one of the Au phonon mode (~ 20 mV) which is present in all spectra. As shown in Figure S9, the spectra for different contacts, although absolutely stable and reproducible for a given contact, vary largely, confirming that several different binding configurations are realized. Nevertheless, there are voltage ranges in which most spectra reveal peaks and which are therefore considered as typical for the specific molecule. For both molecules we observe the vibrations known from hydrocarbons, including C-H stretching and ring vibrations, confirming that the junctions are indeed molecular junctions. For **Fe-cyclam**, the normalized peak heights are systematically smaller than for **Ru-dppe** and pronounced peaks beyond the background are observed only in the range around 90 mV and 280 mV. The former corresponds to ring vibrations, while the latter hallmarks the carbon triple-bond stretching, confirming that the current path includes this bond. For **Ru-dppe**, prominent peaks arise around 150 and 220 mV. The vibrations around 150 mV most likely reflect C-S stretching modes, while the ones around 220 mV cannot be unambiguously identified. We tentatively assign them to a softened Ru-C≡C vibration, since no other mode is known in this range.

## 3 – Self-assembled monolayers: preparation and physical characterizations

**Self-assembled monolayers on Au electrodes.** Ultraflat template-stripped Au surfaces ($^{TS}$Au), with rms roughness of ≈ 0.4 nm, were prepared according to methods already reported.[16,17] In brief, a 300–500 nm thick Au film was evaporated on a very flat silicon wafer covered by its native SiO$_2$ (rms roughness of ≈ 0.4 nm), which was previously carefully cleaned by piranha solution (30 min in 7:3 H$_2$SO$_4$:H$_2$O$_2$ (v/v)); **Caution**: Piranha solution is a strong oxidizer and reacts exothermically with organics), rinsed with deionized water, and dried under a stream of nitrogen. Clean 10 × 10 mm pieces of glass slide (ultrasonicated in acetone for 5 min, ultrasonicated in 2-propanol for 5 min, and UV irradiated in ozone for 10 min) were glued on the evaporated Au film (UV-polymerizable glue, NOA61 from Epotecny), then mechanically peeled off providing the $^{TS}$Au film attached on the glass side (Au film is cut with a razor blade around the glass piece). The self-assembled monolayers (SAMs) of **Ru-dppe** and **Fe-cyclam** were prepared from a 1 mM solution of [Ru(dppe)$_2$(C≡C-Ph-S(EDMS))$_2$] and Fe(cyclam)(C≡C-Ph-S(EDMS))$_2$(OTf) in a mix of THF (50%) and ethanol (50%). The $^{TS}$Au substrates were immersed for 24 hours in this solution, and then cleaned in THF/EtOH with ultrasounds for one minute.

**Ellipsometry.** We recorded spectroscopic ellipsometry data (on ca. 1 cm$^2$ samples) in the visible range using a UVISEL (Horiba Jobin Yvon) spectroscopic ellipsometer equipped with DeltaPsi 2 data analysis software. The system acquired a spectrum ranging from 2 to 4.5 eV (corresponding to 300–750 nm) with intervals of 0.1 eV (or 15 nm). The data were taken at an angle of incidence of 70°, and the compensator was set at 45°. We fit the data by a regression analysis to a film-on-substrate model as described by their thickness and their complex refractive indexes. First, a background for the substrate before monolayer deposition was recorded. We acquired three reference spectra at three different places of the surface spaced of a few mms. Secondly, after the monolayer deposition, we acquired once again three spectra at three different places of the surface and we used a 2-layer model (substrate/SAM) to fit the measured data and to determine the SAM thickness. We employed the previously measured optical properties of the substrate (background), and we fixed the refractive index of the monolayer at 1.50.[18] We note that a change from 1.50 to 1.55 would result in less than a 1 Å error for a thickness less than 30 Å. The three spectra measured on the sample were fitted separately using each of the three reference spectra, giving nine values for the SAM thickness. We calculated the mean value from these nine thickness values and the thickness incertitude corresponding to the standard deviation. Overall, we estimated the accuracy of the SAM thickness measurements at ± 2 Å.[19]

Table S4 provides the evaluation the tilt angle formed by the grafted molecular wires using the measured thicknesses and the optimized geometry length of the [Ru(dppe)$_2$(C≡C-Ph-SH)$_2$] and [Fe(cyclam)(C≡C-Ph-SH)$_2$]$^+$ (see computational results). Since the measured thicknesses are smaller than the length (-S to S-), we assume that the protecting groups were removed (see below XPS measurements). We estimated a tilt angle of ≈ 38° and ≈ 10° with respect to the Au surface normal for **Ru-dppe** and **Fe-cyclam** grafted molecules, respectively. Similarly, the area per molecule is calculated from the nominal diameter of the molecules (distance between the two farthest H atoms, plus twice the H van der Waals radius), which is used to estimate the number of molecules in the SAM contacted by the C-AFM and SThM tips (see below).

Table S4. Measured thickness (t) of the SAMs and comparison with the molecule length (l). The S-to-S length is taken from the geometry optimization in the gas phase shown in the upper row. We also give the tilt angle Θ from the surface normal (cosΘ = t/l), the molecule diameter (d), the projected diameter on the surface (D = d/cosΘ), and the corresponding area per molecule occupied on the surface.

|  | Ru-dppe | Fe-cyclam |
|---|---|---|
| Measured thickness (nm) | 1.5 ± 0.2 | 1.8 ± 0.2 |
| Theoretical length -S to S- (nm) | 1.88 | 1.84 |
| Tilt angle (°) | 38 | 10 |
| Theoretical diameter (nm) | 1.47 | 1.13 |
| Projected diameter (nm) | 1.86 | 1.15 |
| Area per molecule (nm$^2$) | 2.71 | 1.04 |

**Topographic AFM images**. The SAMs were examined by topographic AFM (Figure S10) using a tip probe in silicon, model LprobeTapping20 by Vmicro. The SAM surfaces are homogeneous and flat, they are free of gross defects (neither pinhole nor aggregate, the dark spots are defects (pinholes) in the underlying Au substrate, the white spots are small dusts since the measurements were done in ambient air, both are masked for the roughness analysis). The observed rms roughness value for the surface of the **Ru-dppe** SAM is ≈ 0.7 nm, while the value for the **Fe-cyclam** SAM is ≈ 0.5 nm. Both values are close to the roughness observed for our $^{TS}$Au substrates (ca. 0.4 nm),[20] which indicates the good formation of a monolayer.

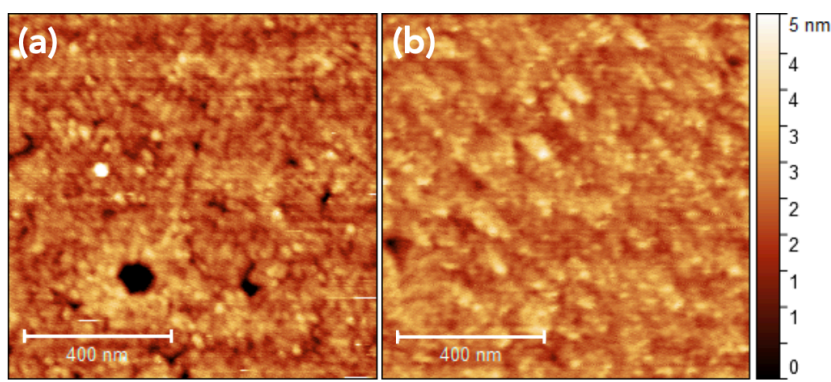

Figure S10. (a) Topographic AFM image of the **Ru-dppe** SAM, rms roughness ≈ 0.7 nm. (b) Topographic AFM image of the **Fe-cyclam** SAM, rms roughness ≈ 0.5 nm. The black holes and white spots are pinholes in the underlying Au substrate and tiny dusts, respectively, and were masked for the roughness analysis. Scale bars: 400 nm. Same height scale for the two images.

**XPS and UPS**. High resolution XPS spectra (Physical Electronics 5600 spectrometer fitted in an UHV chamber with a residual pressure of 3 × 10$^{-10}$ mbar) were recorded with a monochromatic Al$_{K\alpha}$ X-ray source ($h\nu$ = 1486.6 eV), a detection angle of 45° as referenced to the sample surface, an analyzer entrance slit width of 400 μm and an analyzer pass energy of 12 eV. Background was subtracted by the Shirley method.[21] The peaks were decomposed using Voigt functions and a least squares minimization procedure. Binding energies (BE- were referenced to the C 1s BE, set at 284.8 eV. The same equipment was used for UPS characterizations with an ultraviolet source (He1, $h\nu$ = 21.2 eV).

The S2p region for the **Ru-dppe** SAM shows the expected two doublets (S2p$_{1/2}$ and S2p$_{3/2}$) associated to the S bonded to Au, S-Au (S2p$_{1/2}$ at 163.1 eV, S2p$_{3/2}$ at 161.9 eV) and S not coordinated to Au, i.e., S-C (S2p$_{1/2}$ at 164.7 eV, S2p$_{3/2}$ at 163.5 eV) (Figure S11-a). These doublets are separated by ca. 1.2 eV as expected with an amplitude ratio [S2p$_{1/2}$]/[S2p$_{3/2}$] of 1/2. Similar results were obtained for the **Fe-cyclam** SAM, Figure S11-b (S-Au (S2p$_{1/2}$ at 163.0 eV, S2p$_{3/2}$ at 161.8 eV), S-C (S2p$_{1/2}$ at 164.8 eV, S2p$_{3/2}$ at 163.6 eV)). The S-Au bonding is confirmed by the calculated amplitude ratios (integrated peak area) of the non-bonded S atoms and the bonded S to Au [S-C]/[S-Au]. For both SAMs, we obtain [S-C]/[S-Au] = 0.6, slightly lower than the expected value of 1.

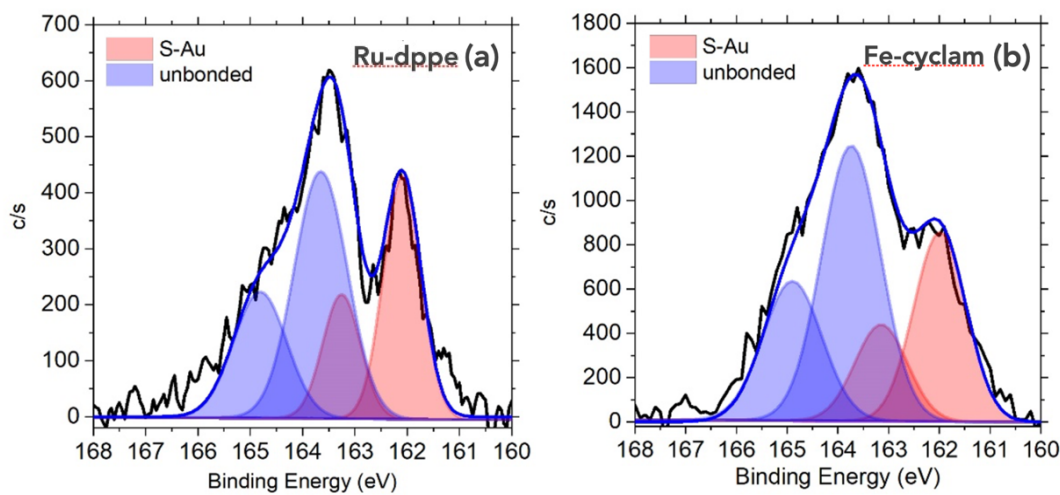

Figure S11. XPS spectra of the S2p region: (a) **Ru-dppe** SAM, (b) **Fe-cyclam** SAM. The dark lines are the measurements. The blue lines are the fits with the two-doublet decomposition shown in orange. Figure S12-a shows the Ru3$d_{3/2}$ peak observed for the **Ru-dppe** SAM. This peak is weak, as also noted in other previous results on **Ru-dppe**-type SAMs.[22,23] In the Fe2p region (Figure S12-b) of the **Fe-cyclam** SAM, we observed a complex signal corresponding to the Fe$^{II}$ and Fe$^{III}$ oxidation states. Each chemical species is fitted with the 2$p_{1/2}$ and 2$p_{3/2}$ doublet constrained to have a 1:2 peak area ratio, the same FWHM for each doublet, and a peak separation of 13.1 eV. The main doublet at 710.0 eV and 723.1 eV is associated to Fe$^{II}$2$p_{3/2}$ and Fe$^{II}$2$p_{1/2}$, respectively. The doublet associated to Fe$^{III}$2$p_{3/2}$ and Fe$^{III}$2$p_{1/2}$ is observed at 712.4 eV and 725.5 eV, respectively. Satellite peaks corresponding to the two oxidation states of iron are also observed at 715.1 eV, 728.7 eV, 719.1 eV and 732.1 eV for Fe$^{II}$2$p_{3/2}$, Fe$^{II}$2$p_{1/2}$, Fe$^{III}$2$p_{3/2}$ and Fe$^{III}$2$p_{1/2}$, respectively.[24,25] From the integrated peak areas, we estimated a ratio [Fe$^{II}$]/[Fe$^{III}$] ≈ 2.1.

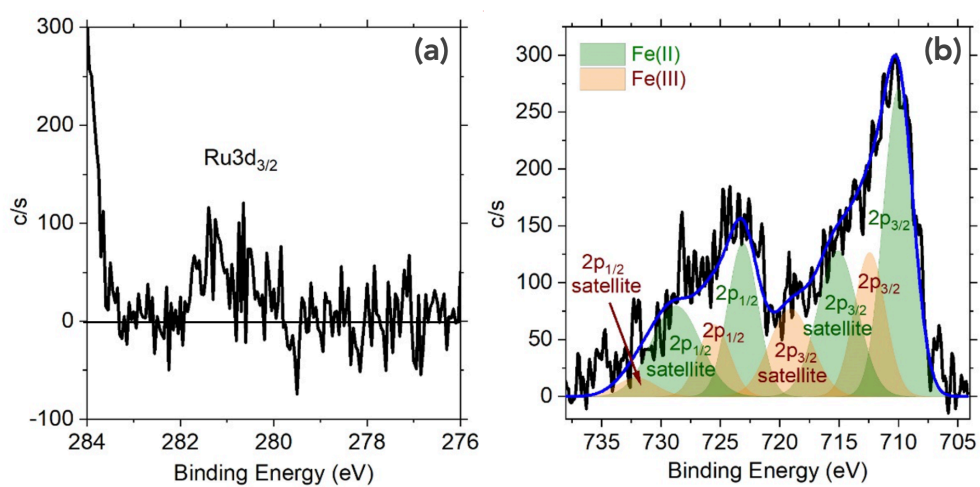

Figure S12. XPS spectra: (a) Ru3$d_{3/2}$ for the **Ru-dppe** SAM, (b) Fe2p region for the **Fe-cyclam** SAM. The dark lines are the measurements. The blue line is the fit taking into account the different contributions from the Fe$^{II}$ and Fe$^{III}$ oxidation states, in green and orange, respectively. See text for details.

The UPS spectra (Figure S13-a) show the onset of the HOMO at ≈ 0.34 eV below the Au electrode Fermi energy for the **Ru-dppe** SAM and at ≈ 0.75 eV for the **Fe-cyclam** SAM. From the secondary electron cutoff (Figure S13-b), we observed a shift of the work function by 0.40 and 0.55 eV for the **Ru-dppe** and **Fe-cyclam** SAMs, respectively.

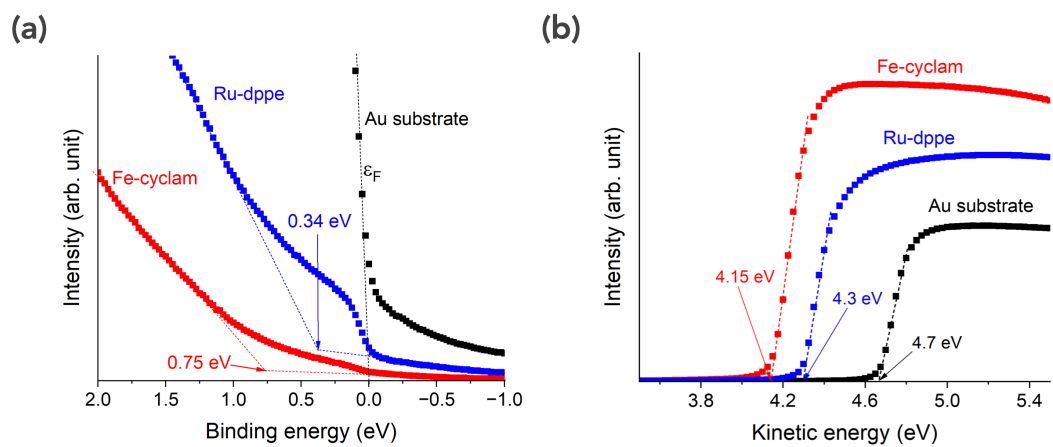

Figure S13. (a) UPS spectra of the **Ru-dppe** and **Fe-cyclam** SAMs at the onset of the HOMO. The Au Fermi energy is set to 0 eV from an UPS measurement on a clean reference Au substrate. (b) UPS spectra at the secondary electron cutoff to determine the work functions.

## 4 - Conductive-AFM

**General procedure**. We measured the electron transport properties at the nanoscale by C-AFM (ICON, Bruker) at room temperature (RT) (air-conditioned laboratory: 22.5°C and a relative humidity of 35-40%) using a tip probe in platinum/iridium (PtIr), model SCM-PIC-V2 from Bruker. We used a "blind" mode to measure the current-voltage (*I-V*) curves and the current histograms: a square grid of 10×10 was defined with a pitch of 50 to 100 nm. At each point, one *I-V* curve is acquired leading to the measurements of 100 traces per grid. This process was repeated several times at different places (randomly chosen) on the sample, and up to several thousands of *I-V* traces were used to construct the current-voltage datasets.

The tip load force was set at ≈ 30 nN for all *I-V* measurements, a lower value leading to too many contact instabilities during the *I-V* measurements. Albeit larger than the usual load force (2-5 nN) used for C-AFM on SAMs, this value is below the limit of about 60-70 nN at which the SAMs start to suffer from severe degradations. For example, a detailed study[26] showed a limited strain-induced deformation of the monolayer (≲ 0.3 nm) at this used load force. The same conclusion was confirmed by our own study comparing mechanical and electrical properties of alkylthiol SAMs on flat Au surfaces and tiny Au nanodots.[27]

**C-AFM contact area.** Considering: (i) the area per molecule on the surface (as estimated from the thickness measurement and calculated geometry optimization, Table S4), and (ii) the estimated C-AFM tip contact surface (see below), we estimated the C-AFM tip contact area and the number, N, of molecules contacted by the tip as follows. As usually reported in literature,[26,28–30] the contact radius, $a$, between the C-AFM tip and the SAM surface, and the SAM elastic deformation, $\delta$, are estimated from a Hertzian model:[31]

$$a^2 = \left(\frac{3RF}{4E^*}\right)^{2/3} \quad \text{Eq. S1}$$

$$\delta = \left(\frac{9}{16R}\right)^{1/3}\left(\frac{F}{E^*}\right)^{2/3} \quad \text{Eq. S2}$$

with $F$ the tip load force (30 nN), $R$ the tip radius (20 nm) and $E^*$ the reduced effective Young's modulus defined as:

$$E^* = \left(\frac{1}{E^*_{SAM}} + \frac{1}{E^*_{tip}}\right)^{-1} = \left(\frac{1-v^2_{SAM}}{E_{SAM}} + \frac{1-v^2_{tip}}{E_{tip}}\right)^{-1} \quad \text{Eq. S3}$$

In this equation, $E_{SAM/tip}$ and $v_{SAM/tip}$ are the Young modulus and the Poisson ratio of the SAM and C-AFM tip, respectively. For the Pt/Ir (90/10) tip, we have $E_{tip}$ = 204 GPa and $v_{tip}$ = 0.37 using a rule of mixture with the known material data.[32] These parameters for the **Ru-dppe** and **Fe-cyclam** SAMs are not known and, in general, they are not easily determined in such a monolayer. Thus, we consider the value of an effective Young modulus of the SAM $E^*_{SAM}$ = 38 GPa as determined for the "model system" alkylthiol SAMs from a combined mechanic and electron transport study.[26] With these parameters, we estimated $a$ = 2.4 nm (contact area = 18 nm$^2$) and $\delta$ = 0.28 nm. With an area per molecule of ca. 2.7 nm$^2$ (**Ru-dppe**) and ca. 1 nm$^2$ for **Fe-cyclam** (see ellipsometry section) and calculating the maximum filling of a 18 nm$^2$ circle by these smaller circles,[33] we infer that about 5 molecules are measured in the SAM/PtIr junction for the **Ru-dppe** SAM and *ca.* 15 for the **Fe-cyclam** SAM.

**Data analysis.** Before constructing the *I-V* datasets shown in Figures 8 and 9 (main text) and analyzing the *I-V* curves with the single energy-level model, the raw set of *I-V* data is scanned and some *I-V* curves were discarded from the analysis (**Figure S14**):

− At high current, the *I-V* traces that reached the saturating current during the voltage scan (the compliance level of the trans-impedance amplifier, typically ≈ 10$^{-7}$ A here (depending on the gain of the amplifier) and/or *I-V* traces displaying large and abrupt steps during the scan (contact instabilities).

− At low currents, the *I-V* traces that reached the sensitivity limit (almost flat *I-V* traces and noisy *I-V*s) and displayed random staircase behavior (due to the sensitivity limit - typically few pA depending on the used gain of the trans-impedance amplifier and the resolution of the ADC (analog-digital converter).

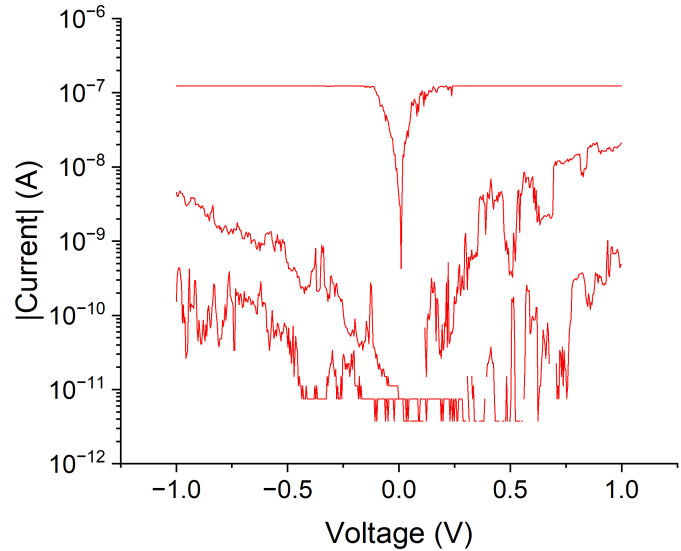

Figure S14. Typical *I-V* discarded from the analysis (from the raw *I-V* dataset of the **Ru-dppe** SAM used for the Figure 8, main text).

**Fits of the *I-V* curves with the analytical single-level model (SLM).** The *I-V* traces of the dataset of the $^{TS}$Au-SAM/C-AFM tip molecular junctions as well as the MCBJ data were fitted with the single-level model (SLM) given by the following analytical expression:[15,34]

$$I(V) = N\frac{8e}{h}\frac{\Gamma_1\Gamma_2}{\Gamma_1+\Gamma_2}\left[\arctan\left(\frac{\epsilon_0+\frac{\Gamma_1}{\Gamma_1+\Gamma_2}eV}{\Gamma_1+\Gamma_2}\right) - \arctan\left(\frac{\epsilon_0-\frac{\Gamma_2}{\Gamma_1+\Gamma_2}eV}{\Gamma_1+\Gamma_2}\right)\right] \quad \text{Eq. S4}$$

with $\epsilon_0$ the energy of the molecular orbital (MO), here HOMO, involved in the transport (with respect to the Fermi energy of the electrodes), $\Gamma_1$ and $\Gamma_2$ the electronic coupling energy between the MO and the electronic states in the two electrodes, *e* the elementary charge, *h* the Planck constant and *N* the number of molecules contributing in the molecular junction (assuming independent molecules conducting in parallel, i.e., no intermolecular interaction[35–37]). For the MCBJ data we used *N* = 1, for the SAM data the value, *N* as determined above (N = 5 for **Ru-dppe** and 15 for **Fe-cyclam**).

This model is valid at 0 K, since the Fermi function of the electrodes is not taken into account. However, it was shown that it can be used to describe data measured at RT for voltages below the transition between the off-resonant and resonant transport conditions (i.e., for $|eV| < 2\epsilon_0$ in the symmetric case $\Gamma_1 = \Gamma_2$) at which the broadening of the Fermi function modifies the *I-V* shape.[38–40] We defined this bias voltage window of confidence by TVS (transition voltage spectroscopy)[41–45] that gives us an estimate of this transition regime (see below). Figure S15-a shows the TVS curves obtained from the mean $\bar{I}$-*V* of the **Ru-dppe** SAM. The maxima (red dashed lines) indicate transition voltages at $V_{T-}$ = -0.49 V and $V_{T+}$ = 0.33 V. Therefore, we fixed the bias window of confidence between -0.5 V and 0.5 V to fit all *I-V*s of the dataset. Figure S15-b shows a typical fit of the SLM model (on the mean $\bar{I}$-$V$ of the **Ru-dppe** SAM. We note that, with these conditions, the two methods give almost the same value for the energy level $\epsilon_0$ (0.32 eV fit with SLM, 0.34 eV by TVS).

The same protocol was applied for the **Fe-cyclam** SAM (Figure S15-c Figure S15-d), using the same bias window range for a valid comparison between the two samples. The fits of the SLM were done with the routine included in ORIGIN software (version 2023, OriginLab Corporation, Northampton, MA, USA), using the method of least squares and the Levenberg Marquardt iteration algorithm.

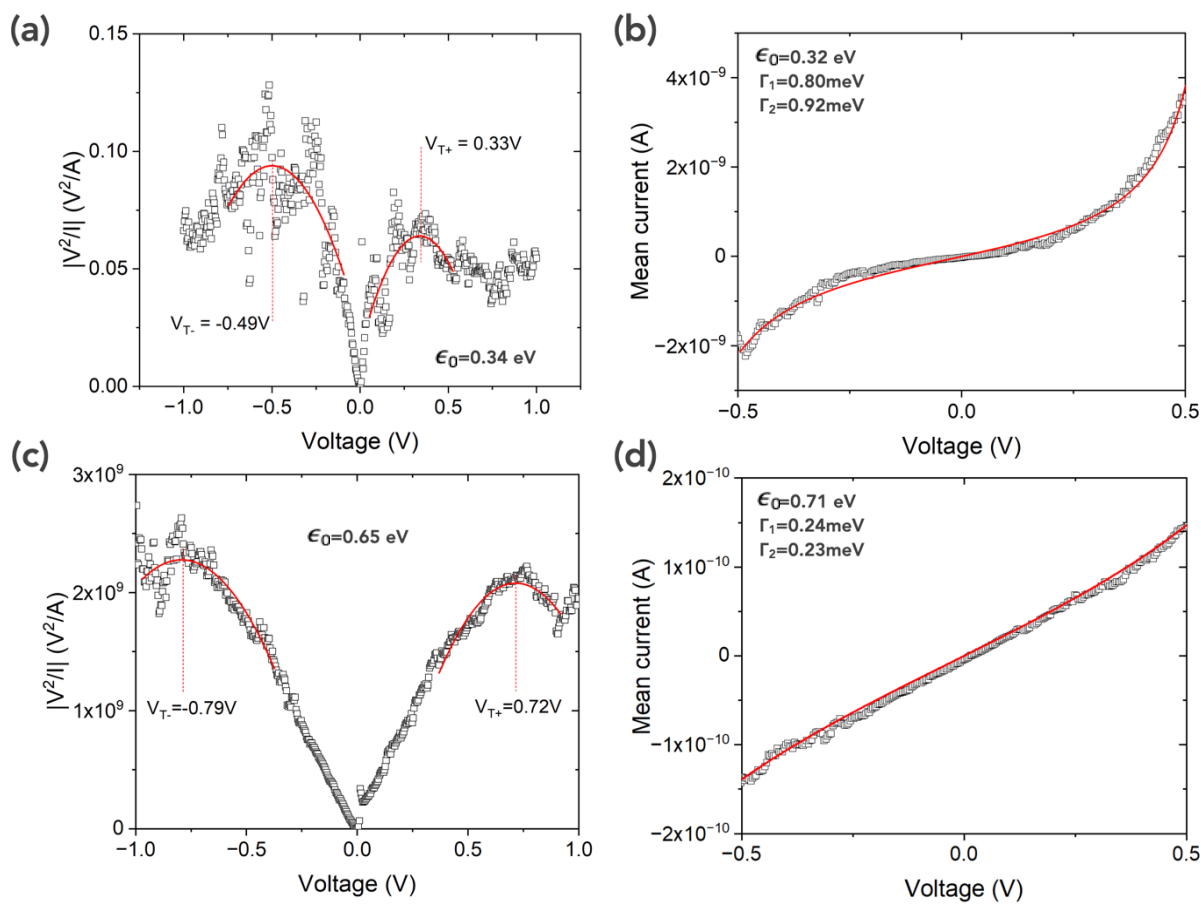

**Figure S15.** (a) Data of the mean $\bar{I}$-$V$ of the **Ru-dppe** SAM plotted as $|V^2/I|$ vs. $V$. The red lines are the fits by a 2$^{nd}$ order polynomial function. (b) Fit of the SLM model on the mean $\bar{I}$-$V$ of the **Ru-dppe** SAM, fit limited between -0.5 and 0.5 V, the fit parameters are given in the panel. (c-d) Similar data for the **Fe-cyclam** SAM. See text.

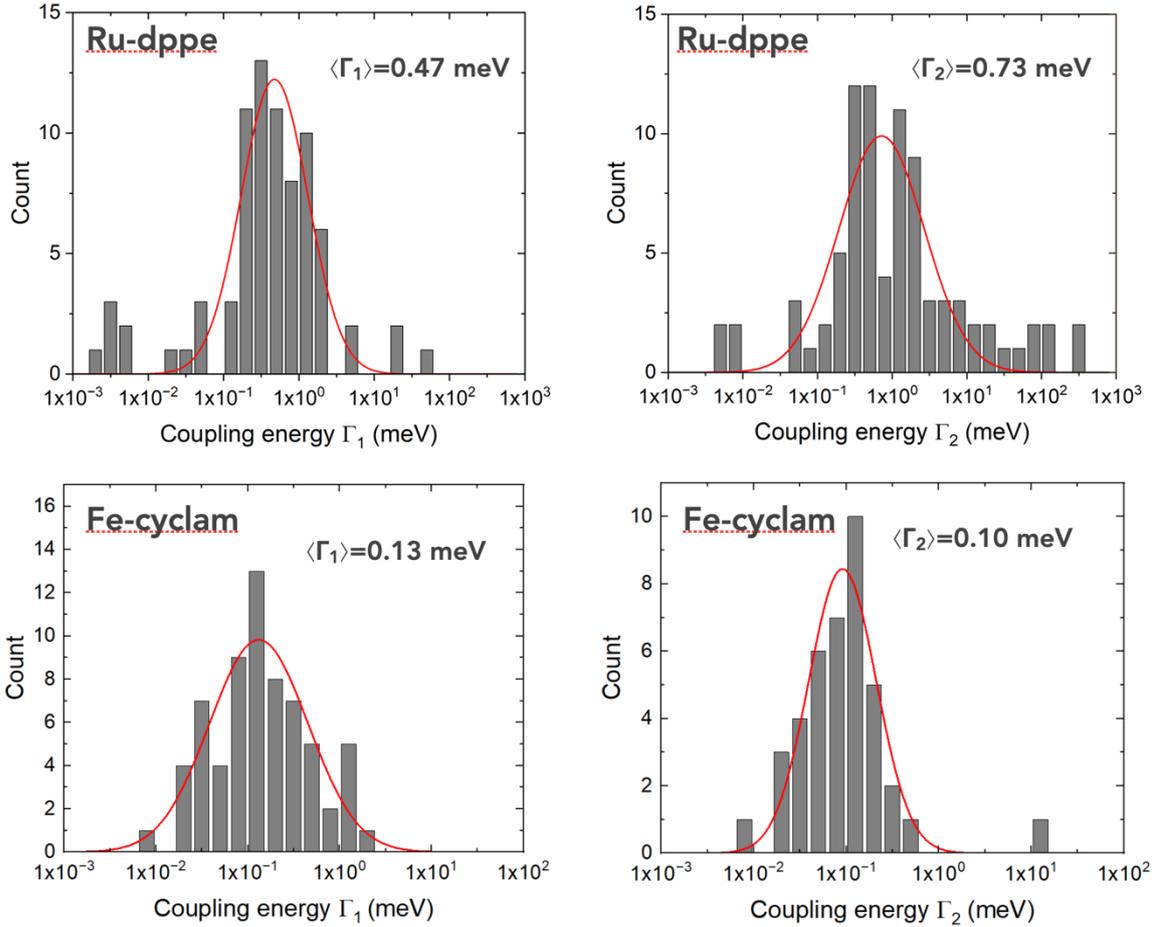

Figure S16. Histograms of the electrode coupling energy, $\Gamma_1$ and $\Gamma_2$, determined from the fit of the SLM (Eq. S4) on the *I-V* data sets of the **Ru-dppe** and **Fe-cyclam** SAMs shown in Figure 5 and Figure 6 (main text), respectively.

In the TVS[41–45] method, the *I-V* datasets are replotted as $|V^2/I|$ vs. $V$ (Figure S15-a and Figure S15-c),[46] and we determine the transition voltages $V_{T+}$ and $V_{T-}$ for both voltage polarities at which the bell-shaped curve is maximum. This threshold voltage indicates the transition between off-resonant (below $V_T$) and resonant (above $V_T$) transport regime in the molecular junctions and can therefore be used to estimate the location of the energy level. In Figure S15, the thresholds $V_{T+}$ and $V_{T-}$ are indicated by the vertical dashed red lines (with values) and determined from the maximum of a 2nd order polynomial function fitted around the maximum of the bell-shaped curves (to cope with noisy curves). The value of $\epsilon_{0-TVS}$ is estimated by:[43]

$$|\varepsilon_{0-TVS}| = 2 \frac{e|V_{T+}V_{T-}|}{\sqrt{V_{T+}^2 + 10|V_{T+}V_{T-}|/3 + V_{T-}^2}} \qquad \text{Eq. S5}$$

## 5 - Null-point scanning thermal microscopy (SThM)

**General procedure.** SThM[47,48] were carried out with a Bruker ICON instrument equipped with the Anasys SThM module and in an air-conditioned laboratory (22.5°C and a relative humidity of 35-40 %). We used Kelvin NanoTechnology (KNT) SThM probes with a Pd thin film resistor in the probe tip as the heating element (VITA-DM-GLA-1). The SThM tip is inserted in a Wheatstone bridge, the heat flux through the tip is controlled by the DC voltage applied on the Wheatstone bridge ($V_{WB}$, typically 0.6-1.1 V). The tip temperature, $T_{tip}$, is obtained by measuring the output voltage of the Wheatstone bridge, knowing the transfer function of the bridge, the gain of the voltage amplifier and the calibrated linear relationship between the tip resistance and the tip temperature.

The null-point SThM[49] was used at selected points on the SAMs. We define a 5 × 5 grid, each point spaced by 10 nm. At each point of the grid, in the z-trace mode (approach and retract), we recorded the tip temperature versus distance curve ($T_{tip}$-z). At the transition from a non-contact (NC, tip very near the surface) to a contact (C, tip on the surface) situation, we observe a temperature jump, $T_{NC}$ - $T_C$, which is used to determine the sample thermal conductivity according to the protocol described by Kim *et al.*[50] (Eq. 1 in the main text, the constriction thermal conductance $G_{th}$ is related to the thermal conductivity $\kappa$ by $G_{th} = 4r_{th}\kappa$). The temperature jump is measured from the approach trace only (to avoid any artifact due to well-known adhesion hysteresis of the retract curve) and averaged over the 25 recorded $T_{tip}$-z traces. This differential method is suitable to remove the parasitic contributions (air conduction...): at the contact (C) both the sample and parasitic thermal contributions govern the tip temperature, whereas, just before physical tip contact (NC), only the parasitic thermal contributions are involved. The plot of the temperature jump, $T_{NC}$ - $T_C$, versus the sample temperature at contact $T_C$ is linear and its slope is inversely proportional to the thermal conductivity (Figure 7 and Figure S23). The tip-sample temperature $T_C$ increases with the supply voltage of the Wheatstone bridge $V_{DC}$ (typically from 0.6 to 1.1 V).

To determine the thermal conductivity from data like in Figure 7 and Figure S23 and using Eq. 1, we calibrated the null-point SThM according to the protocol of Kim e*t al.*[50] The same $T_C$ vs. $T_{NC}$-$T_C$ measurements were done on two materials with well-known thermal conductivity: a glass slide (1.3 W m$^1$ K$^{-1}$) and a low-doped silicon wafer with its native oxide (150 W m$^{-1}$ K$^{-1}$). A new calibration was done for each new sample or new tip to cope with slight changes of the instrument parameters (e.g., wear and tear of the tip, shift of loading force). Figure S17 shows a typical calibration curve. From a linear fit on the data (Figure S18), we get α = 10.12 W m$^{-1}$ K$^{-1}$ and *β* = 10.16 µV in the present case.

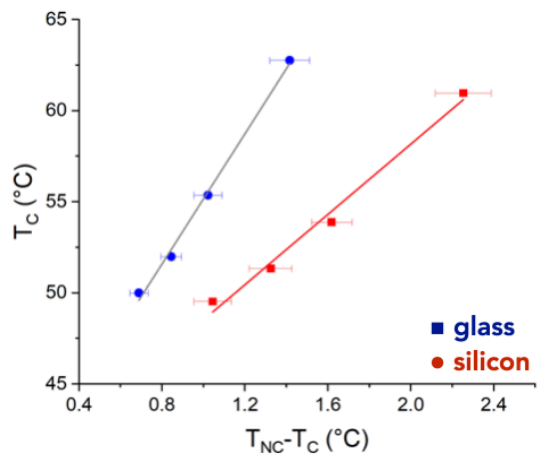

Figure S17. Typical calibration curve.

The thermal contact radius (at the tip/SAM interface) is calculated following the approach reported by Majumdar and collaborators[51] taking into account the mechanical tip radius $r_{tip}$ and the size of the water meniscus at the tip/surface interface. The thermal radius of the thermal contact is estimated by:[52]

$$r_{th} = 2.08\sqrt{\frac{-r_{tip}\cos\theta}{\ln\varphi}} \quad \text{Eq. S6}$$

with $r_{tip}$ = 100 nm (data from Bruker), the relative humidity $\varphi$ = 0.35-0.4 (air-conditioned laboratory, values checked during the measurements) and the contact angle of the concave meniscus between the tip and the surface $\theta \approx 30°$ as measured for π-conjugated molecular crystals.[53] We get $r_{th} \approx 20$ nm. The water meniscus contact angle depends on the surface energy of the sample; however, we cannot perform water contact angle measurements inside the nanometer size tip/SAM interface, and we consider the same literature value of 30° in both cases. Considering the same area per molecule as discussed above for the C-AFM measurements, we estimate by the same method that ca. 400 ($10^3$, respectively) **Ru-dppe** (**Fe-cyclam**, respectively) molecules are contacted during the SThM measurements.

**Correction with the Dryden model.** The SAMs are deposited on a high thermal conducting Au substrate and the measured "effective" conductance $G_{th}$(SAM/Au) contains a contribution from the substrate. The Dryden model[54] allows calculating the constriction thermal conductance $G_{th}$ (or equivalently the thermal conductivity κ, $G_{th} = 4r_{th}\kappa$) when a small thermal spot is contacting a thin layer coating on a substrate. We used this model to calculate the thermal conductance $G_{th}$(SAM) of a very thin film (here the SAM) of thickness $t_{SAM}$ deposited on a semi-infinite substrate (here the thick underlying Au electrode) with a thermal conductance $G_{th}$(Au) = 25.4 µW/K ($\kappa_{Au}$ = 318 W m$^{-1}$ K$^{-1}$ and $r_{th} \approx 20$ nm, see above). In the case $t_{SAM}/r_{th} < 2$ (here $t_{SAM}$ is ca. 1.9 nm, see main text) the model reads:

$$\frac{1}{G_{th}(SAM/Au)} = \frac{1}{G_{th}(Au)} + \frac{4}{\pi G_{th}(SAM)}\left(\frac{t_{SAM}}{r_{th}}\right)\left(1 - \left(\frac{G_{th}(SAM)}{G_{th}(Au)}\right)^2\right) \qquad \text{Eq. S7}$$

Solving this equation for all measured $G_{th}$(SAM/Au) allows determining the SAM thermal conductance $G_{th}$(SAM).

## 6 - Seebeck coefficient measurements

The samples were placed on the heating stage of the ICON C-AFM setup. The substrate temperature was systematically measured by a thermocouple (KM340) and controlled with a precision of ± 1 K. The temperature of the tip was assumed to be at room temperature as also validated in previous experiments with scanning probe microscope.[55,56] We used an intermittent mode to measure the *I-V*, the tip being in contact with the SAM during a short time (~ few seconds) and retracted (see C-AFM section). Then the tip was moved to another location to acquire another *I-V* curve. The tip was connected to a large thermal reservoir (cantilever and support) held at room temperature. We posit that this time is too short to allow the tip temperature to increase significantly during this time lap.[55,56] However, we cannot completely exclude that a partial thermalization of the tip occurs. This means that the reported Seebeck coefficients in this work are underestimated and must be considered as a lower bound. The *I-V*s were measured at low voltage (± 25 mV) with a high gain of the trans-impedance preamplifier ($10^{-11}$ A/V). We used a full-metal Pt tip (12Pt400A from Rocky Mountain Nanotechnology). Hundreds of *I-V*s were acquired as for the usual C-AFM measurements (see above) and the $V_{NC}$ values (voltage at null current) were determined by a python script to construct the $V_{NC}$ histograms (shown in Figures 8-a and 8-c, Figure S18, Figure S19, Figure S20). We first measured the Seebeck coefficient without molecules, i.e., the interface C-AFM tip/Au substrate (Figure S18). A Seebeck coefficient $S_{tip/Au}$ = -$\Delta V_{NC}/\Delta T$ = -5 ± 1 µV/K is obtained. This value is subtracted from the Seebeck measurements of the $^{TS}$Au-molecule/tip MJs to obtain the value of the SAMs.[57,58]

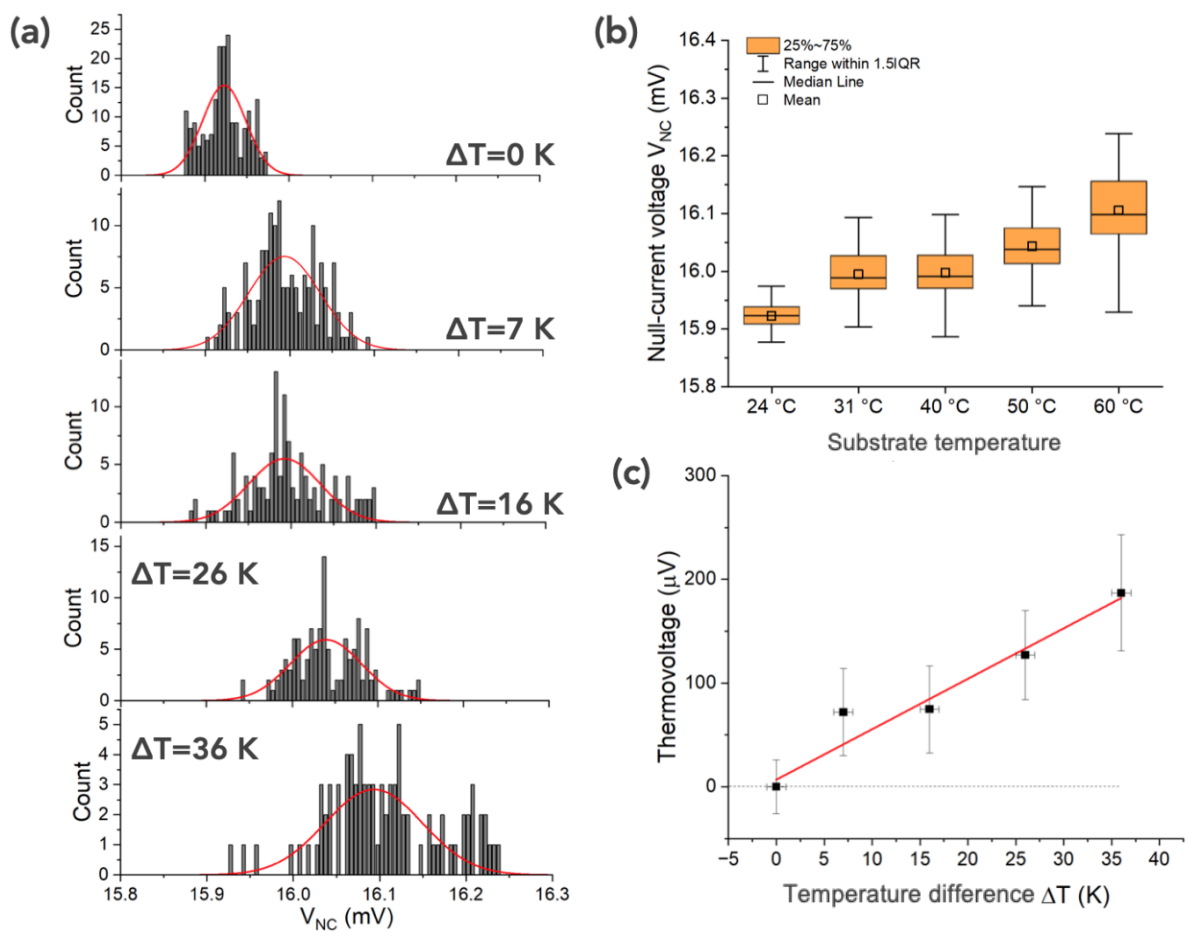

Figure S18. (a) Histograms of the voltage at null current $V_{NC}$ at several temperature differences applied on the SAM/tip without molecules (125 to 200 measurements, depending on temperature). The red lines are fits with a Gaussian distribution. At $\Delta T$ = 0°C, we measured the offset voltage of the equipment. (b) Box plots of $V_{NC}$ measurements showing the mean and median values, the 25-75 % percentiles and the limits at 1.5 IQR (interquartile range). (c) Evolution of the mean thermovoltage (the mean $V_{NC}$ normalized to the value at room temperature; error bar from the standard deviation of the histograms in panel (a)) versus the temperature difference and linear fit, with a slope $\Delta V_{NC}/\Delta T$ = 5 ± 1 µV/K.

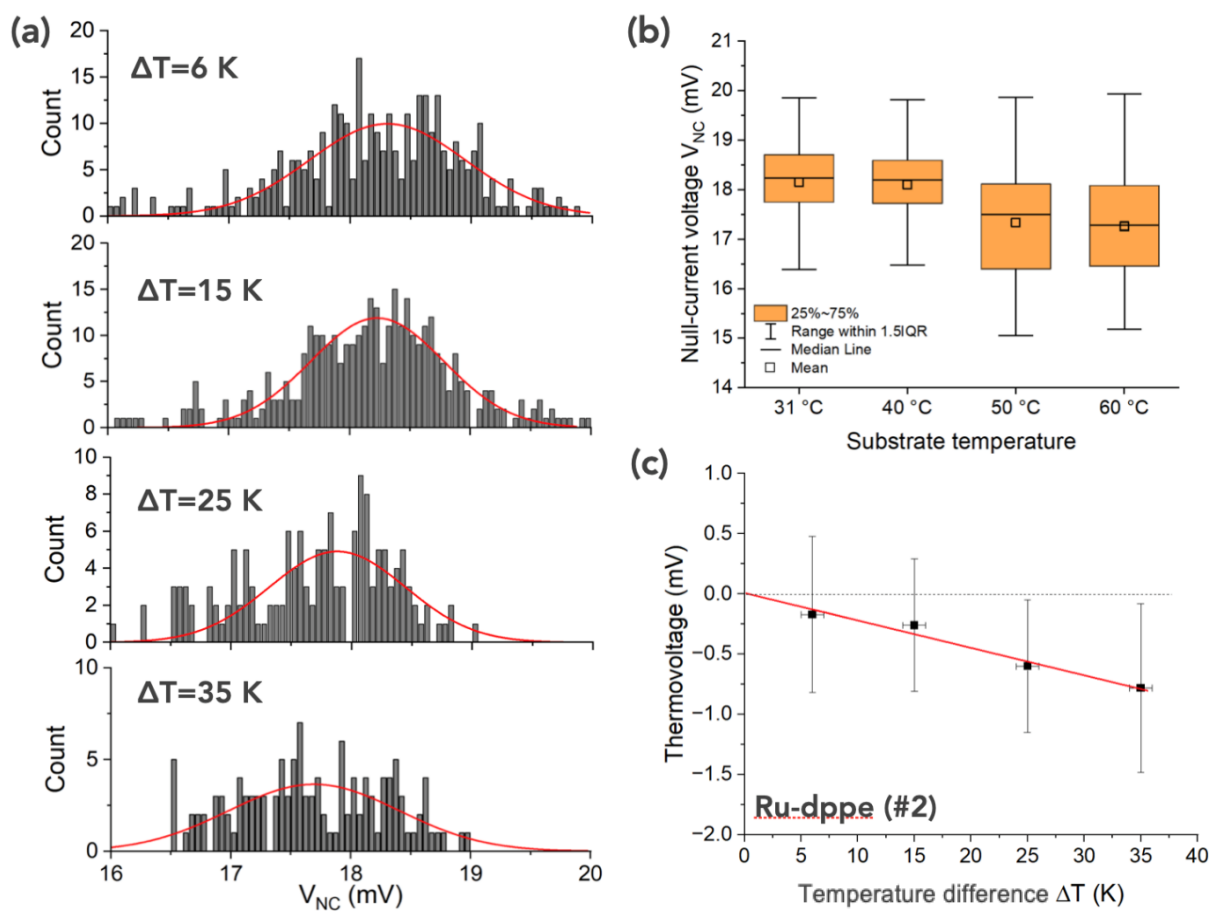

Figure S19. (a) Histograms of the voltage at null current $V_{NC}$ at several temperature differences applied on a second **Ru-dppe** SAM (175 to 350 measurements, depending on temperature). The red lines are the fits with a Gaussian distribution. (b) Box plots of $V_{NC}$ measurements showing the mean and median values, the 25-75 % percentiles and the limits at 1.5 IQR (interquartile range). (c) Evolution of the mean thermovoltage (the mean $V_{NC}$ normalized to the value at room temperature; error bar from the standard deviation of the histograms in panel (a)) versus the temperature difference and linear fit, with a slope $\Delta V_{NC}/\Delta T$ = -23 ± 4 µV/K. The dispersion of the $V_{NC}$ data set is larger for this sample. However, the same trend is observed as for the sample shown in Figure 8 (main text).

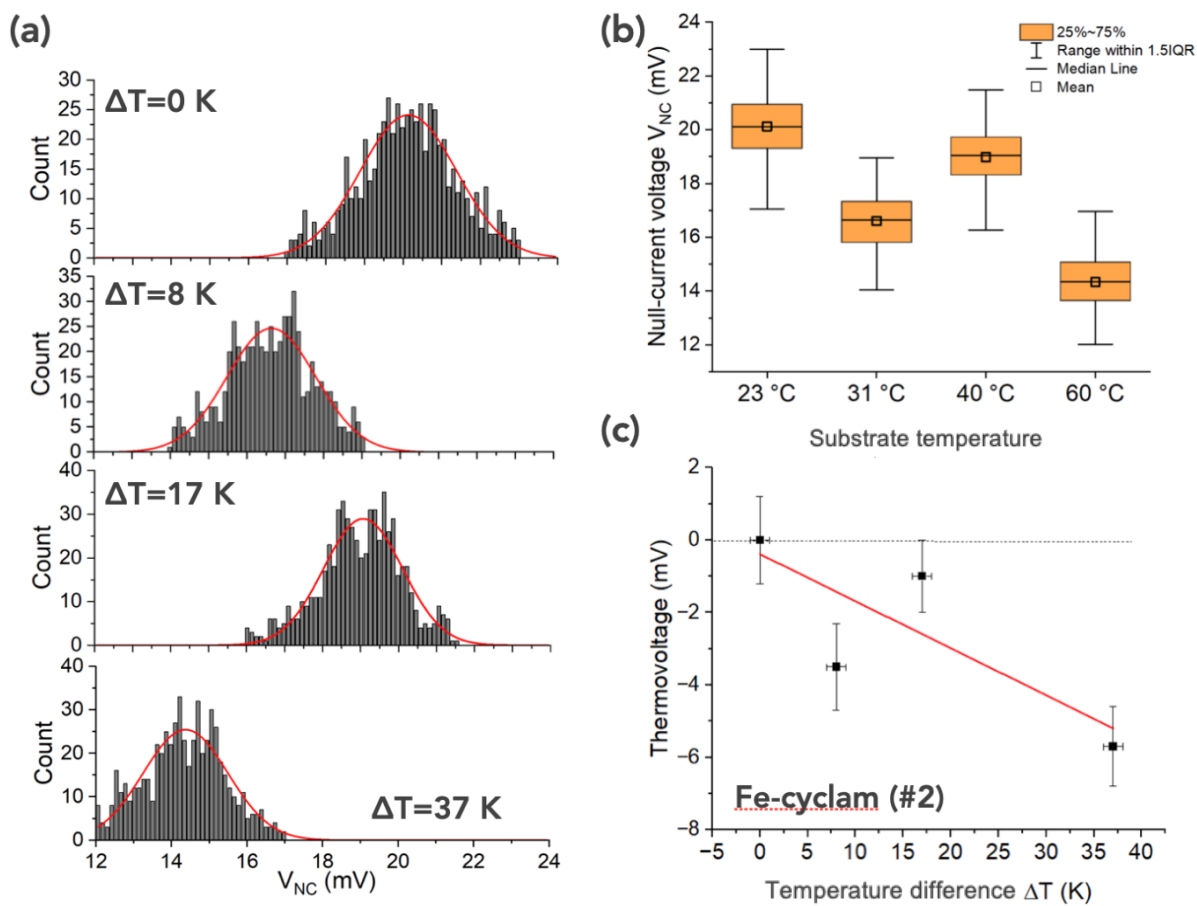

Figure S20. (a) Histograms of the voltage at null current $V_{NC}$ at several temperature differences applied on a second **Fe-cyclam** SAM (580-640 measurements, depending on temperature). The red lines are the fits with a Gaussian distribution. (b) Box plots of $V_{NC}$ measurements showing the mean and median values, the 25-75 % percentiles and the limits at 1.5 IQR (interquartile range). (c) Evolution of the mean thermovoltage (the mean $V_{NC}$ normalized to the value at room temperature; error bar from the standard deviation of the histograms in panel (a)) versus the temperature difference and linear fit, with a slope $\Delta V_{NC}/\Delta T$ = -123 ± 71 µV/K.

The method used for the statistical analysis of these data can influence the value of the Seebeck coefficient.[59,60] We have compared three approaches. We used the mean and the median values from box plots calculated for all $V_{NC}$ distributions and the peak position of the Gaussian fits (this latter neglecting outliers and tails). In our case, Figure S21 shows (**Fe-cyclam**, data from Figure 8-d) that the three analyses give similar values (minor variations significantly smaller than the error bars). The slopes $\Delta V_{NC}/\Delta T$ are -147, -143 and -140 µV/K with the mean, median and Gaussian peak data, respectively. The same conclusion applies to the other data sets.

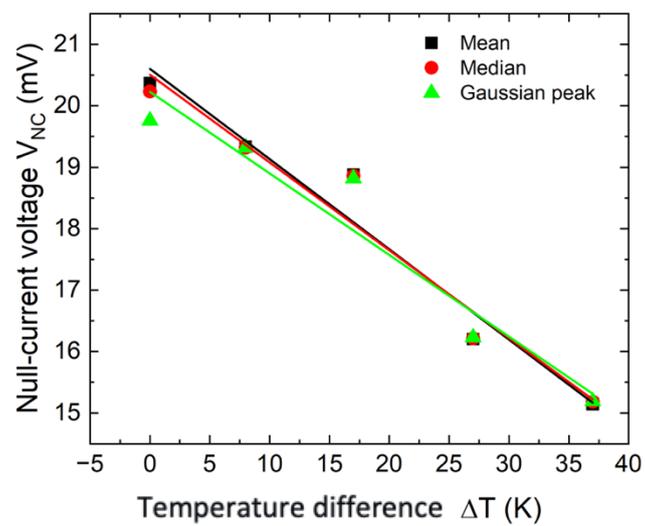

Figure S21. Evolution of the mean, median and Gaussian peak $V_{NC}$ values versus temperature difference and the corresponding linear fits.

## 7 – Additional data

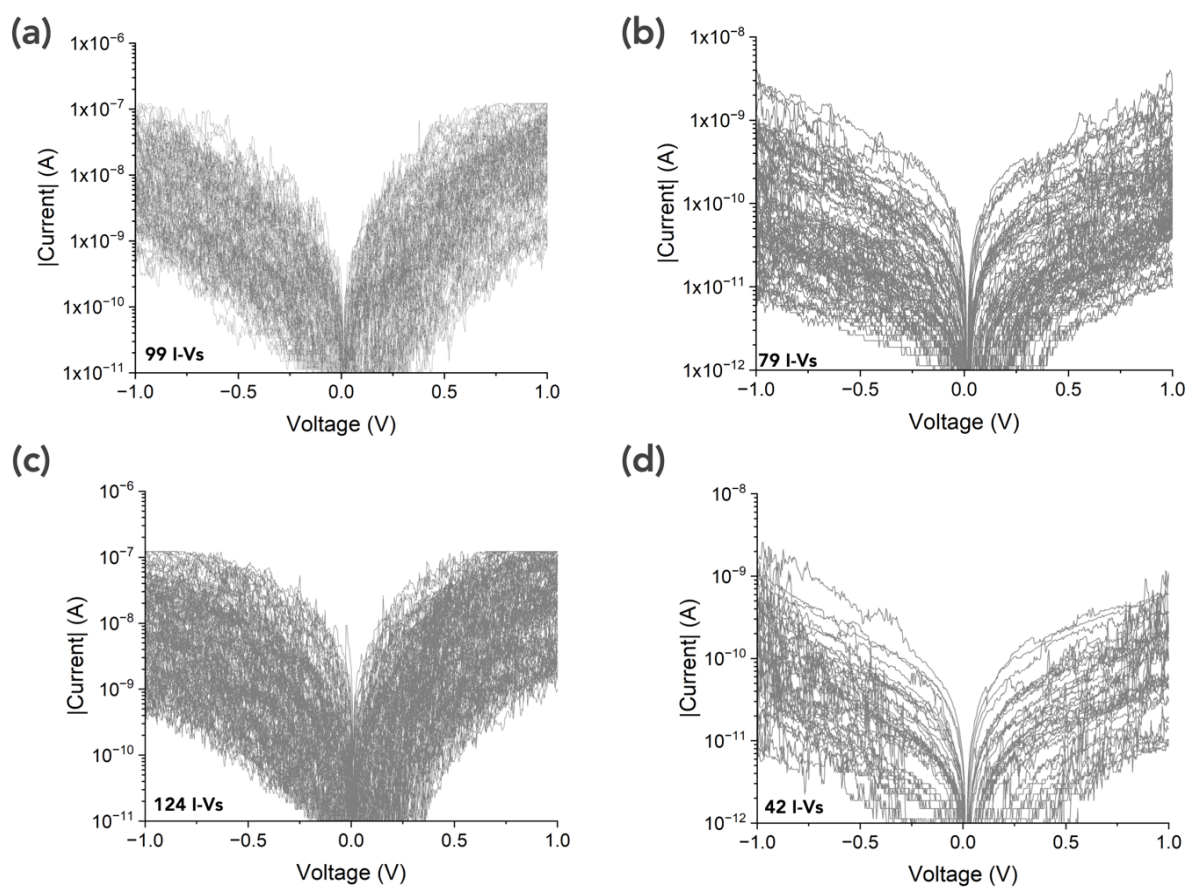

Figure S22. Current-voltage (*I-V*) dataset for other samples from two additional batches: (a) and (c) **Ru-dppe**, (b) and (d) **Fe-cyclam**.

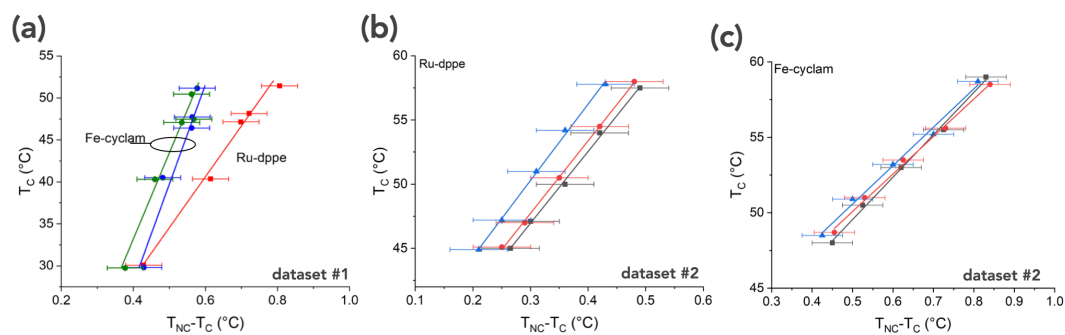

Figure S23. Tip temperature at contact, $T_C$, versus the temperature jump, $T_{NC}$-$T_C$ measured for other **Ru-dppe** and **Fe-cyclam** samples: (a) batch #1, (b-c) batch #2.

## 8 – Computational details

**General procedure.** The QuantumATK-2021.06 software[61] was used to optimize the geometries of some molecular junctions and calculate their electronic transmission spectra at the DFT level (GGA-RPBE functional)[62,63] using the Non-Equilibrium Green's Function (NEGF) framework.[64] The SZP basis set provided in QuantumATK-2021.06 was used for the Au valence electrons and the DZP for the rest of the atoms. The pseudo-potentials of Fritz-Haber Institute (FHI) of Troullier-Martins type were used to treat the core electron shell. The Fast Fourier 2D solver was used in the device configuration with periodic conditions along A and B, and Dirichlet boundary conditions along C. Ghost atoms were added to calculate the electronic configurations and transmission functions. An SCF tolerance convergence of $10^{-5}$ Ha was applied. Spin-polarized calculations were performed for **Fe-cyclam$^{+1}$** junctions. The flat electrodes were built from a cleavage of bulk *fcc* Au along the (111) direction to make a slab of nine Au layers: the sixth firsts are included in the scattering region (Figure S24).[64]

The molecular structural arrangements which are introduced in the junctions are issued from a molecular geometry optimization of thiol-terminated molecules. The S atoms are anchored in a hollow configuration with a distance of 1.76 Å from the Au surface. A full geometry relaxation is not possible for charged molecular junction (see below). The influence of the functional, full geometrical relaxation and electrode shape were evaluated for the systems for which it was possible to check them. These results are presented in the "Evaluation of the computational model on the computed main values" section.

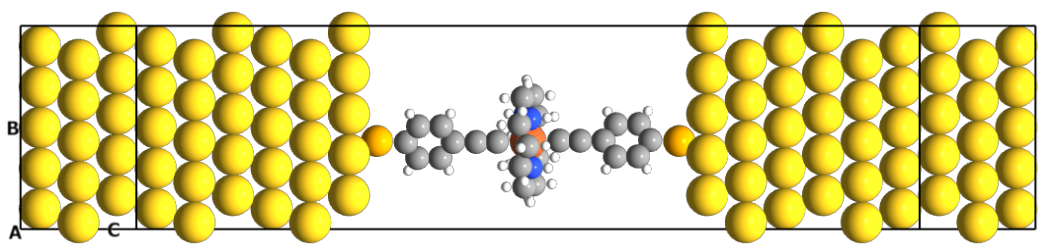

Figure S24. Lateral view of the computational unit cell of the **Fe-cyclam$^0$** molecular junction.

For **Fe-cyclam** junctions, a cell of 5 × 5 Au atoms in the **A** and **B** direction was chosen to minimize interaction between molecules of the neighboring cells. A larger cell was needed for the **Ru-dppe** junction: 7 × 7 Au atoms. A 5 × 5 × 100 grid for k-point sampling and a mesh cutoff for the Poisson equation of 125 Ha were applied in the SCF energy procedure. The electronic transmission function was calculated with a 6 × 6 × 1 k-sampling in the Monkhorst Pack scheme. To maintain a neutral system for the charged molecule **Fe-cyclam$^+$**, we included a neutralizing Br$^-$ ion to play the role of a counterion in the junction. A similar approach provided satisfactory comparison with experiment for charged junctions in a previous study.[65] The Br atom was placed in the position in which the interaction with the rest of the system was the lowest (see "k-sampling and transmission instability section). The negative charge on the Br was induced by the application of a potential shift of -3 eV on the atomic orbitals,[61,65] to which an additional down-energy shift of the Br 4p orbitals (-10 eV) was used to prevent orbital interactions. This allowed us to access to an electronic structure in the HOMO-LUMO energy region for [Fe(cyclam)(C≡C-Ph-SH)$_2$]···Br almost identical to the [Fe(cyclam)(C≡C-Ph-SH)$_2$]$^+$ calculated in the isolated charged molecule (molecule configuration implementation of QuantumATK, see "Simulating charged **[Fe-cyclam]**$^+$ junction" section). The [Fe(cyclam)(C≡C-Ph-SH)$_2$]···Br geometrical arrangement was used to build the corresponding molecular junction. Unfortunately, the gradients are not implemented in QuantumATK for the counterion-included device configuration, thus preventing any geometry optimization. The HSE06 functional could not be used for this configuration either. The effect of these limitations is discussed on the basis of calculations for other systems (see "Evaluation of the computational model on the computed main values" section). Unfortunately, we are limited to 1024 Gb of RAM which is not sufficient to calculate the dynamical matrix necessary to evaluate the phonon thermal conductance. Recent papers treating computationally thermal conductance by phonons of organometallic junctions were published showing similarities with the systems that we are investigating.[66,67] These studies converge in their conclusions and reveal that the Debye frequency of the electrodes and the mismatch between molecular vibrations and electrode vibrational modes are limiting the phononic thermal conductance of this heavy-atom containing systems to values ranging from 15 to 20 pW/K at 300 K. Accordingly, the value of 20 pW/K is used in our calculations for the phononic thermal conductance ($\kappa_{ph}$).

The next sections are detailing the computational procedures implemented and the different evaluation performed to evaluate the accuracy and limitation of the computational approach.

**Construction of device geometries.** To build the device structures, we used the QuantumATK interface which provides a fine control of distances and orientations. Since open boundary conditions are imposed along **C** direction (Figure S24) for electron transport calculations, the left and right electrodes are composed by three Au layers and treated like a bulk. Six layers of Au are included in the scattering region to guarantee the continuity of the electrostatic potential at the bulk/scattering electrode interface (see Figure S25). The three external layers (surface layer + the next two) are optimized upon geometry relaxation. Only translations are permitted for the rest of the Au layers as implemented in QuantumATK ("Rigid Body Constrain"). The molecular parts are introduced between the two leads, with their geometry previously optimized for isolated thiol-terminated compounds. The S atoms are anchored in a hollow configuration with a distance of 1.76 Å from the Au surface. The influence of device optimization was evaluated by optimizing two structures (not the charged molecule), and different shapes of the electrodes were tested (flat, tip). We also tested a **Ru-dppe** junction configuration for which a tilt angle is formed by the **Ru-dppe** molecules with respect to the Au surface normal. This is detailed in the next sections. For sake of consistency in the results, non-optimized flat electrodes configurations are presented in the main text.

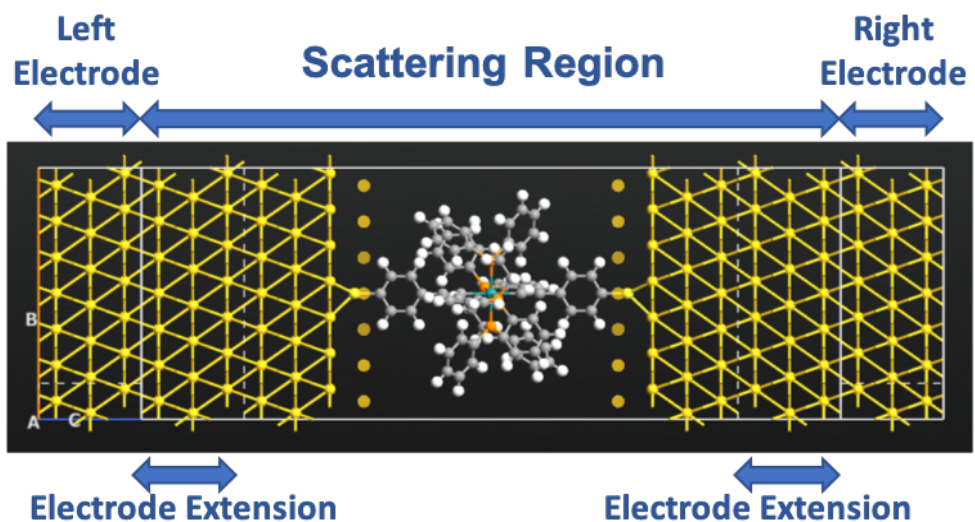

Figure S25. Device geometry of **Ru-dppe** junction with flat Au electrode configuration.

**Evaluation of the computational model on the computed main values**

*Electrode shape*. Additionally to the flat electrode configuration detailed above, a tip-configuration electrode is made of a 9-atom Au cluster on the flat Au(111) surface. The scattering region is composed of the molecule connected to a 9-atom Au cluster and 4-Au(111) layers at both ends. The molecular junctions are built similarly to the molecular junctions in flat configuration. The electronic transmission and the subsequent conductance and Seebeck coefficient are given in Figure S26 and Figure S27, respectively, for tip configuration of **[Fe-cyclam]$^0$** and **Ru-dppe** junctions.

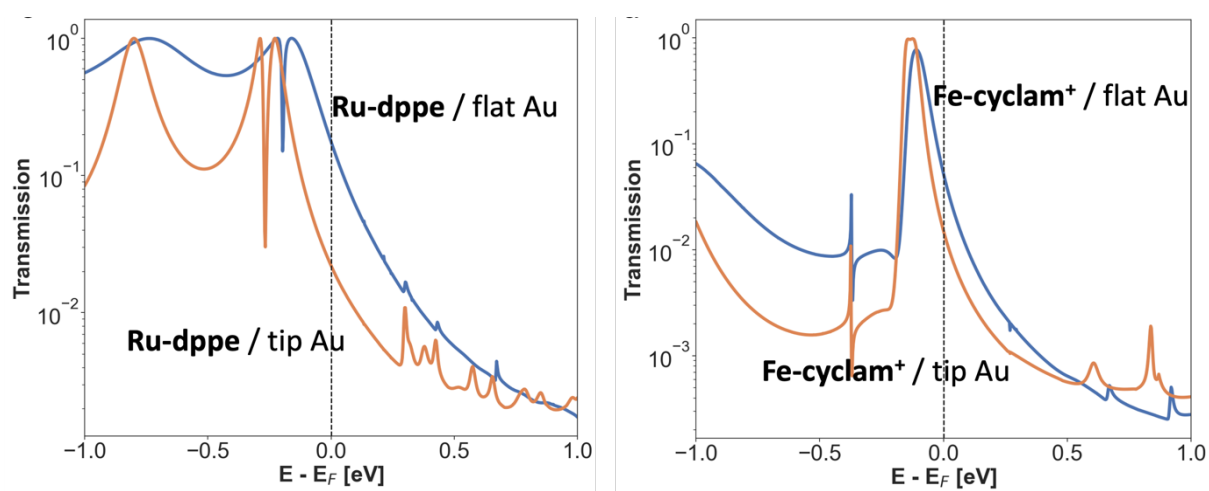

Figure S26. Calculated transmission spectra at zero bias of flat (blue) and tip (orange) configuration-type (left) **Ru-dppe** and (right) **[Fe-cyclam]$^0$** junctions.

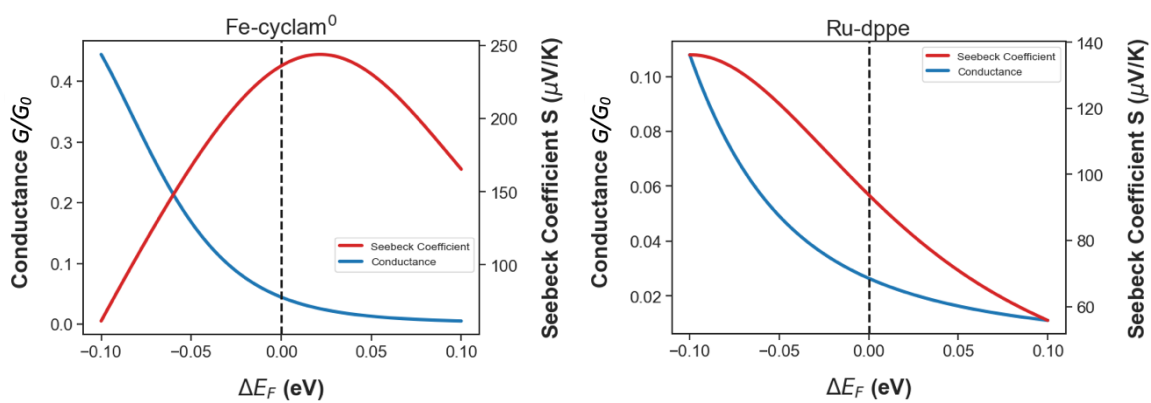

Figure S27. Conductance and Seebeck coefficient at 300 K of tip configuration-type **[Fe-cyclam]$^0$** (left) and **Ru-dppe** junctions (right) as function of energy shift of the transmission spectrum shown in Figure S26 by ΔE$_F$, negatively and positively with respect to the Fermi level.

**[Fe-cyclam]$^0$** and **Ru-dppe** junctions in the non-optimized tip electrode type lead to Seebeck coefficients between 61 $\mu$V/K and 244 $\mu$V/K and between 56 $\mu$V/K and 136 $\mu$V/K, respectively, which can be compared with the results obtained for the non-optimized flat junctions for which Seebeck coefficients are ranging between 2 $\mu$V/K and 178 $\mu$V/K and between 51 $\mu$V/K and 91 $\mu$V/K respectively. This is a direct consequence of the decrease of conductance by about one order of magnitude (Figure S28, Figure S29) linked to the reduction of electronic interaction between the Au levels and the molecular orbitals in the tip configuration.

*Geometry optimization of the molecular junctions*. The geometries of the **[Fe-cyclam]$^+$** junctions could not be optimized because of the absence of implementation of gradient calculation in QuantumATK when using the option "potential shift of orbital". In order to evaluate the effect of geometrical relaxation of the molecular junction on the results, **Ru-dppe** and **[Fe-cyclam]$^0$** junctions in the tip configuration were optimized using the Limited-memory Broyden-Fletcher-Goldfarb-Shanno bound method and a convergence criterion of maximum forces of 0.02 eV/Å. The geometric configurations of the **[Fe-cyclam]$^0$** and **Ru-dppe** molecular junctions were optimized starting from pre-optimized thiol-terminated molecules that are connected to the Au tips in a hollow conformation, as described previously. The electronic transmission and the resulting conductance and Seebeck coefficients are given in Figure S28 and Figure S29 respectively for **Ru-dppe** and **[Fe-cyclam]$^0$**. The instability of the transmission function of **Ru-dppe** found 0.25 eV above the Fermi level is discussed in the section "Work function shift".

**[Fe-cyclam]$^0$** and **Ru-dppe** junctions in this optimized tip electrode type lead to Seebeck coefficients between 38 $\mu$V/K and 228 $\mu$V/K and between 81 $\mu$V/K and 144 $\mu$V/K respectively which can be compared with the results obtained for the non-optimized junctions for which Seebeck coefficients are ranging from 61 $\mu$V/K and 244 $\mu$V/K and between 56 $\mu$V/K and 136 $\mu$V/K for **[Fe-cyclam]$^0$** and **Ru-dppe** junctions respectively. The close similarity between the two sets of data indicates that the optimization of the junction is not mandatory here.

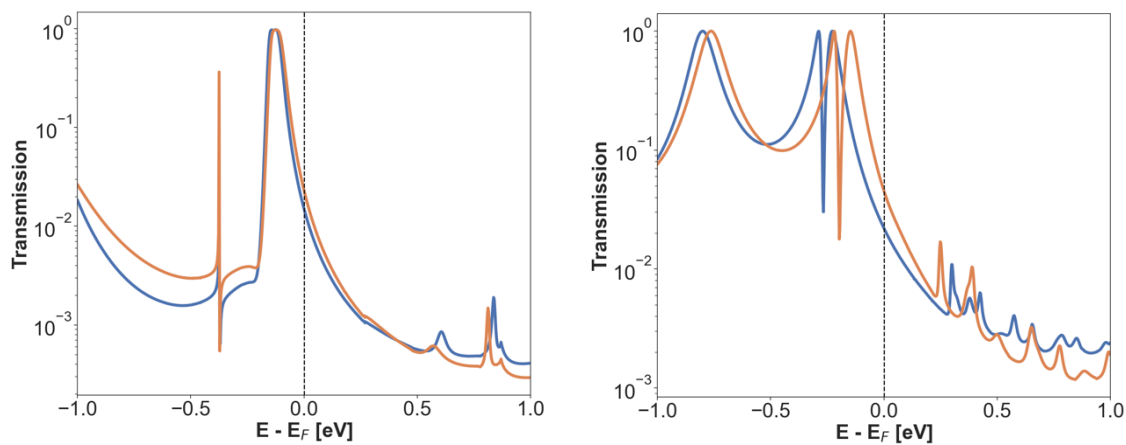

Figure S28. Calculated transmission spectra at zero bias of non-optimized (blue) and optimized (orange) tip configuration-type **[Fe-cyclam]$^0$** (left) and **Ru-dppe** junctions (right).

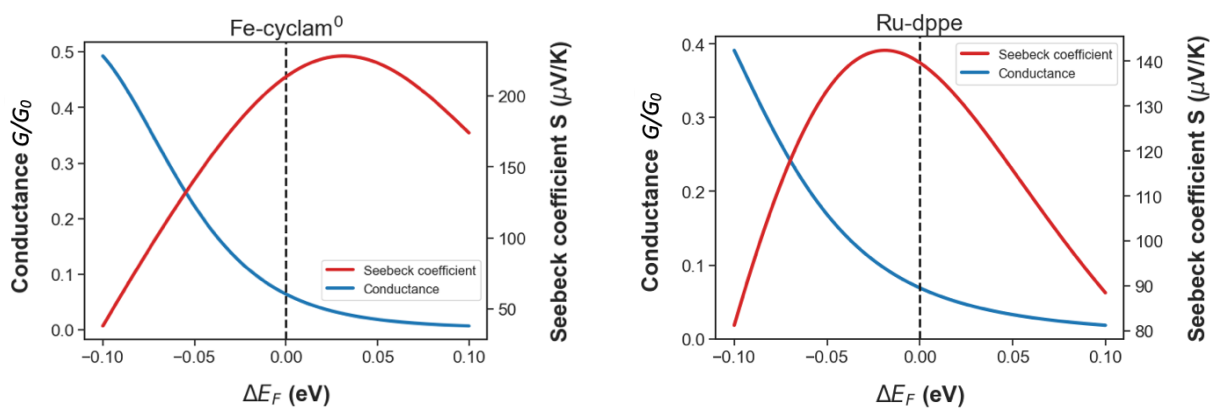

Figure S29. Conductance and Seebeck coefficient at 300 K of optimized tip configuration-type **[Fe-cyclam]$^0$** (left) and **Ru-dppe** junctions (right) as a function of energy shift of the transmission spectrum shown in Figure S28 by ΔE$_F$, negatively and positively with respect to the Fermi level.

A bias voltage of 100 mV was applied to the simulation of tip configuration-type **[Fe-cyclam]$^0$** and **Ru-dppe** to assess the impact on the junction conducting properties. The transmission function profiles are compared to the zero-volt bias results in Figure S30 around the Fermi level. Since the polarization of the molecular levels is negligible at low bias, the transmission is slightly modified, leading to highly similar values of thermoelectric parameters.

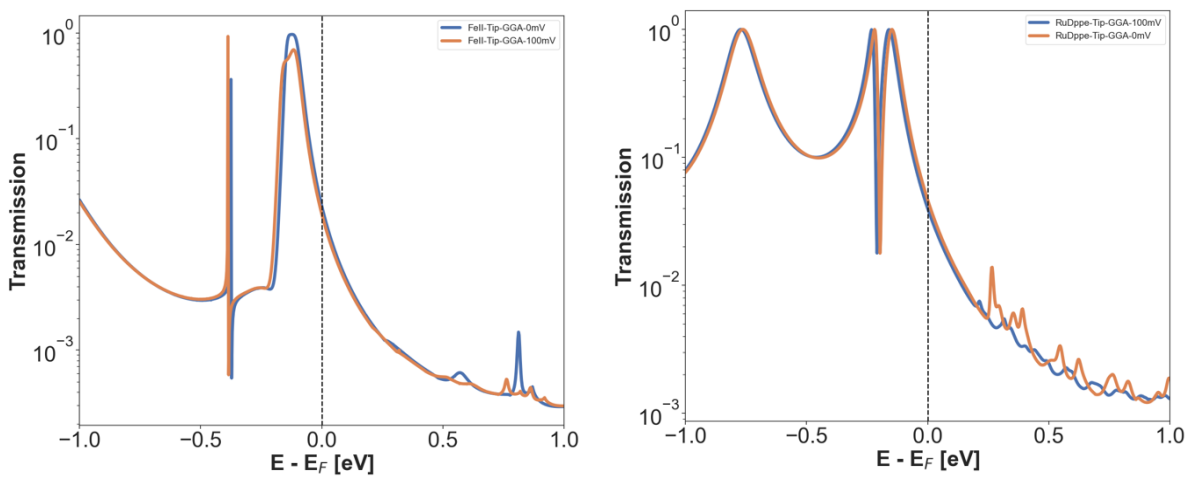

Figure S30. Calculated transmission spectra at zero and 100 mV bias of tip configuration-type **[Fe-cyclam]$^0$** (left) and **Ru-dppe** junctions (right).

*HSE06 functional*. The use of the option "potential shift of orbital" which is compulsory to simulate **[Fe-cyclam]$^+$** junctions is not applicable with the use of the HSE06 functional which give more reliable results than RPBE-GGA functionals.[68] We have checked the consequences of changing the functional to HSE06[63] by computing the transmission function of the optimized tip configuration-type **Ru-dppe** junction used to calculate the transmission in Figure S28. The corresponding transmission function is represented in Figure S31 for both a 6 × 6 × 1 k-sampling and at the Γ point (see section "k-sampling and transmission instability"). As expected, the conductance at the Fermi energy is lowered from 6.85 × 10$^{-2}$ $G_0$ using the RPBE-GGA functional to 2.44 × 10$^{-2}$ $G_0$ using HSE06 due to the reduction of the HOMO-LUMO gap at the hybrid level (Figure S32). The quantum interference peak is less shifted than the conducting peak, leading to a more pronounced Fano peak. However, this does not affect the conducting and thermoelectric properties for which only the transmission function close to the Fermi level is of importance. The Seebeck coefficients are ranging from 66 $\mu$V/K and 150 $\mu$V/K when considering energy shift of the Fermi level ± 0.1 eV which is a bit little more than the values obtained with RPBE-GGA (81 to 144 $\mu$V/K).

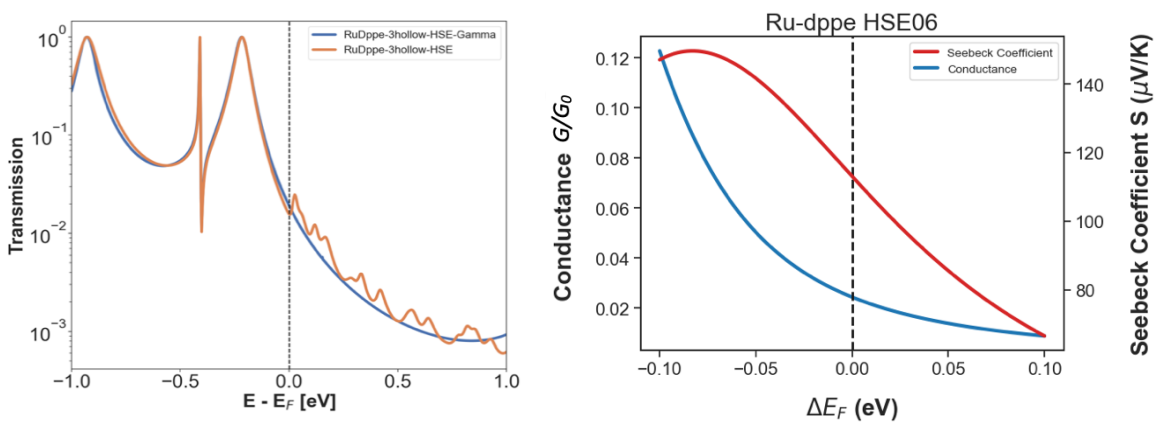

Figure S31. Calculated transmission spectra at zero bias voltage for a 9 × 9 × 1 k-sampling and at the gamma point (right) and corresponding conductance and the Seebeck coefficient at 300 K as a function of energy shift of the transmission spectrum by $\Delta E_F$, negatively and positively with respect to the Fermi level (left) for the flat configuration-type **Ru-dppe** junction.

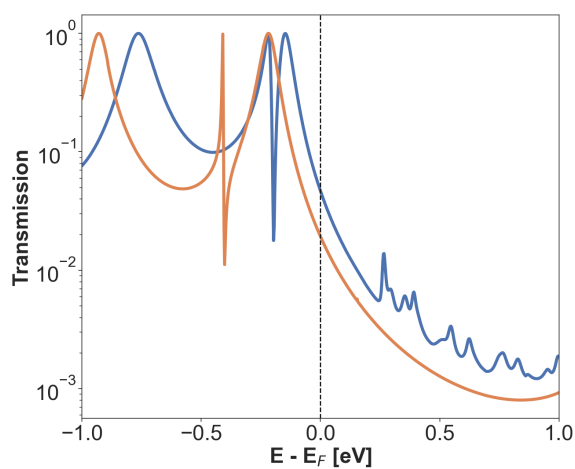

Figure S32. Calculated transmission spectra at zero bias voltage using GGA-RPBE (blue curve) for 9 × 9 × 1 k-sampling and HSE06 at gamma point (orange curve) of the optimized tip configuration-type **Ru-dppe** junction.

The transmission spectra calculated using HSE06 for the tip configuration-type **[Fe-cyclam]$^0$** junction involves the same transmission paths as when using GGA-RPBE at slightly different energies, leading to the small splitting in two of the main transmission peaks (Figure S22). The conductance at the Fermi energy (calculated at 300 K) is slightly increased from $6.44 \times 10^{-2}$ $G_0$ to $9.91 \times 10^{-2}$ $G_0$ by changing the functional from GGA-RPBE to HSE06 due to the non-Lorentzian shape of the transmission. The Seebeck coefficient is globally strengthened ranging from 26 $\mu$V/K to 458 $\mu$V/K when shifting the Fermi level energy.

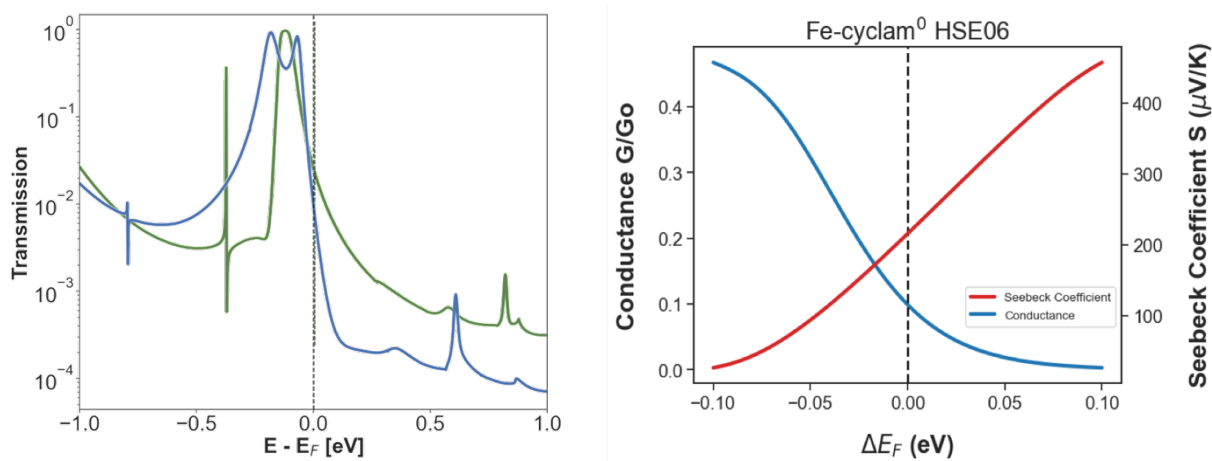

Figure S33. Right: calculated transmission spectra at zero bias voltage using GGA-RPBE in green and HSE06 in blue of the optimized tip configuration-type **[Fe-cyclam]$^0$** junction. The HSE06 conductance

and the Seebeck coefficient at 300 K as a function of energy shift of the transmission spectrum by $\Delta E_F$, negatively and positively with respect to the Fermi level (left)

*From chemical to physical contact.* The molecule/electrode contacts can vary from covalent chemical bond to physical contacts with a wide range of distances between the molecule and the electrodes.[69,70] A wide diversity of conducting properties is thus measured, with the conformational diversity participating also to the spreading of the measurements. Machine learning treatment based on a sufficiently large number of calculations of different configurations could be achieved to reproduce the histogram of conductance.[71] This out of the scope of the present paper to reproduce the experimental features, but it appears interesting to calculate the evolution of the conducting properties when the chemical bond with the tip is not formed. This was done by increasing the distance by +1.5 Å and +2 Å between the S and the Au tip in the tip configuration-type **[Fe-cyclam]$^0$**. Since the S protecting group is removed during the SAM formation, we took into account the radical character of the S while breaking the bond with the electrode. The conductance at 300 K does not evolves linearly, it first decreases from $6.44 \times 10^{-2}$ $G_0$ to $8.52 \times 10^{-3}$ $G_0$ for an elongation of +1.5 Å and then slightly increases up to $1.84 \times 10^{-2}$ $G_0$ when withdrawing the tip by 0.5 Å more (Figure S34). These calculations are in line with the computational deviation with experiment considering that part of the molecules is non-grafted or partly non-grafted in CP-AFM experiments and that the geometry of the molecular junction in CP-AFM and MCBJ is surely differing from the ideal case that we simulate.

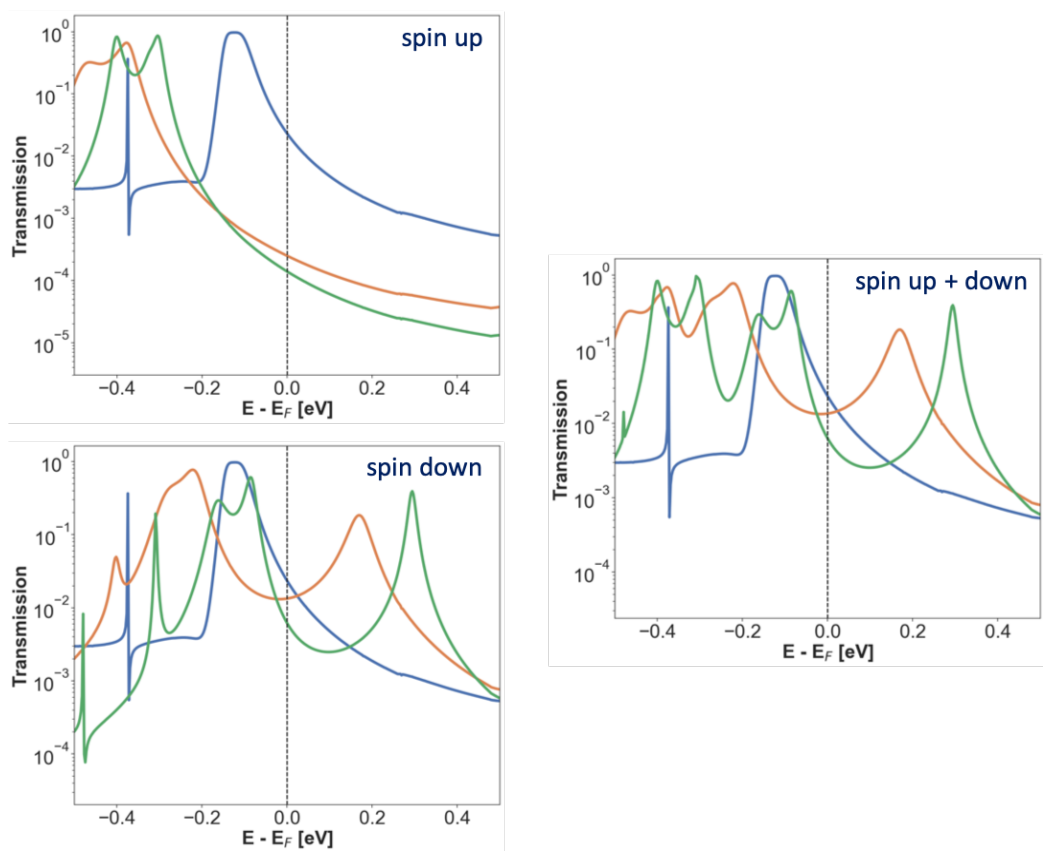

Figure S34. Calculated transmission spectra (spin up, spin down, and sum of spin up and spin down transmissions) for fully connected S-Au bond (blue) and when increasing the distance by + 1.5 Å (orange) and + 2 Å (green) between S and the Au tip in the tip configuration-type **[Fe-cyclam]**$^0$.

*Tilt angle - **Ru-dppe** molecular junctions.* The consequence of the tilting of **Ru-dppe** molecules was evaluated by optimizing a **Ru-dppe** junction in a tilted configuration with flat electrodes. The optimized structural arrangement is represented in Figure S35-a. The mean angle formed by the molecule with respect to the normal of the Au surfaces is 47 ° (left electrode surface: 43°; right electrode surface: 52°). The main change is the S-Au bonding which became on top configuration on the right electrode after optimization. Five H atoms of the phenyl groups on both sides are in interaction with the surfaces (Au-H distances ranging from 2.21 to 3.40 Å). Interestingly, the total energy is hardly affected by the bending of the molecule (+0.010 eV on total energy). The H bonding interactions are probably compensating the less stable on-top anchoring on the right electrode (compare to hollow). The transmission function and associated values for this junction configuration are given in Figure S35 and compared to the linear configuration values. The tilting of the **Ru-dppe** molecule results in an increase of conductance from 0.212 to 0.680 $G_0$ and a decrease of the Seebeck coefficient.

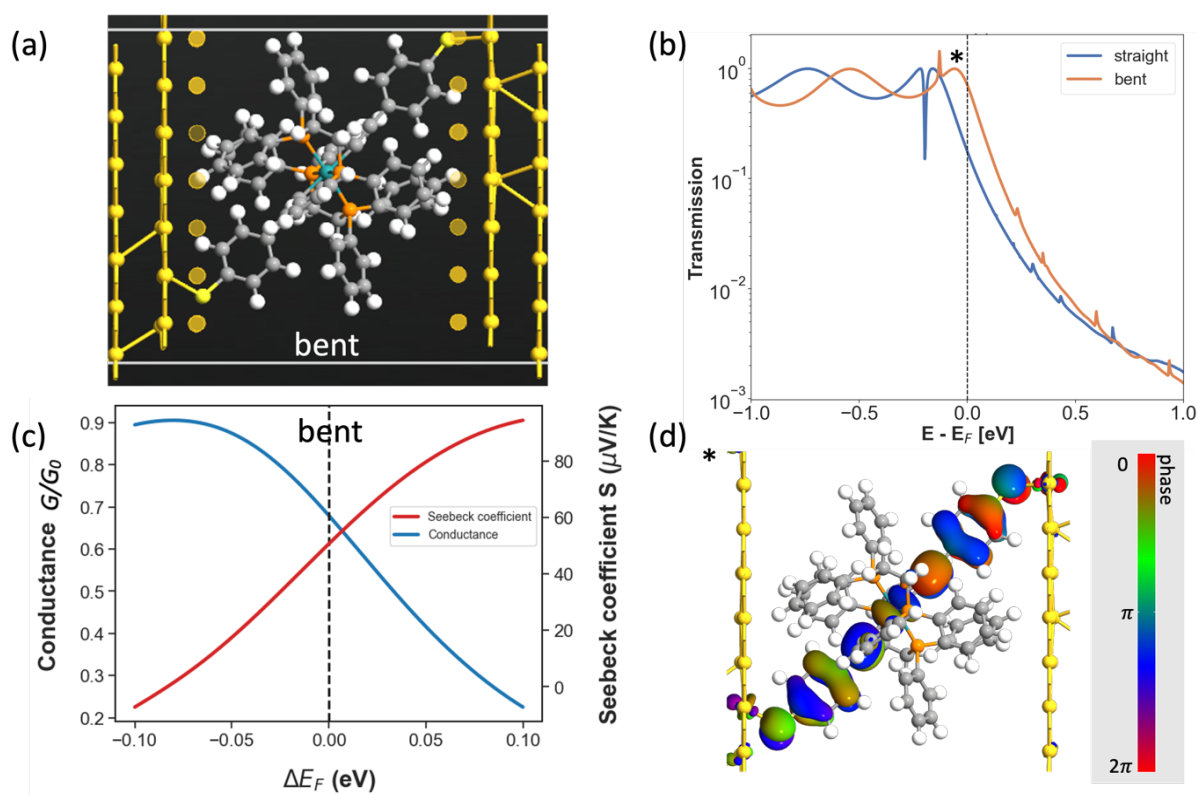

| Ru-dppe junction configuration | $\epsilon_0^{DFT}$ (eV) | $G/G_0$ | min(G), max(G) ($G/G_0$) -0.10-010 eV Fermi shift window | S (µV/K) | min(S), max(S) (µV/K) -0.10-010 eV Fermi shift window |
|---|---|---|---|---|---|
| straight | 0.16 | 0.212 | 6.96 × 10$^{-2}$, 6.30 × 10$^{-1}$ | 90 | 51, 91 |
| bended | 0.05 | 0.680 | 2.26 × 10$^{-2}$, 9.05 × 10$^{-1}$ | 51 | -7, 95 |

Figure S35. (a) Central part of the optimized device geometry of **Ru-dppe** junction in a flat-Au-electrode and 47°-tilted-molecule configuration (bent). (b) Calculated transmission of this bent junction compared to the linear-configuration junction (straight). (c) Conductance and Seebeck coefficient at 300 K as a function of the energy shift of the transmission spectrum with respect to the Fermi level by negative or positive $\Delta E_F$. (d) Phase-colored transmission eigenstate plot issued from the electronic transmission calculations of the bent **Ru-dppe** junction at the energy indicated by a star in the transmission figure. The iso-contour value is 0.01 $(e/Å^3)^{½}$.

**Simulating charged [Fe-cyclam]$^+$ junction.** Studies on charged junctions are sparse, most probably because of the lack of implementation in usual quantum-chemical programs to treat the transmission properties of charged systems out of equilibrium. The introduction of a point charge or a counter-ion using DFT+NEGF is not straightforward since the results are sensitive to their position and to the potential incompleteness of their charging (i.e., leading to a fractional charge on the counter-ion). These aspects are often overlooked in the literature. Following the work of Baghernejad and coworkers, we have introduced a constant energy shift to the diagonal elements of the Hamiltonian localized on the Br ("*atomic shift*" option in QuantumATK[61]) in the scattering region to impose a full -$e$ charge.[65] The value of this energy shift, -3 eV, was optimized to converge to a fully charged Br$^-$. The position of the counterion also affects the electronic structure. We investigated this using the computational results obtained for the isolated [Fe(cyclam)(C≡C-Ph-SH)$_2$]$^+$ as target values, i.e., a +$e$ charge, a magnetic momentum equal to 1 $\mu_B$, and energy and spatial extension of the molecular levels around the Fermi energy. The Br was placed in a symmetrical position with respect to the conjugated backbone, 3.34 Å from the closest H atom of the cyclam ligand. To ensure the preservation of the electronic structure of the isolated [Fe(cyclam)(C≡C-Ph-SH)$_2$]$^+$ around the Fermi energy, a down shift of the p orbitals of the Br was further performed by adding an extra term to the exchange-correlation functional (Hubbard term of -10 eV).[72,73] The comparison of the resulting electronic structure of [Fe(cyclam)(C≡C-Ph-SH)$_2$][Br] compared to that of [Fe(cyclam)(C≡C-Ph-SH)$_2$]$^+$ is shown in Figure S36. This methodology does not rely on the simple inclusion of a counter-ion. It can be described as an electron trapping methodology to generate a +$e$ charge since orbital interactions are prohibited, and is thus equivalent to a calculation on a charged system. The charged background produces a non-zero electrostatic potential in the vacuum region. It has to be noted that XPS measurements do not reveal any presence of counterions in **Fe-cyclam** SAM. Such counterion absence for a charged grafted-molecule surface was previously evidenced.[74]

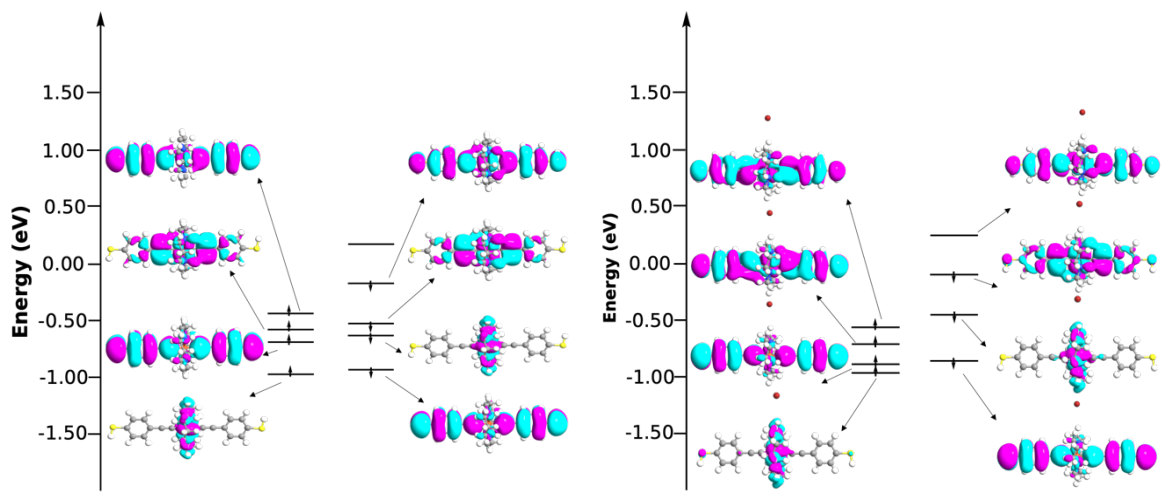

Figure S36. Molecular orbital diagrams of [Fe(cyclam)(C≡C-Ph-SH)$_2$]$^+$ (left) and [Fe(cyclam)(C≡C-Ph-SH)$_2$][Br] (right). The isovalues of the contour plots of the molecular orbitals (MOs) are ±0.02 (e/Å$^3$)$^{½}$.

This protocol was then implemented in the molecular junction configuration leading to the electronic structures shown in Figure S37.

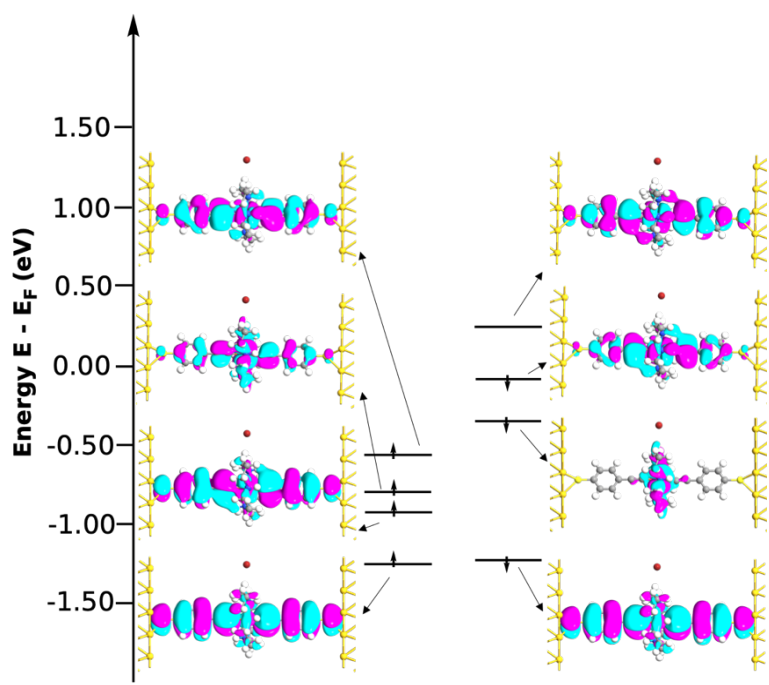

Figure S37. Electronic structure of the **Fe-cyclam$^{+1}$** junction. The isovalues of the contour plots of the obtained from a molecular projected self-consistent Hamiltonian (MPSH) are ±0.02 (e/Å$^3$)$^{½}$.

**k-sampling and transmission instability.** The incompleteness of the k-sampling leads to non-convergence problem of the transmission functions for **Ru-dppe** optimized junctions in tip configuration above the Fermi level (for both RPBE-GGA and HSE06 functionals). To solve the problem, we increased the density mesh cut-off to 200 Ha and k-sampling to (9 × 9 × 50); for the electronic transport simulation, we increased the k-sampling up to the limit of memory of our computers (23 × 23 × 1) without noticeable improvement.

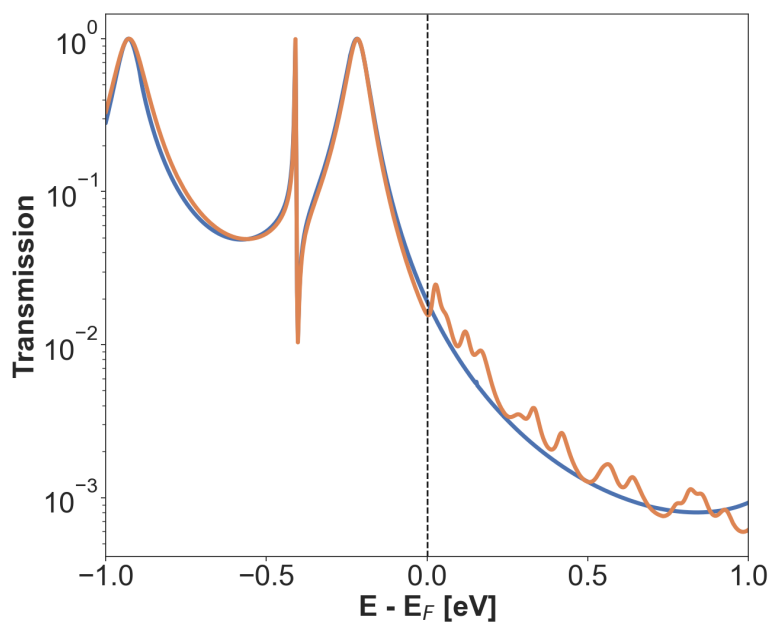

Figure S38. Transmission function for **Ru-dppe** junction using tip Au electrodes using the HSE06 exchange-correlation functional. Orange curve: transmission calculated with a k-point sampling of 9 × 9 × 50. Blue curve: transmission at the $\Gamma$ point.

The electronic transmission depends on the level energy and k-point in the transversal direction A and B of incoming electrons from the electrodes at a given energy ($T(E, k)$), which implies the possible need for a specific *k*-sampling for each energy. The time reversal symmetry allows us to do the calculation for only half of the Brillouin zone. The non-convergence in transmission for the **Ru-dppe** junction with tip-type Au electrodes using HSE06 exchange-correlation functional starts at +0.0263 eV above the Fermi level (Figure S38). In order to check the value of the transmission coefficient at each *k*-point, we performed a 50 × 50 *k*-point mapping that we compared with a 9 × 9 grid, as shown in Figure S39. Interestingly, it reveals that the transmission coefficient at k-points close to the ends of the reciprocal lattice vectors are two orders of magnitude higher than the others. This leads to artefacts when averaging the transmission with same weight for each k point. Indeed, the transmission coefficient at ($E-E_F$) = +0.026 eV is 2.47 x $10^{-2}$ using a 9 × 9 *k*-point grid while it is 1.48 × $10^{-2}$ using a 50 × 50 k-point grid. Similarly, the transmission coefficient at the Fermi energy is identical using only $\Gamma$ point calculation or using a 50 × 50 *k*-point grid.

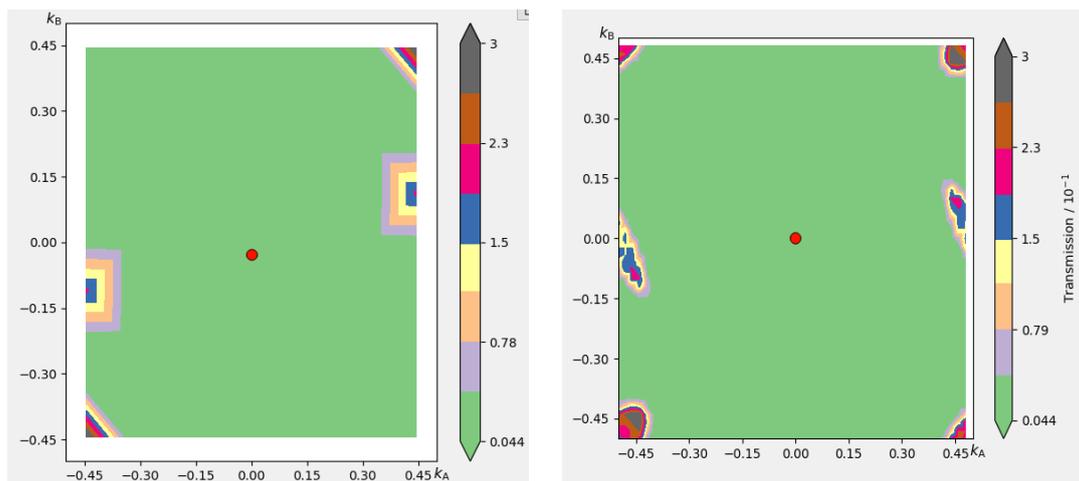

Figure S39. Transmission coefficient k-point maps for transmission of **Ru-dppe** junction using tip Au electrodes and HSE06 exchange-correlation functional at ($E$-$E_F$) = 0.0263 eV with a 9 × 9 *k*-point grid (left) and with a 50 × 50 *k*-point grid (right). The k-points $k_A$ and $k_B$ are given as fractional coordinates. The transmission at the Γ point, indicated by the red dot, is 1.48 × 10$^{-2}$.

**Evolution of thermoelectric parameters upon Fermi level energy shift**

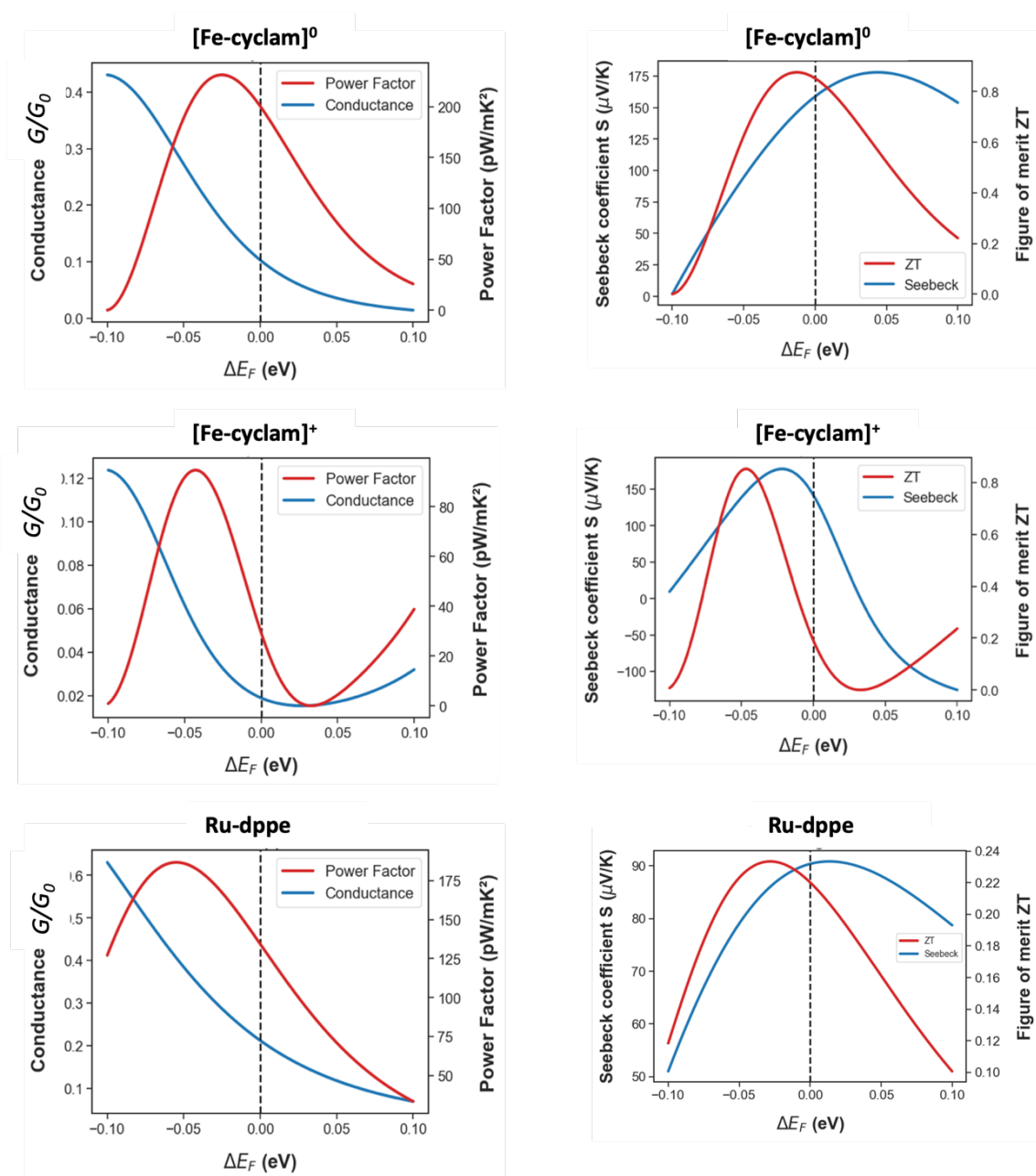

Figure S40. Evolution of the conductance, Seebeck coefficient, power factor, and figure of merit calculated in the flat-type electrode configuration using GGA-RPBE exchange correlation functional as a function of energy shift of the transmission spectrum with respect to the Fermi level by $\Delta E_F$ for **[Fe-cyclam]$^0$**, **[Fe-cyclam]$^+$** junctions ($\kappa_{ph}$ = 20 pW/K), and the **Ru-dppe** junction ($\kappa_{ph}$ = 50 pW/K).

**Work function shift.** The experimental work function value of the Au(111) surface is 5.33 ± 0.06 eV on the basis of Derry and co-workers' critical review of 2015.[75] The grafting of molecules on top of a surface modifies the work function due to the creation of an interface dipole, the intrinsic molecular dipole of the molecule and image charge effects.[76] These shifts were estimated experimentally (see Figure S13-b). The work function shift values associated with the **Fe-cyclam** and **Ru-dppe** SAM were simulated by comparing the values obtained with and without grafted molecules using a slab configuration calculation (Figure S41). The boundary conditions used below the surface are of von Neumann type. Periodic boundary conditions are imposed in the A and B directions (5 × 5 Au atoms for **Fe-cyclam** slabs and 7 × 7 for **Ru-dppe**). In the C direction, the unit cell is formed by 9 Au layers (∼ 10 Å) and separated by 30 Å from the next cell. The calculations were done with the GGA-PBE exchange-correlation functional, SZP basis set for Au and DZP for the other atoms using 7 × 7 × 1 *k*-point sampling and Fast Fourier 2D Poisson Solver. The molecules in their optimized geometry were grafted perpendicularly to the surface. The coordinated S atom is located in a hollow configuration with an initial distance of 1.76 Å from the Au surface. Geometry optimizations were performed for **[Fe-cyclam]$^0$** and **Ru-dppe** SAMs by fixing the Au atoms of the 6 bottom layers and relaxing the 3 top layers. The grafted molecules keep a configuration parallel to the normal of the surface upon optimization. The area of the surface is 1.08 nm$^2$ for a **Fe-cyclam** molecule and 2.52 nm$^2$ for **Ru-dppe** (Figure S41).

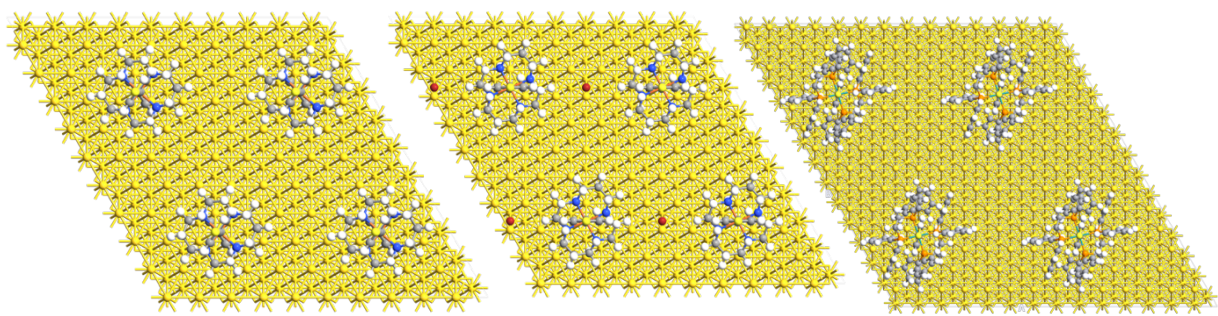

Figure S41. Representation of 2 by 2 simulation cells of (left) **[Fe-cyclam]$^0$**, (middle) **[Fe-cyclam]$^+$** and (right) **Ru-dppe** SAMs.

Ghost atoms covering the surface, including the molecule when present, were added to obtain more reliable work function values following the procedure detailed in the supporting information of the paper of C. Van Dyck and J. Bergreen.[77] The calculated work function for Au was found equal to -5.26 eV in the range of the measured experimental values. The work function shift associated to the intrinsic dipole is negligible for molecules showing a *pseudo-C$_i$* symmetry (at the metal atom center) and because the grafting does not change the charge distribution (no polarization). The charged background implemented in QuantumATK in slab configuration prohibits the calculation of work function. The treatment of counter-ion that we have built for charged molecular junction was thus applied for **[Fe-cyclam]$^+$** slab. We checked that the Hartree difference potential (HDP) zeroes at the boundary of the unit cell above the surface for the three systems to ensure the correspondence between work function and chemical potential (Figure S42). The presence of adsorbed molecules on the surface shifts down the work function by 2.01 eV, 1.05 eV, and 1.34 eV for **[Fe-cyclam]$^0$**, **[Fe-cyclam]$^+$**, and **Ru-dppe**. We can observe that **[Fe-cyclam]$^0$** shifts more the work function of the metal than **Ru-dppe** but it is the reverse for **[Fe-cyclam]$^+$**. Experimentally, the measured shifts are less pronounced but the same trends are observed with 0.40 eV for the **Ru-dppe** junction and 0.55 eV for **Fe-cyclam** junction (composed of 2/3 **[Fe-cyclam]⁰** and 1/3 of **[Fe-cyclam]⁺**, i.e., a rough estimation considering a weighted average value gives a theoretical shift of 1.69 eV).

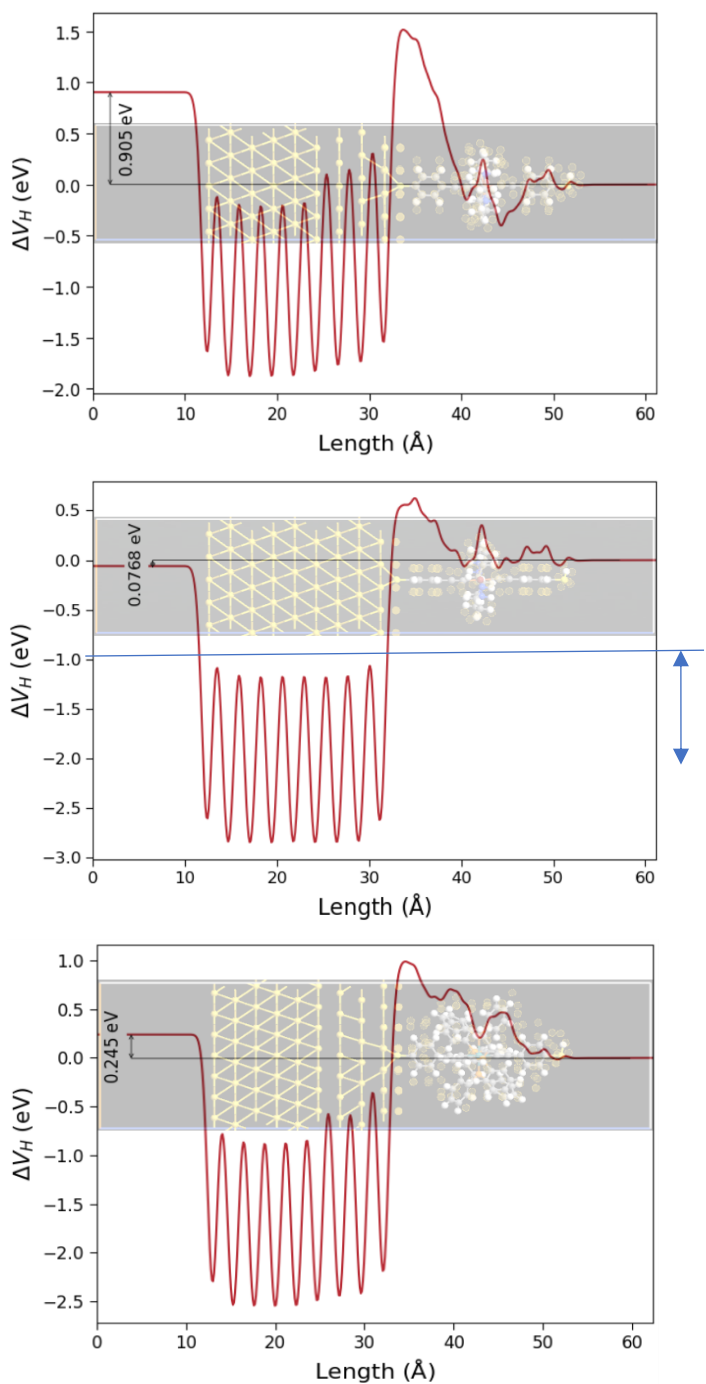

Figure S42. Electrostatic potential profile along the C direction of the cell for **[Fe-cyclam]⁰** (top), **[Fe-cyclam]⁺** (middle), **Ru-dppe** (bottom) slabs.